\documentclass{article}  
\pdfoutput=1
\usepackage[toc,page]{appendix}

\usepackage{float}
\usepackage{subfig}
\usepackage{graphicx}
\usepackage[T1]{fontenc}
\usepackage[utf8]{inputenc}
\usepackage{authblk}
\usepackage{amsmath,bm}
\usepackage{amssymb}
\usepackage[left=3cm,right=3cm,top=2cm,bottom=2cm]{geometry}
\usepackage{verbatim}
\usepackage[dvipsnames]{xcolor}
\usepackage{bbm}
\usepackage[normalem]{ulem}

\graphicspath{{./plots/}}

\renewcommand\footnotemark{}

\newcommand{\rf}[1]{(\ref{#1})}
\newcommand{\beq}{\begin{equation}}
\newcommand{\beql}[1]{\beq\label{#1}}
\newcommand{\eeq}{\end{equation}}
\newcommand{\bea}{\begin{eqnarray}}
\newcommand{\eea}{\end{eqnarray}}
\newcommand{\fmod}{\mathrm{mod}}

\newcommand{\rmd}{\mathrm{d}}
\newcommand{\rme}{\mathrm{e}}
\newcommand{\rmi}{\mathrm{i}}
\newcommand{\rmk}{\mathrm{k}}

\newcommand{\const}{\mathrm{const.}}

\newcommand{\bA}{\mathbf{A}}

\newcommand{\bC}{\mathbf{C}}
\newcommand{\bD}{\mathbf{D}}
\newcommand{\bH}{\mathbf{H}}
\newcommand{\bK}{\mathbf{K}}
\newcommand{\bL}{\mathbf{L}}

\newcommand{\bP}{\mathbf{P}}

\newcommand{\cT}{\mathcal{T}}

\newcommand{\bI}{\mathbbm{1}}

\newcommand{\ph}{\phi}

\begin{document}

\title{Scalar fields in Causal Dynamical Triangulations}
\author[a,b]{J.~Ambj\o rn}
\author[c]{Z.~Drogosz}
\author[c]{J.~Gizbert-Studnicki}
\author[c]{A.~G\"orlich}
\author[c]{J.~Jurkiewicz}
 \author[c]{D.~N\'emeth}
 \affil[a]{{\small The Niels Bohr Institute, Copenhagen University, \authorcr Blegdamsvej 17, DK-2100 Copenhagen Ø, Denmark. \authorcr E-mail: ambjorn@nbi.dk.\vspace{+2ex}}} 

\affil[b]{{\small IMAPP, Radboud University, \authorcr Nijmegen, PO Box 9010, The Netherlands.\vspace{+2ex}}}

\affil[c]{{\small Institute of Theoretical Physics, Jagiellonian University, \authorcr \L ojasiewicza 11, Krak\'ow, PL 30-348, Poland. \authorcr Email: zbigniew.drogosz@doctoral.uj.edu.pl, jakub.gizbert-studnicki@uj.edu.pl, andrzej.goerlich@uj.edu.pl, jerzy.jurkiewicz@uj.edu.pl, nemeth.daniel.1992@gmail.com.}}

\date{{\small (Dated: \today)}}
\maketitle


\begin{abstract}
{\normalsize
A typical geometry extracted from the path integral of a quantum 
theory of gravity might be quite complicated in the UV region.
Even if such a configuration is not physical, it may be of interest
to understand the details of its nature, since some universal 
features can be important for the physics of the model. If the formalism 
describing the geometry is coordinate independent, such understanding 
may be facilitated by the use of suitable coordinate systems. In this 
article we use scalar fields that solve Laplace's equation 
to introduce coordinates 
on geometries with a toroidal topology. 
Using these coordinates we observe what  
we denote as the ``cosmic voids and filaments'' structure, even if no matter is present in the theory. We also show that if the scalar fields we used as coordinates are dynamically coupled to geometry, they can change it
in a dramatic way.}

\end{abstract}

\section{Introduction}\label{sec:Intro}

Lattice approaches based on the path integral formalism constitute an important tool 
with which one can investigate non-perturbative aspects of many quantum field theories. The general idea is the following:  given a continuum field theory with a classical action, one defines a quantum theory via the (lattice regularized) path integral, where the length of lattice links provides a natural ultraviolet (UV) cut-off. A continuum quantum field theory might then be defined if there exists a so-called UV fixed point such that it is possible to keep the physical observables fixed while taking the lattice spacing to zero. Although this idea is quite simple, there is a number of practical issues and open questions which need to be addressed, especially when trying to apply this approach to the quantization of Einstein's General Relativity (GR):

\begin{itemize}
 \item[(1)] GR is perturbatively non-renormalizable. Thus, it is not clear that GR exists as a quantum field theory with a 
 well-defined UV limit.
 \item[(2)] The quantum theory of GR should be formulated in a diffeomorphism-invariant way: so how to define geometric degrees of freedom on the lattice and how then to relate lattice measurements to other, more analytical approaches? 
 \item[(3)] Studies of a lattice theory usually require the use of numerical Monte-Carlo (MC) 
 methods, which is technically possible only in spacetimes with Euclidean signature. 
 Although it is known how to relate correlation functions calculated in \emph{flat} spacetimes with Euclidean and Lorentzian signatures (the so-called Osterwalder-Schrader axioms), nothing like that is known when GR is involved. 
 \item[(4)] A realistic quantum theory of gravity should also include coupling to quantum matter fields – what types of fields can and should be included in this approach? Furthermore, what impact do the matter fields have on the underlying geometric degrees of freedom?
 \end{itemize}
Let us briefly answer these questions.
\begin{itemize}
\item[(1)] It is well known that Einstein's gravity as a perturbative field theory is non-renormalizable \cite{nonrenorm}. However, as suggested by S. Weinberg's asymptotic safety conjecture \cite{weinberg},
it may be renormalizable in a non-perturbative way. A necessity for such a scenario is that the renormalization group flow of the gravitational coupling constants can lead to a nontrivial ultraviolet fixed point (UVFP). Some evidence of such an UVFP is provided by calculations in $2+\epsilon$ dimensions \cite{kawai} and from the use of the so-called exact renormalization group \cite{reuter,reuteretc}, but none of the methods have yet provided us with a generally accepted proof that such a fixed point exists. Thus, one of the aims of studying a lattice theory of quantum gravity is to test the asymptotic safety conjecture. In the lattice formulation, the UVFP should be associated with a second- or higher-order phase transition point. In addition, it should be possible to define the renormalization group flow lines in the lattice coupling constant space leading from an infrared limit to the UVFP. This in general requires finding a region in the lattice coupling constant space where the semiclassical limit (consistent with the classical GR) can be defined, 
together with some physical observables.
These physical observables should be such that keeping their values fixed defines a path in the lattice coupling constant space that allows the interpretation of a decreasing lattice spacing when moving away from the semiclassical region. If the lattice spacing goes to zero at the endpoint of the path, this endpoint will be an UVFP. The Causal Dynamical Triangulations (CDT) approach (described in more detail in Section \ref{sec:CDT}) has at least some of the required features of a successful lattice field theory in the sense described above, i.e., it has a semiclassical region in the lattice coupling constant space \cite{semiclassical,c-phase1,c-phase2}, while some of the boundaries of the semiclassical phase are higher order phase transition lines / points \cite{phasetransition2, scalar}. One can define and measure the renormalization group flow lines \cite{cdtrg} in the lattice coupling constant space, however it has not yet been possible to define a suitable continuum limit; it is not ruled out that it will be possible in the future, using better observables (see \cite{cdtrg1} for a more detailed discussion of this issue). 

Although the existence of the UVFP in a lattice theory of quantum gravity is still a conjecture, it can nevertheless be argued that even if the continuum limit were not to exist, the lattice theory would still be useful in investigating non-perturbative aspects of quantum gravity, treated as an effective theory valid up to some finite energy scale. A simple example of such a situation goes all the way back to the first proof of confinement in a gauge theory, where Polyakov showed that three-dimensional compact $U(1)$ lattice theory contained all the non-perturbative physics responsible for the confinement in the Georgi-Glashow model, despite having itself no such non-perturbative continuum limit \cite{polyakov}. 

\item[(2)] One of the key assumptions of GR is the diffeomorphism invariance, i.e., invariance under arbitrary differentiable coordinate transformations. In his seminal work \cite{regge}, Regge provided a prescription for how to assign local curvature to piecewise linear (simplicial) geometries without the use of coordinates. That formulation is manifestly coordinate free and thus diffeomorphism invariant. In that approach, the geometry of a piecewise linear (simplicial) manifold and the resulting Regge action $S_R$ (the Einstein-Hilbert action $S_{EH}$ for the triangulated manifold) are entirely determined by geometric quantities such as the length of edges (links) and the adjacency relations of the d-dimensional simplices glued together to form the manifold. Regge's idea was to describe simplicial discretizations of classical continuously differentiable manifolds with arbitrary precision in a coordinate-independent way. However, the classical theory of Regge is not easily transferred to the path integral of the corresponding quantum theory \cite{as}. A more suitable lattice path integral over \emph{Euclidean} geometries is known as Euclidean Dynamical Triangulations (EDT).\footnote{The use of EDT goes back to attempts to provide a regularization of the bosonic string theory \cite{bosonic-string}, which can be viewed as 2D gravity coupled to Gaussian fields. It was first used in the context of higher dimensional gravity in \cite{3dEDT,4dEDT}. } In this approach, the simplicial manifolds used in the path integral are obtained by gluing together identical four-simplices whose links have length $a$, the UV cut-off in the lattice theory. The geometry of such a manifold is the piecewise linear geometry defined by Regge, and the action associated with such a configuration is the Regge action associated with the piecewise linear geometry. An important feature of the EDT formalism is that each triangulation in the EDT ensemble corresponds to a different geometry, and the basic assumption is that as the link distance $a \to 0$, the EDT ensemble of geometries becomes dense in some suitable way in the set of continuous geometries that appears in the continuum path integral. This seems to be true in two-dimensional quantum gravity where both the continuum theory and the lattice theory can be solved analytically and they agree (see \cite{book} for a review). In higher-dimensional quantum gravity, we do not know if this is true since the continuum path integral has not been rigorously defined and the EDT theory of gravity can only be studied via numerical simulations. If the asymptotic safety scenario discussed above is valid, one should in principle be able to shrink the lattice spacing (the size of the elementary simplicial building blocks) to zero, and thus to get rid of the discretization and recover the continuum limit of the putative quantum theory of gravity. In this limit one could in principle compute expectation values of correlators of some physical observables, although they are not so easily 
defined in a theory of quantum gravity without matter fields. One ``problem'' is that the EDT formalism is ``coordinate free''. While this seems a major achievement from a GR point of view, it comes with its own issues. One of these is that it makes it difficult to relate the results obtained in the lattice theory to more analytical approaches where coordinate systems are used (even if physics of course should be independent of a specific coordinate system). The issue of reintroducing suitable coordinate systems in the lattice theory of gravity has been extensively studied recently by our group \cite{coordinates1,coordinates2}, and in this article we will discuss a new promising way of doing it by using scalar fields – see Section \ref{sec:class}.
 
\item[(3)] 
The formulation of the EDT lattice field theory of (Euclidean) 
quantum GR is simple. The path integration over continuous Euclidean 
geometries is replaced by the summation over the EDT piecewise linear 
geometries. If we consider GR in $d$ dimensions, each such piecewise 
linear geometry is described by an abstract triangulation, and we thus 
obtain a summation over abstract $d$-dimensional triangulations, each 
with the Boltzmann weight given by the Regge action of the corresponding 
piecewise linear geometry. Thus we write 
\beql{PI0}
{\mathcal Z}_{QG} = \int\mathcal{D}_\mathcal{M}[g_L]\;\rme^{\rmi S_{EH}[g_L]} \to
\int\mathcal{D}_\mathcal{M}[g_E]\;\rme^{-S_{EH}[g_E]} \to {\mathcal Z}_{EDT} 
= \sum_\mathcal{T} \rme^{- S_R[\mathcal{T}]},
\eeq
where the first path integral is over geometries with Lorentzian signature and the second path integral is over geometries with Euclidean signature. $S_{EH}[g]$ denotes the Einstein-Hilbert action, and $S_R[\mathcal{T}]$ is the Regge action of the triangulation $\mathcal{T}$. While it is easy to define $Z_{EDT}$, it can be calculated analytically ``only'' in two dimensions. As mentioned above, the very encouraging outcome is that the continuum limit can be taken, and the resulting theory agrees with the continuum two-dimensional Euclidean quantum gravity theory (the so-called quantum Liouville theory), which can also be solved analytically. In higher dimensions the best one can do is to study the theory using Monte Carlo simulations. The model has been studied extensively in three and four dimensions \cite{3dEDT,4dEDT}, together with generalizations where matter fields were added to the action \cite{matterEDT}. However, no suitable UVFP was found \cite{firstorderEDT}.\footnote{Recently attempts have been made to find higher order transitions in generalized EDT models \cite{jack-us}, but so far with no clear success.} This failure led to a reformulation of the model, with the Lorentzian starting point of GR taken more seriously \cite{fundamentalcdt}. In this approach, denoted Causal Dynamical Triangulations (CDT), the starting assumption is that the continuum path integral should include only Lorentzian geometries that are globally hyperbolic. To regularize the path integral, a discretization based on building blocks ($d$-dimensional simplices), similar in spirit to EDT, is introduced. Now each $d$-dimensional simplex has space- and timelike links. Moreover, it is possible to perform a Wick rotation of each simplex to an ``Euclidean'' simplex, and the triangulation built from Lorentzian simplices is then analytically Wick-rotated to an Euclidean triangulation, with the Regge action of the triangulation changed accordingly. The change from Lorentzian geometries alluded to in (\ref{PI0}) thus becomes a real analytical continuation, and we can write
\beql{PI}
{\mathcal Z}_{QG} = \int\mathcal{D}_{\mathcal{M}_H}[g_L]\; \rme^{\rmi S_{EH}[g_L]}
\to
{\mathcal Z}_{CDT} = \sum_{\mathcal{T}_L} \rme^{\rmi S_R[\mathcal{T}_L]} \to \sum_{\mathcal{T}_E} \rme^{- S_R[\mathcal{T}_E]},
 \eeq
where $\mathcal{M}_H$ denotes globally hyperbolic geometries, $\mathcal{T}_L$ a corresponding Lorentzian triangulation, and $\mathcal{T}_E$ the Wick-rotated Euclidean triangulation. When we talk about $Z_{CDT}$ below, we will always have in mind the summation over Euclidean triangulations in (\ref{PI}), but contrary to the situation in EDT shown in (\ref{PI0}) there is now a clear relation between the Lorentzian and the Euclidean theory. However, it comes at the price of introducing a preferred foliation of the triangulated manifolds, which may be incompatible with general 4D spacetime diffeomorphism invariance.\footnote{In this case full 3D (spatial) diffeomorphism invariance remains, but the time direction is distinguished and treated on a special footing.} The question remains whether introducing such a foliation can be treated as a specific gauge choice in a quantum version of GR or if it would rather make CDT fall into some other universality class of quantum gravity theories, e.g., Ho\v{r}ava-Lifshitz gravity \cite{horava}. {Ho\v{r}ava-Liftshitz-gravity is indeed
 a natural candidate for a continuum limit of CDT, since also in this theory there is a time foliation. One can show analytically that two-dimensional CDT corresponds to a quantum version of two-dimensional Ho\v{r}ava-Lifshitz gravity \cite{agsw}, but for higher-dimensional gravity the 
 situation is much less clear since the Ho\v{r}ava-Lifshitz gravity in higher dimensions contains important 
 action terms that are not GR-terms and are
 not included in the CDT action. In three dimensions there
 is some evidence that the physics of the CDT model does not 
 depend in a crucial way on the existence of a time 
 foliation \cite{rj1}. In four dimensions it has not yet been possible to address this question. However, one step 
 in this direction is at least to be able to talk about 
 different time-foliations of the same CDT four-geometry, and to
 check if and how the results depend on the choice of foliation.}
In Section \ref{sec:foliations} we make a first step towards this goal by showing how to use scalar fields to define alternative spacetime foliations for the CDT triangulations. 

\item[(4)] Last but not least, a realistic theory of quantum gravity should not only describe the pure gravity sector but also investigate the impact of quantum matter coupled to geometric degrees of freedom. There are no technical problems associated with the introduction of bosonic matter coupled to the geometry in CDT. That was done already in EDT, as mentioned above \cite{matterEDT}, and the same discretized prescriptions as used there can be applied in CDT. While matter did not have a great impact in EDT, the situation is potentially much more interesting in CDT, where there {\it are} second order phase transitions and thus probably some kind of continuum physics of geometry, which could be influenced in important ways by matter and vice versa. So far, interesting results were obtained for simple 2D CDT models coupled to scalar \cite{2dscalar} and gauge \cite{2dgauge} fields, where matter fields seemingly have a significant impact on the geometry. As regards the more interesting but also more complicated four-dimensional CDT model, we have recently analyzed systems with (multiple copies of) massless scalar fields coupled to the geometry \cite{scalar}, and we have also studied point particles (mass lines).\footnote{Results of the mass line studies will be published in a separate article.} Disappointingly, our previous results did not show any substantial impact of the scalar field(s) on spacetime geometry nor the position of phase transition lines in the CDT coupling constant space. In the present study, we investigate the impact of introducing nontrivial boundary conditions for the scalar field(s), such that the field jumps on the boundary of a periodic elementary cell, which in our setup can be defined. Our formulation is topological, i.e., the matter action does not depend on the specific (unphysical) position of the boundary but just on the value of the jump. Such systems seem to undergo a new type of phase transition where spacetime geometry dramatically changes for large values of the jump vs the (almost pure gravity) geometry observed for small values of the jump; see Section \ref{sec:dynamical} for details.
\end{itemize} 
The remaining part of the article is organized as follows: in Section \ref{sec:CDT} we outline the CDT approach to quantum gravity; in Section \ref{sec:class} we discuss how classical scalar fields can be used to define coordinates in fixed simplicial geometries, and how 
they in turn help better to understand the geometric structures observed in CDT triangulations; in Section \ref{sec:foliations} we describe how the classical scalar fields can serve as a tool to define alternative proper-time foliations of the CDT manifolds; finally in Section \ref{sec:dynamical} we analyze the impact of dynamical scalar fields with non-trivial boundary conditions. 

\section{Causal Dynamical Triangulations}\label{sec:CDT}

As already mentioned in the introduction, CDT is a background-independent and diffeomorphism-invariant lattice field theory aiming at 
providing a non-perturbative definition of quantum gravity. Below we provide for completeness a short description of the 
actual lattice construction of the geometries. For a complete account, we refer the reader to the review \cite{physrep} (and to \cite{lollreview,toroidalreview} for an update on the recent results). CDT provides a definition of the (formal) continuum gravitational path integral appearing in \rf{PI} as a sum over an ensemble of triangulations $\cal T$ constructed from several types of elementary simplicial building blocks. The edge lengths of the simplices are assumed to be fixed\footnote{In computer simulations we set the length of (spatial) links to be one (in abstract lattice units), and then by performing measurements of certain observables and relating them to a continuous theory we measure the effective lattice spacing in physical units, say Planck lengths ${ \ell}_{Pl}$. For a given set of parameters (CDT bare couplings), the lattice spacing is constant and fixed, but it does change from one point to another in the parameter space (see e.g. \cite{LatSpacing} for more details).} and act as the UV cut-off of the lattice theory. The geometries appearing in 
the formal path integral (\ref{PI}) are by assumption globally
hyperbolic, and the piecewise linear geometries represented by the triangulations are constructed to reflect it: they have spatial hypersurfaces of constant ``lattice time'' $t$, and the construction is such that it is actually possible to perform an analytic continuation in the lattice time $t$ to piecewise linear geometries with Euclidean signature, as alluded to in (\ref{PI}) (see \cite{physrep} for a detailed discussion of the analytic continuation). In the four-dimensional case, which is the one we are the most interested i
n, a spatial 3D geometric state with a given fixed topology in a slice with integer (lattice) time coordinate $t$ is constructed by gluing together equilateral tetrahedra (with fixed length of all edges / lattice links: $a_s$). Similarly, an independent
3D geometry with the same topology is constructed in the spatial slice at time $t+1$. These two 3D geometries are now connected
by 4D simplices filling out the four-dimensional ``slab'' between the two hypersurfaces. This is done by introducing two types of 4D simplices – the $(4,1)$ and the $(3,2)$ simplices\footnote{The $(i,j)$ simplex has $i$ vertices in a spatial slice with integer (lattice) time coordinate $t$ and $j$ vertices in the neighboring spatial slice with $t\pm 1$. } – whose timelike edges (links) have a fixed length $a_t$. In the Lorentzian setting, $a_t^2 = - \alpha a_s^2$, with the asymmetry parameter $\alpha > 0$. The rotation to an Euclidean four-simplex is performed by rotating $\alpha$ to the negative real axis in the lower complex plane (for restrictions on the 
value of $\alpha$ on the negative real axis see \cite{physrep}). Since the four-dimensional simplices are glued together in such a way that no topological defects are introduced in the slab between the three-dimensional triangulations at $t$ and $t+1$, it is possible 
to assign non-integer time and piecewise linear 3D geometries to spatial hypersurfaces between $t$ and $t+1$. This construction is analogously extended to hypersurfaces $t+2$, $t+3$, etc. and the corresponding slabs in between. In the path integral ${\cal Z}_{CDT}$ in (\ref{PI}), the summation is performed over all 3D geometries (of the given topology) at $t=1,2,\ldots$ and all 4D slab geometries connecting them as described. All four-simplices (and their subsimplices) are assumed to be flat (their interior being a fragment of either Minkowski or Euclidean spacetime, depending on whether or not we have performed the analytic continuation). In the Regge prescription, the nontrivial spacetime curvature of the four-dimensional triangulation is localized on the two-dimensional subsimplices, i.e., triangles, and depends on the number of four-simplices sharing a given triangle. Using the Regge prescription \cite{regge}, one can derive the Einstein-Hilbert action for such simplicial geometries, the Regge action $S_R$ mentioned above, which for CDT takes a very simple form after the rotation to Euclidean signature has been made (see e.g.\ \cite{physrep}):
\beql{bareaction}
S_{R}[{\cal T}]=-\left(\kappa_{0}+6\Delta\right)N_{0}+\kappa_{4}\left(N_{4,1}+N_{3,2}\right)+\Delta N_{4,1},
\eeq
\noindent where $N_{i,j}$ denotes the number of four-simplices of the type $(i,j)$ (see above), and $N_0$ is the number of vertices in the triangulation $\cal T$. $\kappa_{0}$, $\Delta$ and $\kappa_{4}$ are bare dimensionless coupling constants, related to Newton's constant, the cosmological constant, and the asymmetry parameter $\alpha$ (see above), respectively. In principle, one could choose some fixed initial (at $t=1$) and final (at $t=T$) 3D geometric states, but for the purpose of this 
article it is convenient instead to impose time-periodic boundary conditions such that a 3D spatial geometry at time $t$ is identified with the geometry at time $t+T$. At present, the only tool we have available to investigate four-dimensional CDT is Monte Carlo simulations. This is a method to generate configurations with a probability distribution in accordance with the Boltzmann distribution dictated by the action of the system. However, to function, it requires a real probability distribution. This is why we have to rotate to geometries with Euclidean signatures in (\ref{PI}), as described. More precisely, our rotation of a configuration ${\cal T}_L \to {\cal T}_E$ is such that $\rmi S_R[{\cal T}_L]~\to~- S_R[{\cal T}_E]$, which implies that the Boltzmann weight $ \rme^{\rmi S_R[{\cal T}_L]} \to \rme^{-S_R[{\cal T}_E]}$. With this analytic continuation to an ensemble of geometries $\{ {\cal T}_E \}$, we can now view ${\cal Z}_{CDT}$ in (\ref{PI}) as a statistical theory of random geometries with Euclidean signature. A special feature of the gravity system is that the volume of spacetime is not fixed but instead is a dynamical variable. In our simulations, this implies that the number of four-simplices is not fixed. For a positive cosmological constant $\Lambda$, the corresponding term in Euclidean Einstein-Hilbert action, $\Lambda \int \rmd^4x \sqrt{g(x)}$, will try to force the spacetime volume to be as small as possible. The same term is present in the discretized Regge action \rf{bareaction}, and it will appear with a Boltzmann weight $\rme^{-\kappa_4N_4 ({\cal T})} $, where $N_4(\cT)=N_{4,1}+N_{3,2}$ is the number of four-simplices in the triangulation ${\cal T}$. This seems to hint that for a positive dimensionless coupling constant $\kappa_4$ there should be very few four-simplices. However, there are many triangulations with a given number $N_4$ of four-simplices. In fact, up to the leading order, the number grows exponentially \cite{aj-expo}, approximately like $\rme^{\kappa_4^c N_4}$. In the MC simulations, we are interested in as large $N_4$'s as possible, and this is achieved by fine-tuning $\kappa_4$ to $\kappa_4^c$ from above. From a practical point of view, it is convenient to keep $N_4$ or $N_{4,1}$ fixed when measuring observables and then to perform the measurements for different values. In addition, this allows us to use powerful techniques of finite-size scaling, borrowed from the study of critical phenomena in statistical physics, to evaluate the behavior of systems of infinite size from those of finite size. It is such techniques that we use to determine the phase diagram and the corresponding phase transitions (for details we refer to the review \cite{physrep}).
\begin{figure}
\centering
\includegraphics[width=0.59\textwidth]{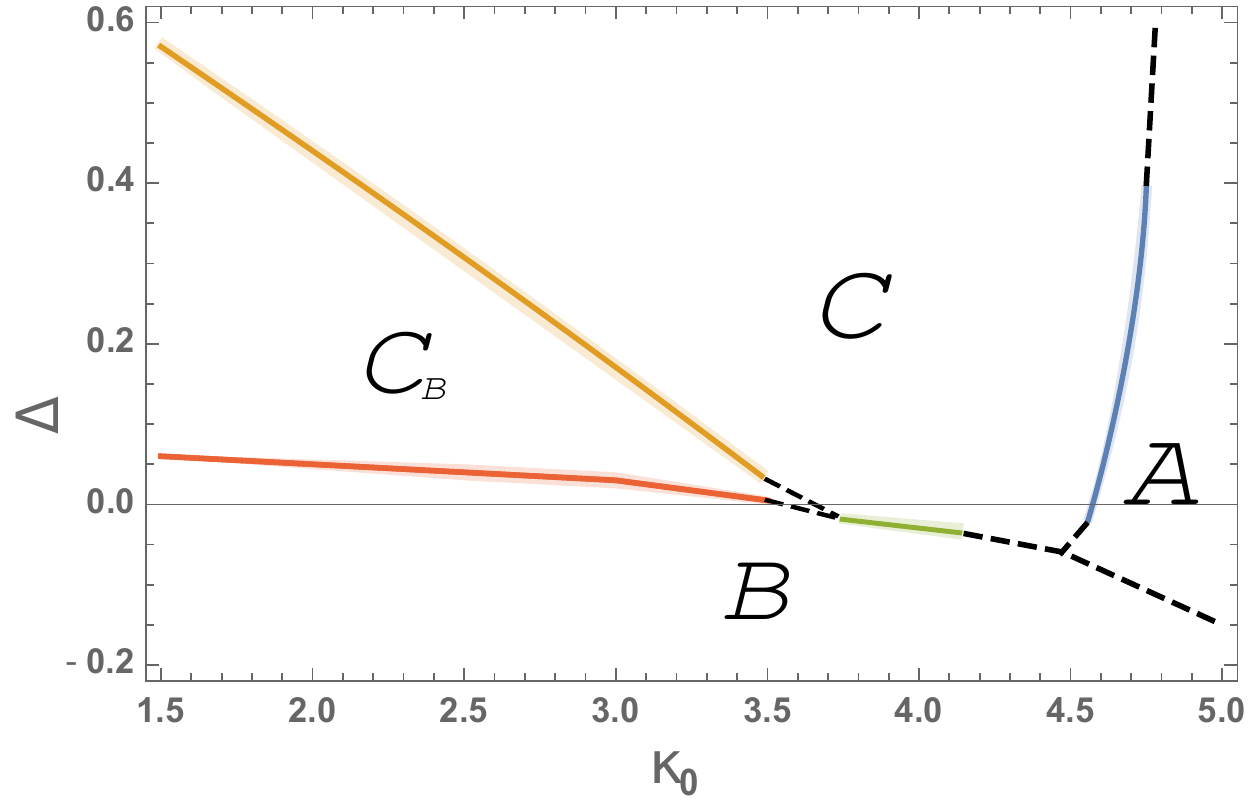}\\
\caption{
The phase structure of four-dimensional CDT in the $(\kappa_0, \Delta)$ parameter space. Blue color denotes first-order and red color higher-order phase transition lines. See footnotes 10-11 for additional remarks.}
\label{fig:phases}
\end{figure}

\begin{figure}
\centering
\includegraphics[width=0.54\textwidth]{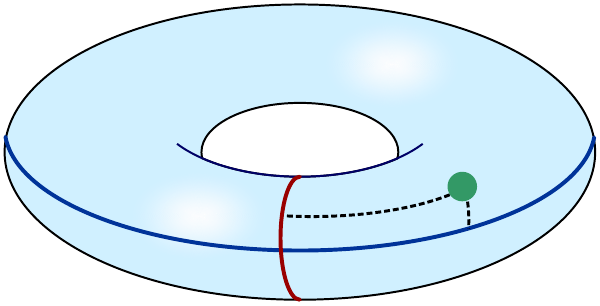}\\
\includegraphics[width=0.27\textwidth]{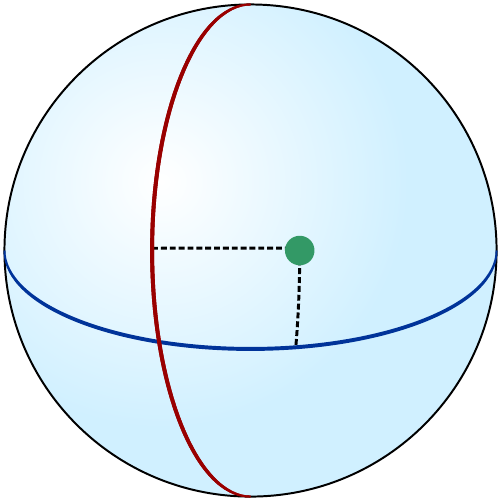}
\includegraphics[width=0.27\textwidth]{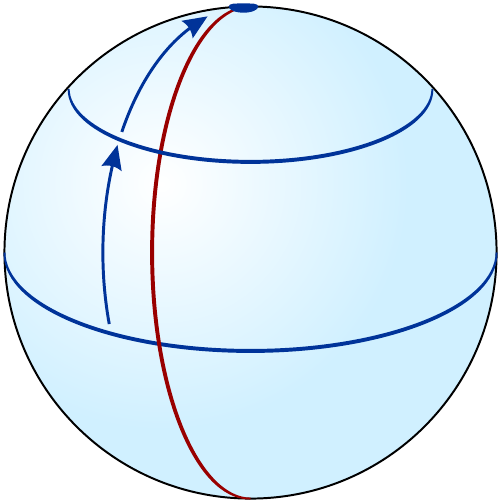}
\caption{ In the 2D toroidal case two orthogonal non-contractible loops can be constructed and used to define coordinates (top chart). This is not possible in a spherical case, where all loops are contractible to a point (bottom chart). 
}\label{fig:spheretorus}
\end{figure}

\begin{figure}
\centering
\includegraphics[width=0.49\textwidth]{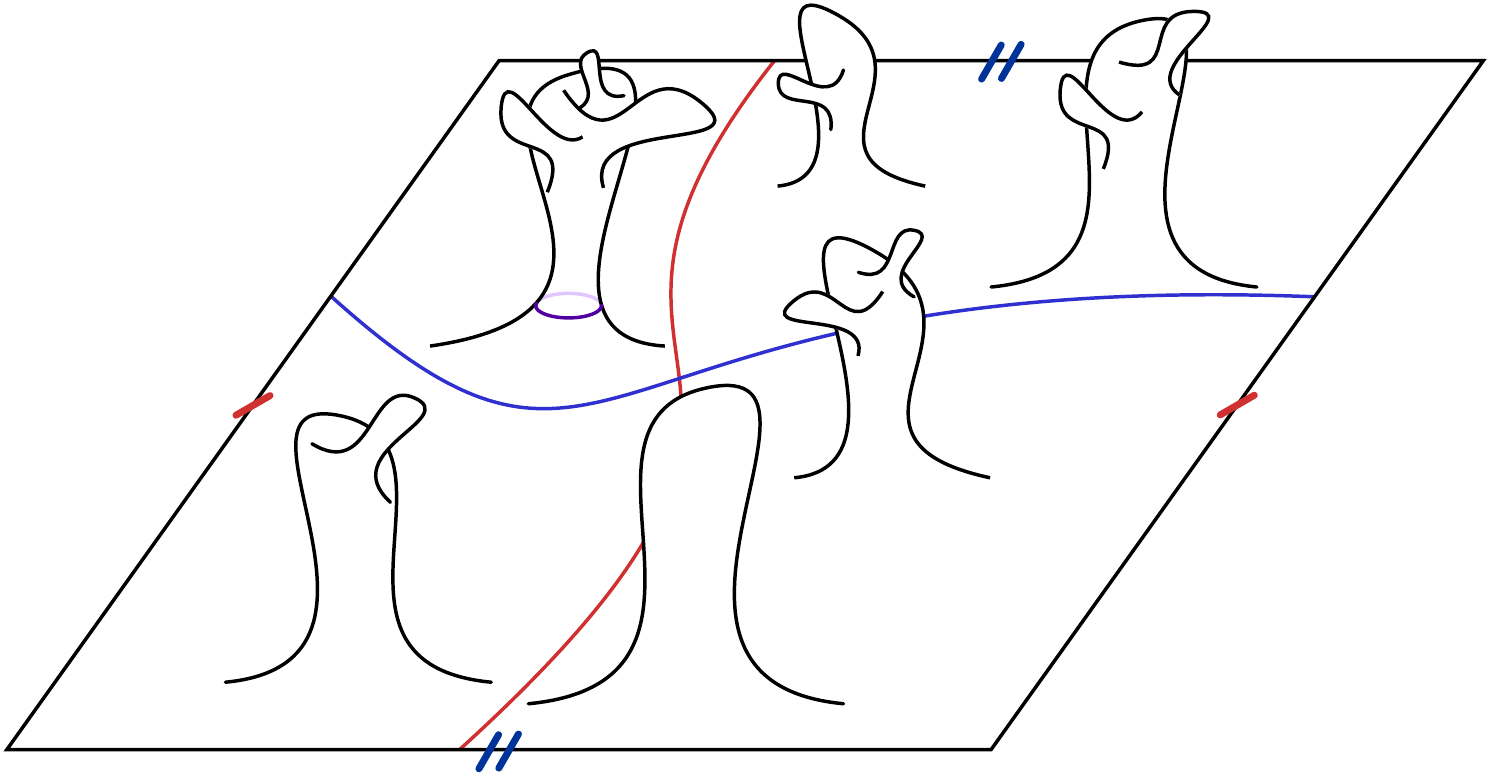}
\includegraphics[width=0.49\textwidth]{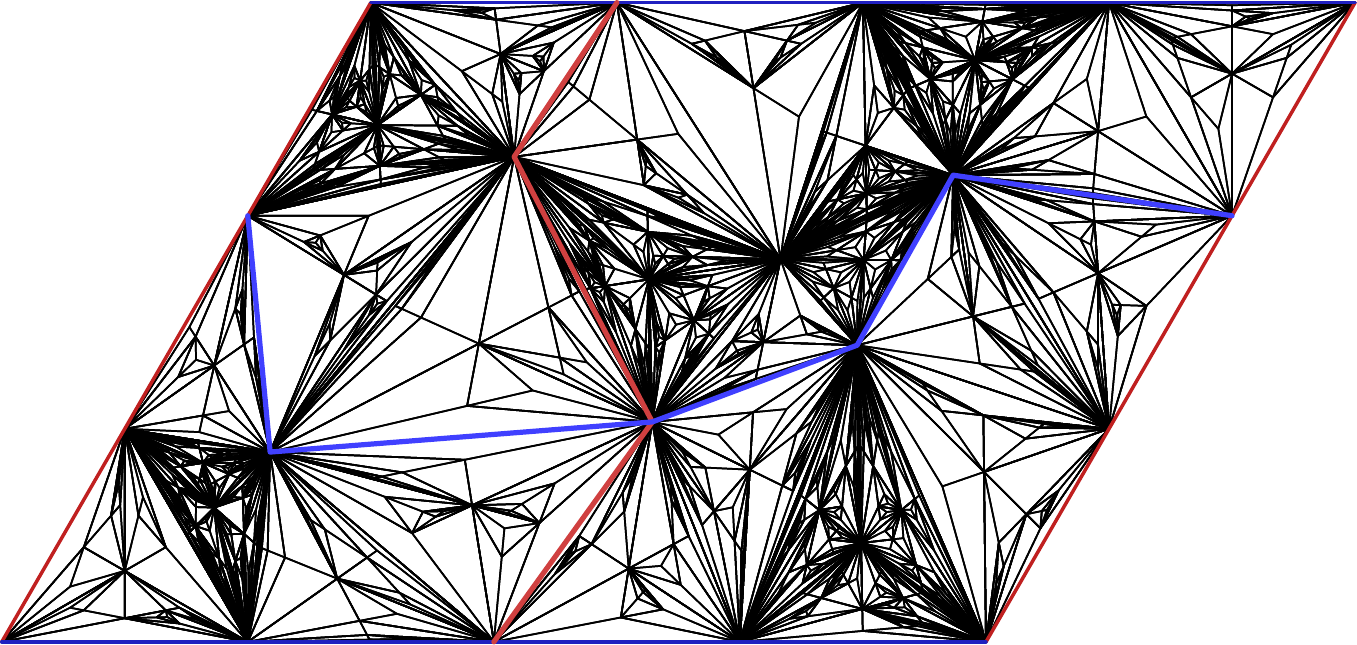}
\caption{
Left: a 2D visualization of a fractal structure of a quantum manifold with sizable outgrowths 
originating from the toroidal center (the boundaries of the rectangular cell are pairwise identified, making it a topological torus). Right: a visualization of 2D toroidal triangulation with outgrowths. In CDT all triangles are assumed to be identical, but a triangulation can be transformed by a conformal map to the regular square lattice with non-identical triangles. The quantum outgrowths are represented by denser regions. For similar pictures coming from  ``real'' computer simulations of 2D quantum gravity see \cite{ab}. }
\label{fig:outgrowths}
\end{figure}

Below we briefly summarize the most important CDT results; for more details we direct the reader to the review articles \cite{physrep,lollreview,toroidalreview}. Despite the relative simplicity of its formulation and the fewness of its parameters (three coupling constants), CDT has a surprisingly rich phase structure, which seems to be independent of the spatial topology choice \cite{phasestorus}.\footnote{So far we have investigated only two cases, namely the spherical $S^3$ and the toroidal $T^3$ topologies.} Four phases of quantum geometry with distinct physical features have been observed for various combinations of the bare coupling parameters $(\kappa_0, \Delta)$; see the phase diagram in figure~\ref{fig:phases}.\footnote{In the Monte Carlo simulations of CDT, the parameter $\kappa_{4}$, which is proportional to the cosmological constant, is tuned so that the infinite-volume limit can be taken (as described above), which effectively leaves a two-dimensional coupling constant space.} At this point it is worth reminding the reader that no background geometry is introduced by hand. So even if the building blocks are four-dimensional simplices, a priori it is in no way clear what kind of geometries will be observed. The experience from the old four-dimensional EDT simulations was that it was close to impossible to obtain something that even vaguely resembled four-dimensional universes. From that point of view it is non-trivial and very encouraging that in one of the phases, the so-called $C$-phase (also called the semiclassical or de Sitter phase) we observe what looks like a four-dimensional universe where the scale factor admits a semi-classical description \cite{semiclassical,c-phase1,c-phase2}. This is different in other phases, called $A$, $B$ and $C_b$, which most likely do not have a good semiclassical interpretation.\footnote{Phases $A$ and $B$ may be realizations of some exotic geometries not observed in the real Universe, and phase $C_b$, also called the bifurcation phase, may be a realization of a quantum spacetime with a singularity, however it has not been proven rigorously.} The four phases are separated by first- ($A-B$, $A-C$ and $B-C$)\footnote{The $B-C$ transition was examined only in CDT with toroidal spatial topology as in the spherical topology it could not be analyzed because of technical issues. It has some properties that may indicate a higher order phase transition and some suggesting a first order transition. This issue has not been completely resolved. The $A-B$ transition is currently examined in CDT with toroidal spatial topology, and it is most likely a first-order transition.} and higher-order ($B-C_b$ and $C-C_b$)\footnote{The order of $C-C_b$ transition was measured only in CDT with spherical spatial topology; in the toroidal case we observe a strong hysteresis in the transition region which may suggest that the order of the transition has changed because of the topology change, but it can be an algorithmic issue as well.} phase transition lines \cite{phasetransition2, phasetransition1},  meeting in two ``triple'' points, which are natural candidates for the UV fixed point of quantum gravity, if it exists. 
A key issue in CDT is how to define good observables, whose expectation values or correlation functions can be measured in the Monte Carlo simulations. One example is the spatial volume distribution in (lattice) proper time. Using this observable, we were able to measure the effective action for the scale factor of CDT, which in phase $C$ is consistent with the (discretized) minisuperspace action of GR \cite{semiclassical,c-phase1,c-phase2}. Some progress towards defining new coordinate-free observables in CDT has recently been made \cite{ObservablesCDT}, but in general it would be beneficial to have a notion of coordinates not only in time but also in spatial directions. They would, for example, be instrumental in measuring a more general effective action of CDT, taking into account not only the scale factor but also the spatial degrees of freedom. They would also help better to understand 
properties of the $C_b$ phase, where spatial homogeneity is strongly broken by very nontrivial geometric structures appearing in generic triangulations. Therefore, we have recently started a research program aimed at restoring spatial coordinates in CDT, whose formulation is ab initio (space-)coordinate free. The choice of a toroidal spatial topology seems convenient for this purpose. In the toroidal CDT, conversely to the spherical case, one can define three (or four, including the time direction) families of 3D surfaces, called boundaries, which are orthogonal to each other and non-contractible in spatial directions; see figure~\ref{fig:spheretorus} for a lower-dimensional visualization.\footnote{In our approach we also require the volume of each such boundary to be (locally) minimal, which seems to lead to three universal boundaries, one in each spatial direction; see \cite{coordinates1} for details.} These boundaries are nonphysical, and their position does not affect the underlying geometries (triangulations) in any way. One of the possibilities is then to use the boundaries as reference frames and to define coordinates by geodesic distances from them \cite{coordinates1}. Such a proposal has some drawbacks as the coordinates are in general dependent on the position of nonphysical boundaries, but it led nevertheless to a better understanding of generic CDT geometries, which in phase $C$ can be described as a semiclassical torus with a number of quantum fractal outgrowths; see figure~\ref{fig:outgrowths}. Another way of analyzing such geometric structures was proposed in \cite{coordinates2}, where the boundaries were used to define the shortest loops (starting at any four-simplex) with nontrivial winding numbers in all three spatial directions and in the time direction. The length of such loops measured in a given geometry (triangulation) is ``topological'' as it does not depend on the position of the boundaries. These concepts led us to the proposal introduced in \cite{LetterClassical}, and discussed in detail in Section \ref{sec:class} below, of using scalar fields as spatial coordinates. 

\section{Classical scalar fields as coordinates in CDT}\label{sec:class}

\subsection{Classical scalar fields}\label{sub:fixed}
The idea of introducing matter fields as coordinates (dynamical reference ``clock-and-rods'' fields) and using them to define relational observables (as functions of the reference fields) is already present in many approaches to gravity \cite{RelationalObs}. Now we want to use a similar concept in CDT. Our CDT configurations come from the path integral. Usually, in the continuum, in order to perform the path integral, we would choose a coordinate system, for instance $x^\mu$, on the manifold defining the whole setup, and we would talk about the equivalence classes of metrics $[g_{\mu \nu}(x)]$ defining the geometry, which would promote the manifold to a Riemannian manifold. In the EDT and the CDT formalism (except for the time-coordinate in CDT), the situation is in a way purified from the GR point of view. No coordinate system is given, only the relations between vertices (belonging to the same link or not, belonging to the same triangle or not,  etc.), and from those data one can reconstruct a coordinate system and, in addition, the geometry. While beautiful from the GR point of view, the lack of a coordinate system has sometimes been quite cumbersome and not very enlightening from the point of view of 
understanding the basic characteristics of the geometries encountered in the path integral. To explore the geometric characteristics of a ``typical'' quantum CDT configuration, i.e., a configuration coming from the path integral, it would be beneficial to have a coordinate system which is ``natural'' for the given geometry. This is what we want to achieve below for typical CDT configurations. The coordinate systems will thus be different for different configurations, contrary to the situation described above, where $x^\mu$ was given from the beginning. To discuss the general principles going into the construction of a coordinate system using scalar fields on a given CDT configuration, let us for a moment use a continuum notation. The topology of the CDT configurations we extract from our MC simulations will be that of $T^4=S^1\times S^1\times S^1 \times S^1$. In principle, we know the geometry of each configuration since we view it as a piecewise linear manifold ${\cal M}$, and from the knowledge of the connectivity of the graph representing the configuration we can reconstruct all distances between points on ${\cal M}$. Let us consider ${\cal M}$ as a Riemannian manifold with the geometry given by some metric $g_{\mu\nu}$ and $T^4$ as a Riemannian manifold ${\cal N}$ with the trivial, flat metric $h_{\alpha\beta}$.
We want to use as our coordinates a ``good'' nontrivial harmonic map ${\cal M} \to {\cal N}$. To define one, we can use four scalar fields $\phi^\alpha$, $\alpha =1,2,3, 4$, $\phi^\alpha (x)$ being a map ${\cal M} \to S^1$ minimizing the action
\beql{SMcont}
S_M[\ph]	= \frac{1}{2} \int \rmd^{4}x \sqrt{g(x)} \;g^{\mu \nu} (x) \; h_{\alpha \beta}(\phi^\gamma(x)) \;\partial_\mu \ph^\alpha(x) \partial_\nu \ph^\beta(x).
\eeq
The choice of the trivial metric $h_{\alpha\beta}$ on ${\cal N}$ reduces equation \rf{SMcont} to four decoupled
equations for the scalar fields $\phi^\alpha$, so for the moment let us concentrate on the scalar field $\phi(x)$ that minimizes \rf{SMcont} and is thus a harmonic map ${\cal M} \to S^1$. The minimization of \rf{SMcont} yields the Laplace equation 
\beql{jan20}
\Delta_x \phi (x) =0, \quad 
\Delta_x = \frac{1}{\sqrt{g(x)}} \;\frac{\partial}{\partial x^\mu} 
\Big(\sqrt{g(x)} \,
g^{\mu\nu}(x)\Big) \frac{\partial}{\partial x^\nu},
\quad \phi(x) \in S^1.
\eeq
If $\phi(x)$ were a scalar field taking values in $\mathbb{R}$, then the constant mode would be the only solution to $\Delta_x \phi (x) =0$ on a compact manifold ${\cal M}$. Thus, here it is important that $\phi(x) \in S^1$. Let the circumference of $S^1$ be $\delta$. One way to force $\phi(x) \in S^1$ is to let $\phi(x)$ take values in $\mathbb{R}$ but to identify $\phi(x)$ and $\phi(x) + n \cdot \delta$. 
We thus write
\beql{jan21}
\phi(x) \equiv \phi(x) + n \, \delta,\quad n \in \mathbb{Z}.
\eeq
The map 
\beql{jan22}
\phi \to \psi = \frac{\delta}{2\pi}\;\rme^{2\pi \rmi \phi/\delta},
\eeq
which maps $\phi$ to a circle in the complex plane, is unchanged by this equivalence. Of course, it is mainly of interest 
in the situation where we have a function $\phi(x)$ that is continuous on the interval $[0,\delta]$ except for a number of jumps that are multiples of $\delta$ as for such a function the corresponding function $\psi(x)$ will be a continuous function on the unit circle provided also $\phi(\delta) - \phi(0) = n \cdot \delta$. The constant mode is still a trivial harmonic map $\phi(x)$ from ${\cal M}$ to $S^1$, but that is clearly an uninteresting choice if we want $\phi(x)$ to act as a coordinate on ${\cal M}$. However, because $\phi(x)$ belongs to $S^1$, we now have other possibilities. Let us illustrate this in the simplest case where ${\cal M}$ is also $S^1$. Then we are considering maps $S^1 \to S^1$, and a solution to \rf{jan20} which winds $k$ times around $S^1$ is simply 
\beql{jan21a}
\phi_k(x) = k \cdot x+c,\quad x\in [0,\delta], \quad k\in \mathbb{Z}.
\eeq
Solutions with different $k$ cannot be deformed continuously into
each other.
Since ${\cal M}$ has the topology of $T^4$, 
we seek a solution to \rf{jan20} with winding number one, and we want
the points $x \in {\cal M}$ satisfying $\phi(x) =c$ to constitute 
hypersurfaces $H(c) \subset {\cal M}$ whose union for $c$ varying in a range of length $\delta$ covers ${\cal M}$.
We now turn to the implementation of this program for triangulations ${\cal T}$ that describe our piecewise linear manifolds ${\cal M}$. \\

In all our previous studies of CDT and also in all cases discussed in the present study, we consider the field $\ph_i$ to be located in the four-simplices and, for the sake of simplicity, we do not distinguish between different simplex types. Therefore, we consider the following discrete counterpart of the continuous action \rf{SMcont} or, more precisely, one of its  components in a given ''direction'':
\beql{SMdis}
S_M^{CDT}[\{\ph\},\cT]	= \frac{1}{2} \sum_{i \leftrightarrow j} (\ph_i - \ph_j)^2 = \sum_{i,j} \ph_i L_{ij} \ph_j \equiv \ph^T \bm L \ph, 
\eeq
where the first sum is over all pairs of neighboring four-simplices and the second sum is over all four-simplices in the triangulation $\cT$. $\bm L$ is the discrete Laplacian matrix. For every four-dimensional triangulation, there are an associated graph and a corresponding five-valent dual graph\footnote{Each four-simplex in a four-dimensional triangulation has exactly 5 neighbors (CDT forbids topological defects, and four-simplices are glued together along all their five 3D faces).} where a vertex corresponds to a four-simplex in the triangulation, and a link denotes a connection between two adjacent four-simplices, i.e., it can be viewed as connecting the centers of the four-simplices across the tetrahedron they share. Given such a dual graph, one can define the $N_4 \times N_4$ symmetric adjacency matrix $\bA$,
\beql{def:adjacency}
A_{ij} =
\begin{cases}
	1 & \textrm{if (the link } i \leftrightarrow j) \in \textrm{dual lattice},\\
	0 & \textrm{otherwise},
\end{cases}
\eeq
where $N_4$ is the number of vertices in the dual lattice or, equivalently, the number of simplices in the original triangulation. Using the dual lattice notation, the Laplacian matrix $\bm L$ in equation \rf{SMdis} can be expressed as 
\beql{lapl}
 \bL = 5 \bI - \bA,
\eeq
where $\bI$ is the $N_4 \times N_4$ unit matrix.
Let us first treat $\phi_i$ as a field taking values in $\mathbb{R}$. Then, a field $\ph_i$ which minimizes the action \rf{SMdis} satisfies the discrete Laplace equation
\beql{eq:laplace}
	\bL \ph	= 0.
\eeq
For any finite triangulation of a compact manifold without boundary, there is a trivial solution: 
\beql{trivialphi}
\ph_i = \const
\eeq
If we project out this zero mode, we can invert the Laplacian matrix (or, in the continuum, the Laplace operator). Thus, if $\phi_i$ is a field taking values in $\mathbb{R}$, the solutions \rf{trivialphi} are the only type of field configurations that minimize \rf{SMdis}. However, as discussed above, we are really interested in fields $\phi_i$ minimizing the action \rf{SMdis} under the constraint that $\phi_i \in S^1$ and that $\phi_i$ winds around $S^1$ once, which allows for new solutions examplified by \rf{jan21a}. Of course, a concept such as the winding number is not strictly defined in our discretized version, but as we will show, we can obtain $\phi_i$ configurations that approximate it well. We thus define the discretized analogue of \rf{jan21}:
\beql{ja10}
\phi_i \equiv \phi_i + n \cdot \delta, 
\quad n \in \mathbb{Z} \quad \forall i \in {\cal T},
\eeq
where $S^1$ has ``circumference'' $\delta$.
In the following, for convenience we will take $\delta =1$, except in Section \ref{sec:dynamical}. With this definition, \rf{eq:laplace} has solutions $\phi_i$ that can serve as coordinates. There are four independent non-contractible loops winding around the toroidal CDT triangulation $\cT$. Let us choose one of them and a no-boundary hypersurface that intersects the loop only once. For a description of how to actually choose such hypersurfaces for our CDT triangulations, we refer to \cite{coordinates1,coordinates2}. Let the field $\phi_i$ jump by $\delta=1$ when crossing the hypersurface. This is precisely what happened in the continuum solution \rf{jan21a}, and viewed as belonging to $S^1$ it does not jump at all. However, to solve the equations for $\phi_i$ it is convenient temporarily to view it as an ordinary scalar field in $\mathbb{R}$ with a jump at the hypersurface. As we will show below, this ensures that we have a unique solution to \rf{ja10} orthogonal to the constant mode, which by definition is ``strechted'' by $\delta=1$ moving around the manifold along the (or any) non-contractible loop intersecting the hypersurface. Although it seems that we have introduced a discontinuity of the field $\phi_i$ along the chosen hypersurface, we want again to emphasize that this is not the case when we view $\phi_i$ as a field belonging to $S^1$, and thus the hypersurface does not have any physical reality since we cannot identify it if we only know $\phi_i$ expressed as a field with values in $S^1$.\footnote{In Appendix 1 we show that if we view $\phi_i$ as a field taking values in $\mathbb{R}$ rather than in $S^1$, the hypersurface represents indeed a physical surface. In the language of electrostatics, it is a dipole sheet with constant dipole density.} We want to apply this construction also to the three other independent non-contractible loops in our triangulation ${\cal T}$ so that we have four scalar fields $(\ph_i^{(x)}, \ph_i^{(y)}, \ph_i^{(z)}, \ph_i^{(t)})$, which provide us with a map from ${\cal T}$ to $S^1\times S^1 \times S^1 \times S^1$, and which we can use (with some modifications) as coordinates for ${\cal T}$. We now turn to the precise description of how to do that.

\subsection{Scalar fields as coordinates with values on $S^1$}\label{sub:jump}

\noindent{\bf The jump condition} 

\vspace{6pt}

\begin{figure}[h]
\centering
\includegraphics[width=0.59\textwidth]{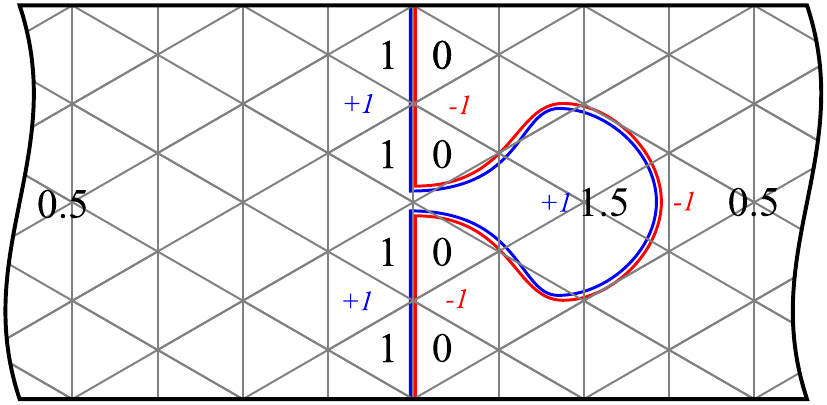}\\
\includegraphics[width=0.59\textwidth]{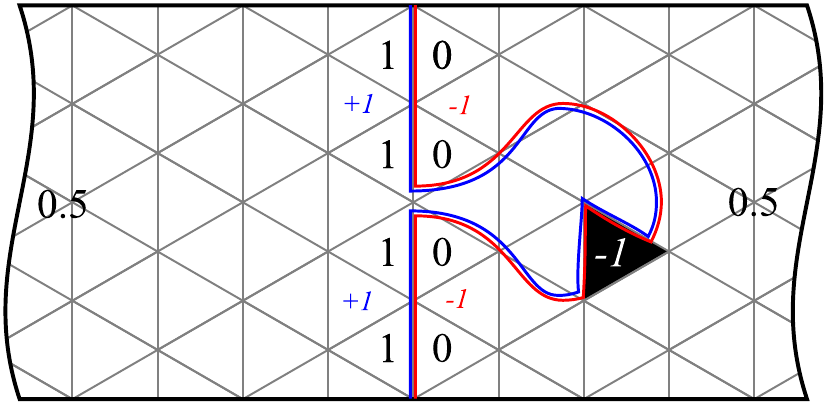}
\caption{Top: an example boundary with a bubble,
for which the field values do not fit into an interval of width $1$. Bottom: a step of the boundary redefinition procedure. The black triangle is flipped to the other side of the boundary. Its field value is decreased by $1$. } 
\label{fig:redefinition}
\end{figure}

We will now discuss how to implement the jump and solve the corresponding 
discretized Laplace equation.
Suppose we have a given {\it oriented} boundary or hypersurface (again, see \cite{coordinates1,coordinates2} for explicit constructions), defined as a non-contractible (in a given spatial or time direction) connected subset of 3D tetrahedral faces of four-simplices or, equivalently, as a subset of links on the dual lattice. The field $\ph_i$ in a simplex $i$ adjacent to the boundary will perceive the value of the field $\ph_j$ in a simplex $j$ on the other side of the boundary as shifted by $\pm \delta$ (the sign depends on the orientation of the boundary); see figure~\ref{fig:redefinition} for a 2D illustration.
Since the classical scalar field solution will trivially scale with the jump magnitude $\delta$, in the following {\it we will assume} $ \delta=1$ (as already noted above), { but we can always release this
assumption and change $\phi_i\to \delta \phi_i$, depending on possible physical
requirements.}\footnote{We release this assumption in Section \ref{sec:dynamical} where we discuss dynamical scalar fields coupled to geometric degrees of freedom. 
The jump magnitude $\delta$ will have an important impact on the underlying generic geometries.}
One can define an antisymmetric \textit{jump matrix}
\beql{def:jump_matrix}
B_{ij} =
\begin{cases}
	+1 & \textrm{if the dual link $i \rightarrow j$ crosses the boundary in the {\it positive} direction},\\
	-1 & \textrm{if the dual link $i \rightarrow j$ crosses the boundary in the {\it negative} direction},\\
	0 & \textrm{otherwise}
\end{cases}
\eeq
and a {\it boundary (jump) vector}
\beql{def:jump_vector}
b_i = \sum_j B_{ij}.
\eeq
The  three-volume (i.e., the number of tetrahedra) of the boundary is then given by:
\beql{eq:bvol} 
V = \frac{1}{2} \sum_{ij} B_{ij}^2 = \frac{1}{2}\sum_i | b_i |, 
\eeq
as the boundary vector $b_i$ is integer-valued in the range $-5 \leq b_i \leq 5$ and measures the number of tetrahedral faces a particular four-simplex $i$ has on the boundary.\footnote{$b_i$ will later be used to find a position of a (redefined) boundary. The sign depends on the flow of the winding number, i.e., whether the four-simplex is on the positive or negative side of the oriented boundary.}
To accommodate to the jump $\delta = 1$, we modify the scalar field action to \beql{SMdisJump}
S_M^{CDT}[\{\ph\}, \cT]	= \frac{1}{2} \sum_{i \leftrightarrow j} (\ph_i - \ph_j - B_{ij})^2 = \sum_{i,j} \ph_i L_{ij} \ph_j - 2\sum_i \ph_i b_i + V \equiv \ph^T \bL \ph - 2 \ph^T b + V,
\eeq
where we used definitions \rf{def:jump_vector} and \rf{eq:bvol}.
The action \rf{SMdisJump} is invariant under a constant shift in the scalar field values (the Laplacian zero mode) and, as we will argue below, it is also invariant under a shift of the boundary, provided that one also modifies the field values in a trivial way that is compatible 
with the equivalence definition \rf{ja10}. Thus, it follows that, viewed as taking values on $S^1$, the field is not changed at all, and  the classical solution is then independent of the specific choice of boundaries which can be ``continuously'' (in a sense defined suitably for the lattice) deformed into each other. 
\\

\noindent{\bf The classical solution} 

\vspace{6pt}

A classical solution for $\ph_i$ that minimizes the action \rf{SMdisJump} will now satisfy the discrete Laplace\footnote{Even though the equation \rf{eq:poisson} formally looks like a Poisson equation, we will call it the Laplace equation since $b$ is not a source term when we view the field as a field with values in $S^1$.} equation with a boundary term:
\beql{eq:poisson}
	\bL \ph	= b.
\eeq
Formally, the solution to equation (\ref{eq:poisson})
is given by $\ph = \bL^{-1} b$. However, as already discussed, the Laplacian matrix $\bL$ is not invertible as it has a zero mode ($ \bL e^{(0)} = 0, \ \mathrm{where} \ e^{(0)} = [1, 1, \dots, 1]^{T} $ is a constant eigenvector). Equation (\ref{eq:poisson}) {is still solvable} since the jump vector $b$ is orthogonal to the zero mode ($e^{(0)} \cdot b = \sum_i b_i = 0$), which is due to the translational symmetry of the action (the action is invariant under a constant shift of the field).
For the sake of simplicity, we shift the field values so that for some simplex (labeled $i_1$) $\ph_{i_1} = 0$.
This can be done by adding a term $\varepsilon \cdot \ph_{i_1}^2$ to the action \rf{SMdisJump}, where $\varepsilon$ is positive (not necessarily small).
The modification can then be absorbed into the Laplacian matrix,
\begin{align}
\label{eq:laplacian_mod}
	L_{i j} \ \longrightarrow \ L_{i j} + \varepsilon \delta_{i i_1} \delta_{ji_1},
\end{align}
and one obtains a unique solution: 
\beq\label{jan1}
\bar \ph = \bL^{-1} b,\qquad \bar\ph_{i_1} = 0.
\eeq
All other solutions to the original Laplace equation \rf{eq:poisson} with the zero mode are thus given by translations $\ph_i =\bar\ph_i + \const$ Computing the classical solution numerically is itself a technical challenge since the Laplacian matrix is large ($N_4 \times N_4$, where $N_4\sim10^5-10^6$). Nevertheless, we managed to construct numerical algorithms that solve this problem with machine precision in relatively short computer time. Technicalities are discussed in Appendix 2. The classical solution $\bar\ph = \bL^{-1} b$ has the property
\beql{eq:phi_mean_b}
\bar\ph_i = \frac{1}{5} \left( b_i+ \sum_{j \to i}\bar \ph_j \right).
\eeq
This is just a discretized version of the mean value property of continuous harmonic functions, where at the boundary one should view the field as taking values in $S^1$ rather than in $\mathbb{R}$. An interesting consequence of eq.~\rf{eq:phi_mean_b} is that the field condensates in the fractal outgrowths observed in CDT triangulations. This is because the (artificial) local boundary surrounding an outgrowth is typically small in size, and therefore the field changes only a little on that local boundary, leaving the field values almost constant in all simplices building the geometric outgrowth. The condensation is observed in all spatial and time directions and for each of the four scalar fields ($\ph_i^{(x)}, \ph_i^{(y)}, \ph_i^{(z)}, \ph_i^{(t)}$). Consequently, if one represents each simplex $i$  by a point with coordinates ($\ph_i^{(x)}, \ph_i^{(y)}, \ph_i^{(z)}, \ph_i^{(t)}$), the fractal outgrowths will constitute dense clouds of points. Examples of such maps are presented in figures \ref{fig:cosmic_C} - \ref{fig:cosmic_A}. The maps (or at least 2D projections) will therefore qualitatively resemble the conformal map in figure~\ref{fig:outgrowths} discussed above, where dense regions are also fractal outgrowths.\\

\noindent{\bf Boundary redefinition}

\vspace{6pt}

The scalar field action with a jump at the boundary \rf{SMdisJump} is invariant under a local shift of the boundary (such that one simplex, labelled $i$, is transferred from one to the other side of the boundary) with a simultaneous change of the scalar field value $\ph_i \to \ph_i \pm \delta$ (the sign depends on whether the simplex is shifted from the negative to the positive side of the oriented boundary or vice versa). This is illustrated by a simple 2D example triangulation with a boundary presented in figure~\ref{fig:redefinition}. 
Let us consider repeated changes in the position of the boundary, which preserve its nature as a hypersurface with the topology of $T^3$, and at the same time the corresponding changes in the field $\phi_i$. Clearly the field $\phi_i$ viewed as a field with values on $S^1$ is not changed at all; nevertheless, it is convenient to think about such a change of the boundary and the field $\phi_i$. The reason is that 
the solution $\bar{\phi}_i$ given by \rf{jan1}
need not be constant on the hypersurface with the jump $\delta \ (=1)$ 
nor does it necessarily take values in the range $[0,1]$ (as illustrated in figure~\ref{fig:redefinition}), even after adjusting the global constant.
Let us now argue that we can deform the hypersurface of the field jump and correspondingly change $\bar{\phi}_i$ so that $\bar{\phi}_i$ is zero on one side of the modified hypersurface and takes the value 1 on its other side. We apply the following procedure to the original classical field solution $\bar\ph_i$:
\begin{enumerate}
	\item Shift all field values by a constant so that the smallest value is $0$.
	\item Choose a simplex with the largest field value. As follows from the maximum principle for a harmonic function, the simplex has to touch the boundary with at least one face.
	\item Modify the boundary so that the simplex is flipped to its other side and
		decrease the corresponding field value by $\delta =1$.
	\item Repeat steps 2-3 until the maximal field value is below $1$.
\end{enumerate}

\begin{figure}[h]
\centering
\includegraphics[width=0.75\textwidth]{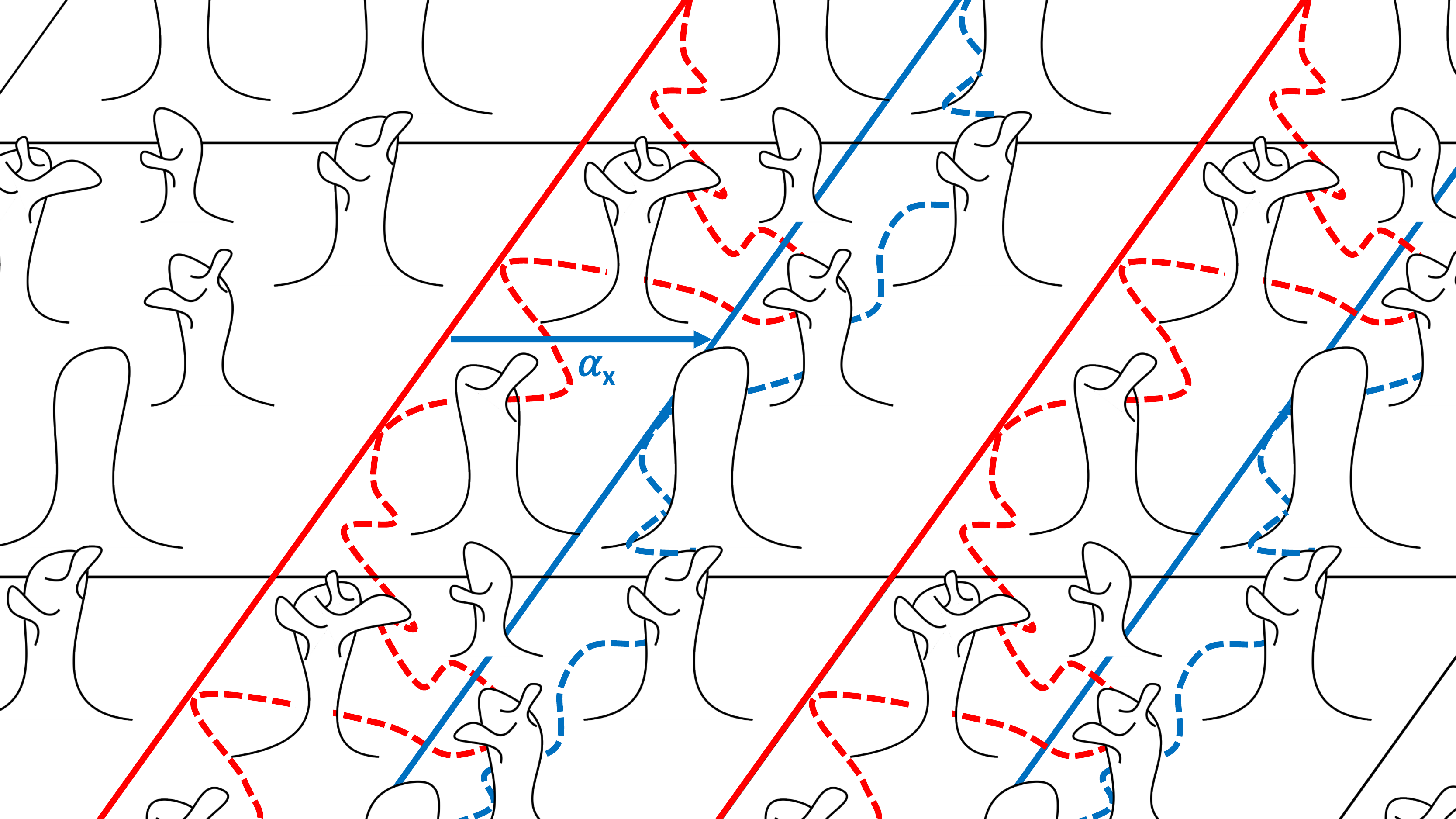}
\caption{A 2D visualization of the (toroidal) periodic geometric structure.
The solid red and blue lines are drawn to guide the eye. The dashed red lines show the periodic structure starting out with the hypersurface corresponding to, say,  $\alpha_x =0$. The dashed blue lines show the periodic structure starting out with the hypersurface corresponding to some other $\alpha_x$.} 
\label{fig:puzzle}
\end{figure}

The argument above shows, using the fact that $\bar{\phi}_i$ is a 
discrete harmonic function, that it is possible to find a hypersurface such that the (new) $\bar{\phi}_i$ defined by it takes values in the range $[0,1]$. One could obtain such a surface ``in one go'' by defining a new field 
\beql{jan30}
\tilde \ph_i(0) = \fmod(\bar \ph_i,1).
\eeq
This removes the original hypersurface and replaces it with the one where $\bar{\phi}_i$ passes through 0 (or an integer $n \in \mathbb{Z}$), at the same time ensuring that the range of $\tilde \ph_i(0)$ is $[0,1]$. Literally mapped to a circle of circumference 1 in the complex plane,
\beql{jan40}
\psi_i =\frac{1}{2\pi} \exp \big(2\pi \rmi \bar{\phi}_i\big)= 
\frac{1}{2\pi} \exp \big(2\pi \rmi \tilde{\phi}_i (0)\big),
\eeq
which illustrates again that from an $S^1$ perspective the hypersurfaces play no role (as long as they are ``continuously'' deformable to each other). We have now achieved our goal of finding a harmonic map from the triangulation ${\cal T}$ to $S^1$ with winding number 1.
The hypersurfaces $H(\alpha)$ in ${\cal T}$ characterized by being mapped to a fixed point $\rme^{\rmi 2 \pi \alpha}/2\pi$ on the circle of circumference 1 cover ${\cal T}$, and $\alpha$ can serve as the coordinate in ${\cal T}$ ``orthogonal'' to these hypersurfaces. Thus, 
\beql{jan41} 
H(\alpha) = \{ i \in {\cal T} \;| \; \psi_i = \rme^{2\pi \rmi \alpha}/2\pi\}.
\eeq
$H(0)$ is precisely the hypersurface where $\tilde \ph_i (0)$ jumps from 0 to 1 constructed above, and we can generalize this construction to find $H(\alpha)$ explicitly. Define
\beql{alpha}
\tilde \ph_i(\alpha) = \fmod(\bar \ph_i - \alpha,1), \quad 
0 \leq \alpha < 1.
\eeq
Again, the original hypersurface of the jump in $\bar{\phi}_i$ is removed and replaced by the new hypersurface where $\bar{\phi}_i$ passes though $\alpha$ (or $\alpha$ plus an integer $n \in \mathbb{Z}$), i.e., where $\tilde{\ph}_i(\alpha)$ jumps from 0 to 1. By construction we have
\beql{jan42}
\psi_i = \rme^{2\pi \rmi \alpha}\, \rme^{2\pi \rmi \tilde \ph_i(\alpha)}/2\pi,
\eeq
so $H(\alpha)$ is indeed the hypersurface with the described property. Since $\tilde{\phi}_i (\alpha)$ is still a solution to eq.~\rf{eq:poisson}, we can explicitly find $H(\alpha)$ by using eq.~\rf{eq:phi_mean_b} to reconstruct the boundary jump vector from $\tilde{\phi}_i (\alpha)$:
\beql{bi}
b(\tilde\ph_i(\alpha)) = {5} \tilde\ph_i(\alpha) - \sum_{j \to i} \tilde\ph_j(\alpha) = \sum_j L_{i j} \tilde\ph_j(\alpha).
\eeq
As already mentioned, the (integer) value of $b(\tilde\ph_i(\alpha))$ counts the  number of faces (tetrahedra) the simplex $i$ shares with the boundary (the value is 0 for no boundary faces shared, or either positive or negative depending on which side of the boundary the simplex is located, as described above). Thus, knowing $b(\tilde\ph_i(\alpha))$, we know $H(\alpha)$. There are several issues related to the hypersurfaces $H(\alpha)$, which we will discuss below: are they really hypersurfaces? How do they change with $\alpha$ ($\bar{\phi}_i$ is a set of discrete variables, and $\alpha$ is a continuous parameter)? What is the size of a typical hypersurface $H(\alpha)$? Is $\alpha$ really a {\it good} coordinate for a typical path integral configuration? We will address these questions in Section \ref{sec:foliations}. Assuming that the issues mentioned have satisfactory answers, let us return to our original problem: for a given toroidal triangulation we have defined in some way (see \cite{coordinates1,coordinates2}) four independent non-contractible boundaries which we can label with $x,y,z,t$, and we want to use the corresponding classical solutions $\bar{\phi}_i^{\mu}$, $\mu =x,y,z,t$ as coordinates, but without any explicit reference to the chosen boundaries and the specific range of these solutions. We have managed to do that by introducing the coordinate system $(\alpha_x, \alpha_y, \alpha_z, \alpha_t)$ where $\alpha_\mu \in [0,1]$ and the corresponding scalar fields $\tilde{\phi}_i^\mu(\alpha_\mu)$ are characterized by being solutions to the Laplace equations that jump from 0 to 1 at the $\alpha_\mu$-hypersurface. Sometimes, it can be convenient to represent the torus as a periodic structure on $\mathbb{R}^4$. If we choose to let the jumps of $\tilde{\phi}_i^\mu(\alpha_\mu)$ define the periodic structure, we can turn the functions $\tilde{\phi}_i^\mu(\alpha_\mu)$ into functions without a jump by adding $\pm 1$ to them when they cross the boundaries where they jump. We can also label the new regions we enter in $\mathbb{R}^4$ by corresponding integer labels that tell us how many multiples of $\pm 1$ we should add to the corresponding functions $\tilde{\phi}_i^\mu(\alpha_\mu)$ in that particular region in order to ensure it is a ``continuous'' function (i.e., a function without the jumps) on $\mathbb{R}^4$. We have tried to illustrate this in figure~\ref{fig:puzzle}, where we show how different choices of $\alpha$ lead to different representations of the torus on $\mathbb{R}^4$. With the choice of the coordinate system given by $(\alpha_\mu)$, we are interested in the volume density $\sqrt{g(\alpha)}$ defined as 
\beq\label{jan2}
\rmd V(\alpha) = \sqrt{g(\alpha)} \prod_\mu \Delta \alpha_\mu =
\# \textrm{ simplices in volume element} \prod_\mu \Delta \alpha_\mu.
\eeq
The easiest way to obtain an idea of the volume density is to fix a coordinate point $\alpha_\mu^0$ and calculate the four scalar fields $\tilde{\phi}_i^\mu(\alpha_\mu^0)$. If we implement $\tilde{\phi}_i^\mu(\alpha_\mu^0)$ on $\mathbb{R}^4$ as described above (without any jumps), then by definition (since the $\alpha$-hypersurfaces are the hypersurfaces of constant $\bar{\phi}_i$ or, equivalently, of constant $\tilde{\phi}_i^\mu(\alpha_\mu^0)$) the density of simplices around a simplex $i$ where $\tilde{\phi}_i^\mu(\alpha_\mu^0) = \alpha_\mu$,\footnote{For clarity of presentation we have made this discussion a little imprecise, treating the simplices $i$ as points in a continuum so that there is locally a one-one map between $i$ and its coordinates $\alpha_\mu(i)$.} measured using the scalar fields $\tilde{\phi}_i^\mu(\alpha_\mu^0)$, will agree with the density $\sqrt{g(\alpha)}$ defined in \rf{jan2}.
We can thus write:
\beq\label{jan3}
\rmd V(\tilde{\phi}_i^\mu(\alpha_\mu^0)) = \sqrt{g(\tilde{\phi}_i^\mu(\alpha_\mu^0))} \prod_\mu \Delta \tilde{\phi}_i^\mu(\alpha_\mu^0).
\eeq
Now we turn to the measurement of $\sqrt{g(\tilde{\phi}_i^\mu(\alpha_\mu^0))} $.

\subsection{Density measurements for generic geometries in various CDT phases}\label{sub:maps}

Below we present the results of scalar fields measurements for generic triangulations observed in all the four phases ($C$, $C_b$, $B$ and $A$) of CDT with the toroidal spatial topology and a periodic time coordinate. The time period used was either $T=4$ or $T=20$, and the $N_{4,1}$ volume was set to fluctuate around $160\rmk$ and $720\rmk$ simplices, respectively. In each case, we picked just one typical configuration and solved for the classical scalar fields $(\tilde\ph^{(x)}(\alpha_x),\tilde\ph^{(y)}(\alpha_y),\tilde\ph^{(z)}(\alpha_z),\tilde\ph^{(t)}(\alpha_t))$
in such a way that the field values are within the range $[0,1]$ (we put $\delta=1$), and the elementary cell boundaries are set at $\tilde\ph^{\mu}(\alpha_\mu) = 0,1$ as described above. We chose $\alpha_\mu$ such that in each direction the field values are centered around $0.5$. 

\subsubsection{Density maps in $\bm{ \tilde \ph}$ coordinates}\label{3.3.1}

\begin{figure}[h]
\centering
\includegraphics[width=0.44\textwidth]{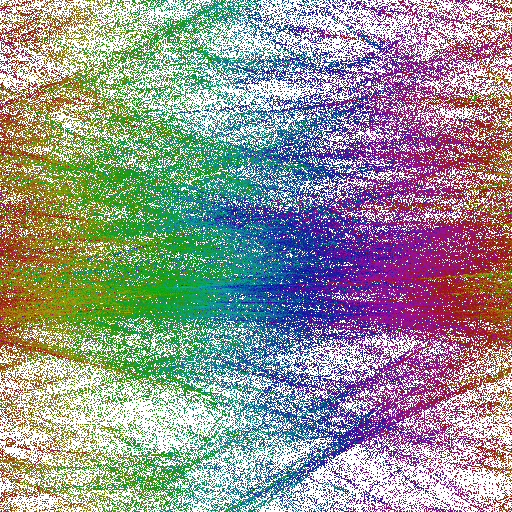}
\includegraphics[width=0.44\textwidth]{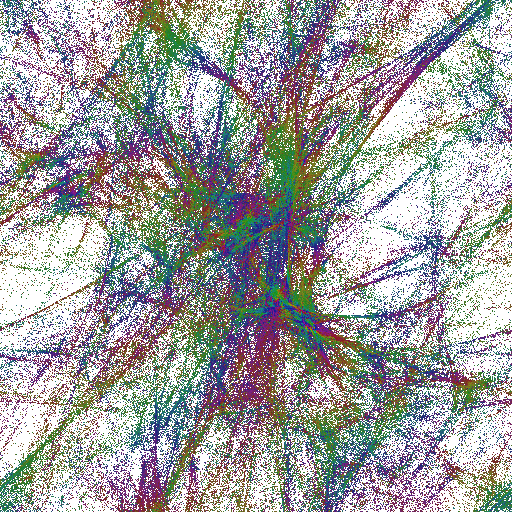} \\
\includegraphics[width=0.44\textwidth]{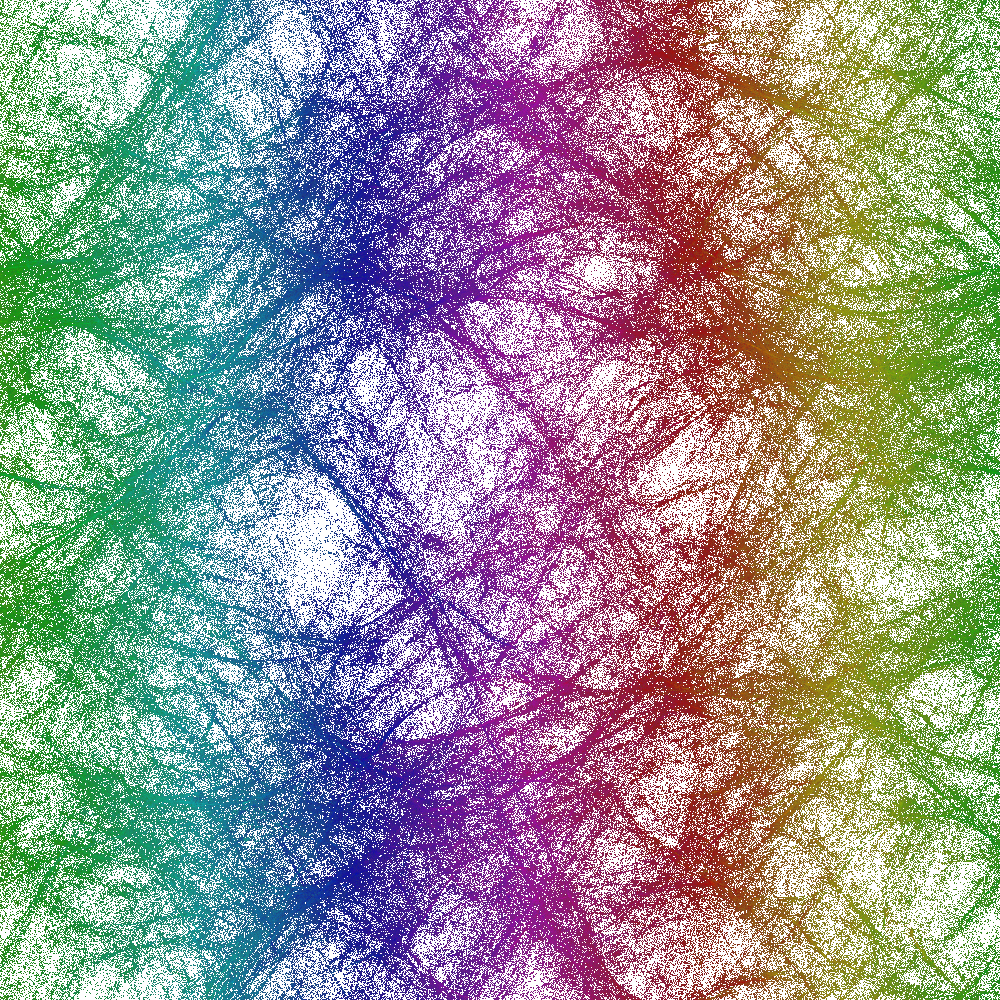}
\includegraphics[width=0.44\textwidth]{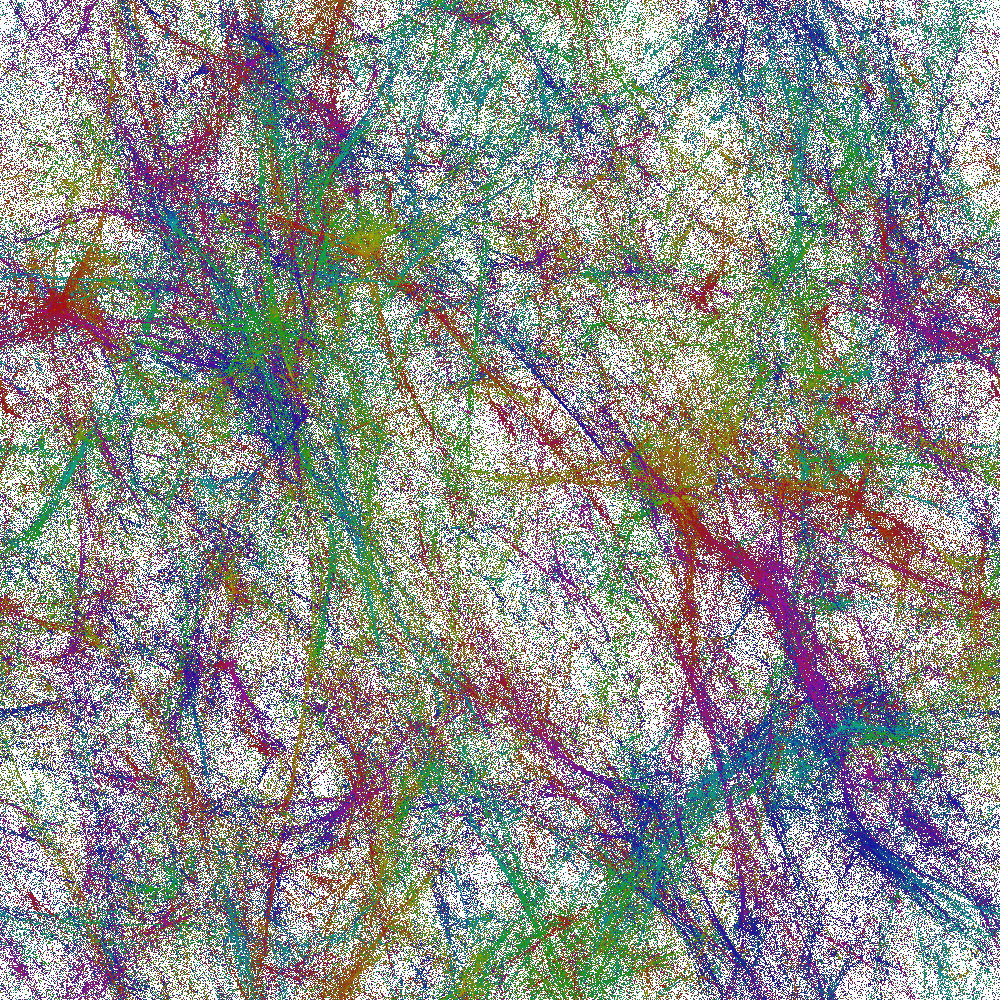}\\
\caption{Cosmic voids and filaments for configurations in phase $C$. Top:  a configuration with $T = 4$ ($\kappa_0=2.2$, $\Delta=0.6$). Bottom: a configuration with $T = 20$ ($\kappa_0=3.0$, $\Delta=0.2$). The left-hand side charts are projections on the $t - x$ plane, the right-hand side charts are projections on the $x-y$ plane. Notice that for the two bottom plots the period $T$ is larger than that for the upper plots, which also explains why the observed structures are more dense. }
\label{fig:cosmic_C}
\end{figure}

\begin{figure}[h]
\centering
\includegraphics[width=0.44\textwidth]{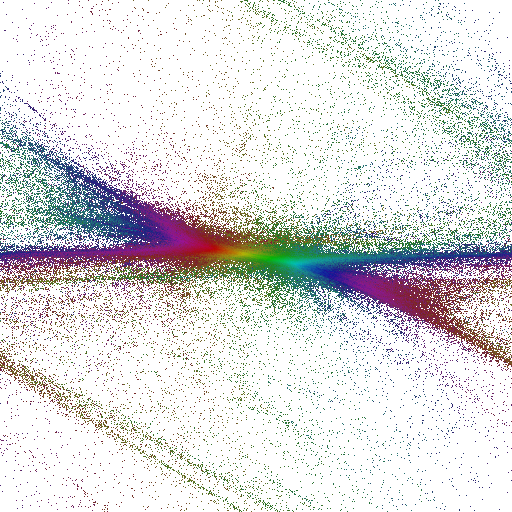}
\includegraphics[width=0.44\textwidth]{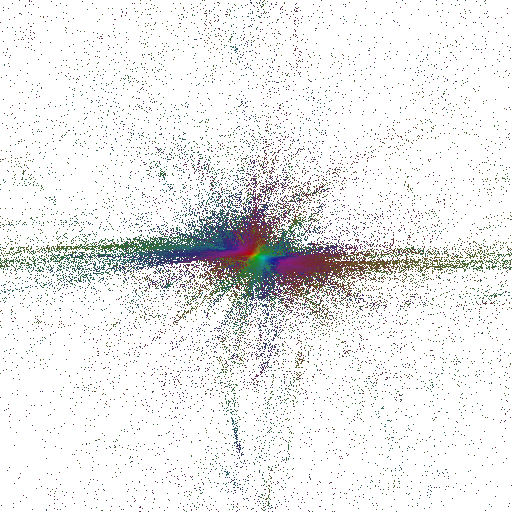}\\
\includegraphics[width=0.44\textwidth]{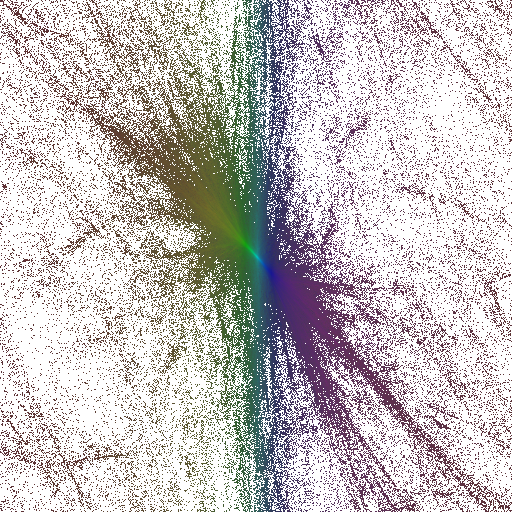}
\includegraphics[width=0.44\textwidth]{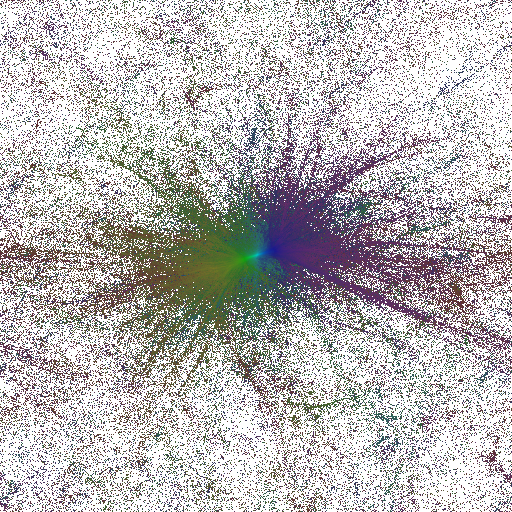}
\caption{Configurations in phase $C_b$. Top: a configuration with $T = 4$ ($\kappa_0=2.0$, $\Delta=0.1$). Bottom: a configuration with $T = 20$ ($\kappa_0=2.5$, $\Delta=0.2$). The left-hand side charts are projections on the $t - x$ plane, the right-hand side charts are projections on the $x-y$ plane.}\label{fig:cosmic_Bif}
\end{figure}

\begin{figure}[H]
\centering
\includegraphics[width=0.44\textwidth]{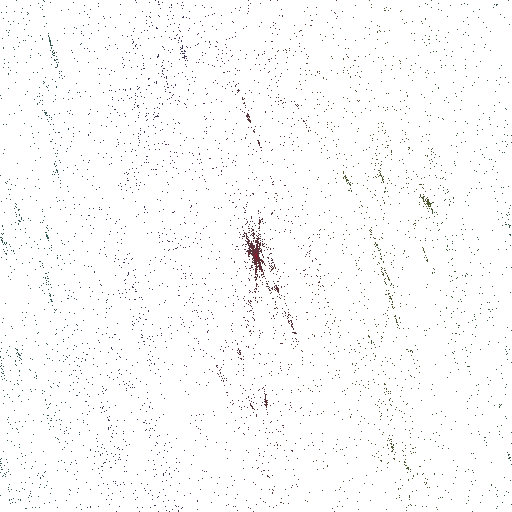}
\includegraphics[width=0.44\textwidth]{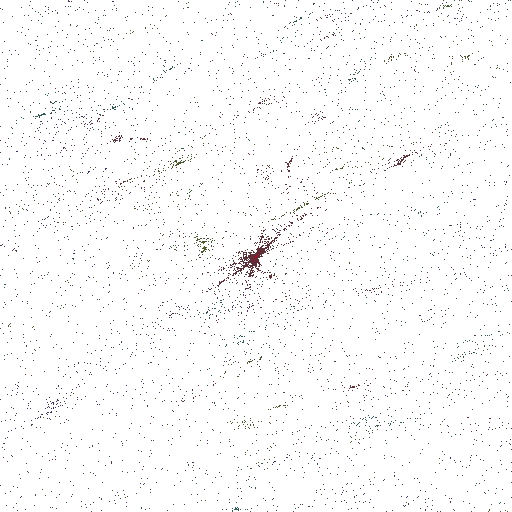}
\caption{A configuration in phase $B$ with $T = 4$ ($\kappa_0=4.4$, $\Delta=-0.7$). The left-hand side chart is a projection on the $t - x$ plane, the right-hand side chart is a projection on $x-y$ plane.}\label{fig:cosmic_B}
\end{figure}

\begin{figure}[H]
\centering
\includegraphics[width=0.44\textwidth]{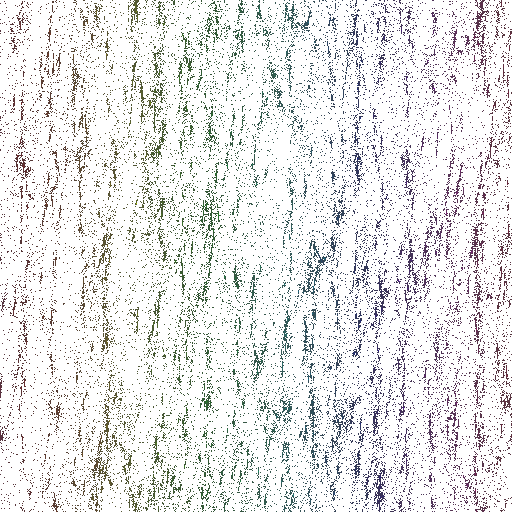}
\includegraphics[width=0.44\textwidth]{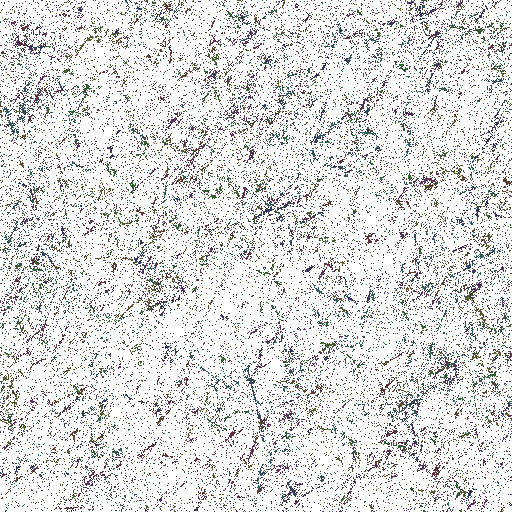}
\caption{A configuration in phase $A$ with $T = 20$ ($\kappa_0=5.0$, $\Delta=0.2$). The left-hand side chart is a projection on the $t - x$ plane, the right-hand side chart is a projection on the $x-y$ plane.}\label{fig:cosmic_A}
\end{figure}

In principle, a density plot of
$(\tilde\ph^{(x)}_i(\alpha_x),\tilde\ph^{(y)}_i(\alpha_y),\tilde\ph^{(z)}_i(\alpha_z),\tilde\ph^{(t)}_i(\alpha_t))$
would provide us with the desired quantity $\sqrt{g(\tilde{\phi}_i^\mu(\alpha_\mu))}$. However, this distribution depends on four fields and is difficult to visualize. We have thus opted to plot in figures \ref{fig:cosmic_C} - \ref{fig:cosmic_A} the periodic 2D projections (in various directions), where each dot represents a simplex with coordinates determined by the classical scalar field solution $(\tilde\ph^{\mu}(\alpha_\mu),\tilde\ph^{\nu}(\alpha_\nu))$. Thus, in a given small area 
\beql{jan4}
\rmd A_{\mu \nu} =\Delta \tilde\ph^{\mu}(\alpha_\mu) \Delta\tilde\ph^{\nu}(\alpha_\nu)
\eeq
we count the total number of four-simplices $i$ with coordinates $(\tilde\ph^{\mu}_i(\alpha_\mu),\tilde\ph^{\nu}_i(\alpha_\nu))$ in the region $\Delta \tilde\ph^{\mu}(\alpha_\mu) \Delta\tilde\ph^{\nu}(\alpha_\nu)$.
With the $(\tilde\ph^{\mu}(\alpha_\mu),\tilde\ph^{\nu}(\alpha_\nu))$-plane serving as a photographic plate, all points 
above and below are projected on it and leave a mark. In terms of the original $\sqrt{g(\tilde{\phi}_i^\mu(\alpha_\mu))}$, we can write (in continuum notation), instead of \rf{jan3},
\beql{jan5}
\rmd V_{\mu\nu} = \left( \int_0^1 \int_0^1 \rmd\tilde{\phi}^\kappa(\alpha_\mu) \rmd\tilde{\phi}^\lambda(\alpha_\nu) \;
\sqrt{g(\tilde{\phi}_i^\rho(\alpha_\rho))}\right)\;\rmd A_{\mu\nu},
\qquad \kappa,\lambda \neq \mu,\nu.
\eeq
Since we have the original coordinate $t$ freely at our disposal, we have chosen to include this information in 
the plots by a color code. The color of each point thus depends on the position of a given simplex in the original proper-time foliation $t$. To each $(4,1)$ simplex with four vertices (a spatial tetrahedron) in $t$ and one vertex in $t+1$ we assign an integer time coordinate $t$. As going from such a simplex to a simplex of the same type in the next $t+1$ layer requires at least 4 steps: $(4,1) \to (3,2) \to (2,3) \to (1,4) \to (4,1)$, we assign non-integer time coordinates $t+\frac{1}{4}, t+ \frac{1}{2}$ and $ t+ \frac{3}{4}$ to the $(3,2)$, $(2,3)$ and $(1,4)$ simplices, respectively. Thus, we have in total $4 \times T$ various time coordinates (and the corresponding colors), and we can trace the location of each simplex in the (original) time foliation. In figure~\ref{fig:cosmic_C} we show configurations measured in the semiclassical phase $C$ for $T=4$ (top charts) and $T=20$ (bottom charts), respectively.
 The left-hand side charts are projections on the $t-x$ plane, while the right-hand side charts are projections on the $x-y$ plane. One can easily see that the scalar field with a jump in the time direction follows the original time slicing (depicted by colors) quite closely, whereas the new coordinates defined by the scalar fields are smeared around the original proper-time slicing. The large-scale structure is quite isotropic in all spatial directions, i.e., it looks qualitatively the same for all $x-y$, $x-z$ and $y-z$ projections (in the plots we show just the $x-y$ projection). This is also the case for the time direction when both $T=20$ and $N_{4,1}=720\rmk$ are large, i.e., the $t-x$ (and also $t-y$ and $t-z$) projection looks qualitatively similar to the $x-y$ projection.\footnote{For $T=4$ the correlation length in the time direction is larger than the fixed time period, and thus the system is too small to allow for the full structure formation in this direction, but this is simply a finite size effect.} For the larger triangulation, the large-scale geometry is also quite homogeneous in all directions, in the sense that shifting all coordinates by constants will produce pictures looking qualitatively the same. Summing up, in the semiclassical phase $C$ one observes a homogeneous and isotropic geometry on large scales. This large-scale homogeneity and isotropy is broken on smaller scales, with sparse regions representing the ``central'' toroidal part and dense regions showing fractal outgrowths. The outgrowths are very non-trivially correlated, forming the characteristic cosmic voids and filaments structure. Remarkably, even though we analyze the pure gravity case (i.e., the classical scalar fields do not impact the CDT geometry in any way), and the measured ``universes'' are only a few Planck lengths in diameter \cite{physrep}, they qualitatively reproduce the basic features of the real Universe, including the large-scale cosmic voids and filaments structure observed in nature. From this perspective, one can imagine that the geometric fractal outgrowths serve as ``seeds'' of some matter field condensations (this is indeed the case for quantum scalar fields coupled to geometry, discussed in Section \ref{sec:dynamical}), leading to nontrivial structure formation caused by quantum gravity effects.

Similar analysis can be performed for geometric configurations measured in the other CDT phases. In figure~\ref{fig:cosmic_Bif} we plot 2D projections of the density maps measured in the bifurcation phase $C_b$ for $T=4$ (top charts) and $T=20$ (bottom charts). Here again, at least for the large $T=20$ and $N_{4,1}=720\rmk$ configuration, the geometry appears quite isotropic in all directions (we will return to this in the next subsection) but is no longer homogeneous. The lack of homogeneity in the time direction is well explained by the nonuniform spatial volume distribution in the proper-time coordinate as the volume profile in this phase is blob-like rather than flat as in phase $C$ (the effect is visible only for large $T$). It is equally well known that the characteristic feature of generic phase $C_b$ triangulations is the emergence of dense volume clusters around high-order vertices observed in every second spatial slice, which makes the spatial volume distribution inhomogeneous also in the spatial directions. In the $C_b$ phase maps in figure~\ref{fig:cosmic_Bif}, unlike in the $C$ phase, no nontrivial structure of fractal outgrowths can be observed as the geometry viewed from any direction seems to concentrate in just one large outgrowth. This effect is even more pronounced in phase $B$; see figure~\ref{fig:cosmic_B}, showing a configuration with $T=4$. In this case, the geometry in all directions becomes effectively compactified to a point. Thus, time and spatial homogeneity are both maximally broken. This, again, was expected from the previous analyses of geometric configurations observed in this phase. Finally, figure~\ref{fig:cosmic_A} shows a generic phase $A$ configuration, with $T=20$. In that case, the dense regions, i.e., the geometric outgrowths, are separated and uncorrelated, and they do not form any nontrivial structures. This kind of behavior was previously noticed in the time direction, but now it can also be observed in the spatial directions. As a result, a generic configuration measured in phase $A$ is highly homogeneous and isotropic on both large and small scales.

\subsubsection{Density maps in alternative $\bm \beta$ coordinates}

\begin{figure}[H]
\centering
\includegraphics[width=0.44\textwidth]{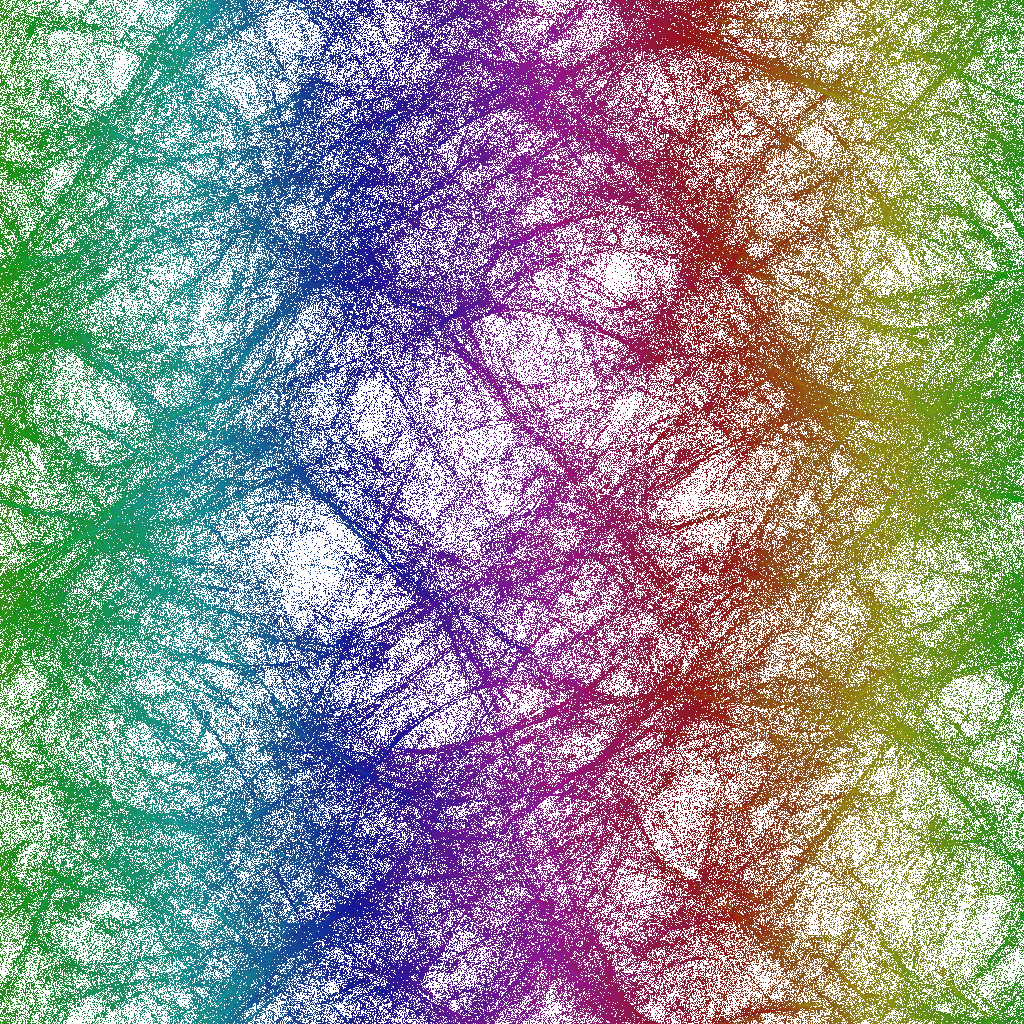}
\includegraphics[width=0.44\textwidth]{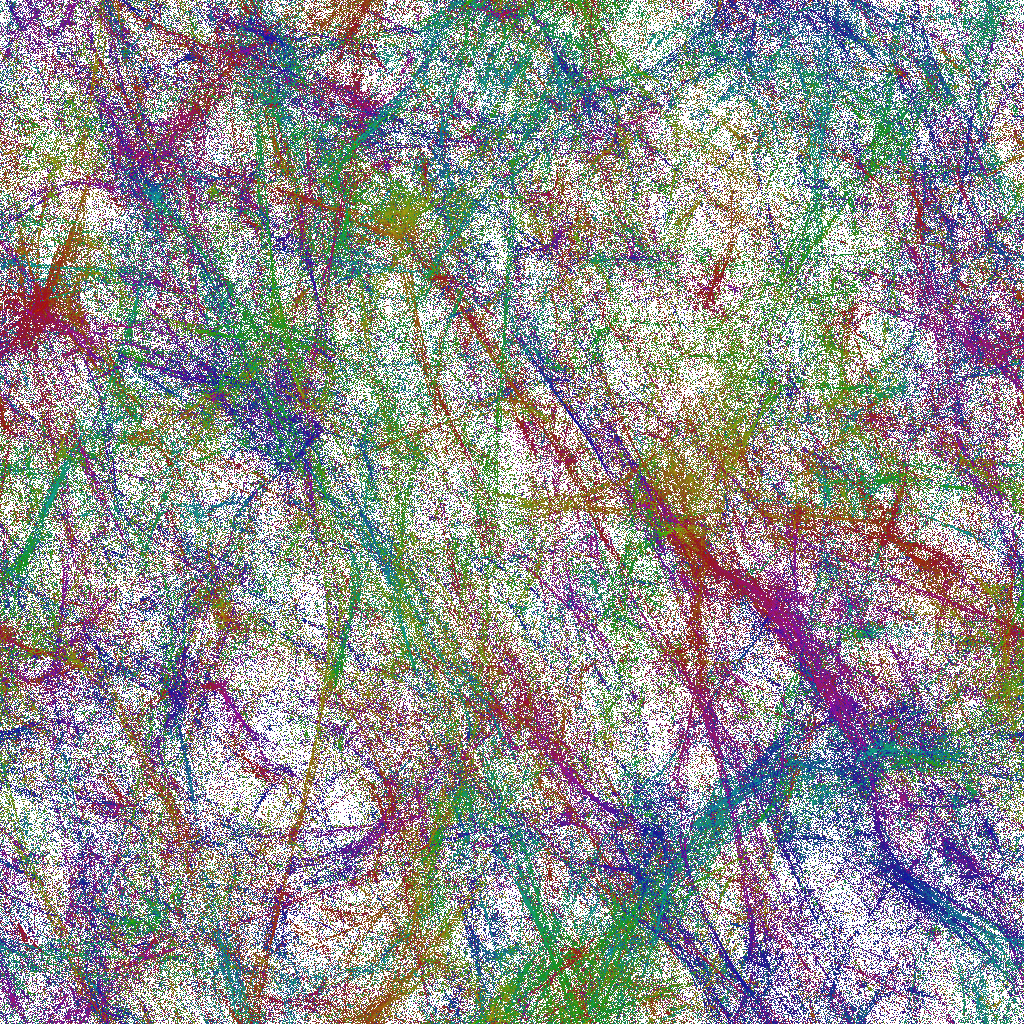}
\caption{A configuration in phase $C$ with $T = 20$ ($\kappa_0=3.0$, $\Delta=0.2$) in $\beta$ coordinates. The left-hand side chart is a projection on the $t - x$ plane, the right-hand side chart is a projection on the $x-y$ plane.}\label{fig:beta_C}
\end{figure}

\begin{figure}[H]
\centering
\includegraphics[width=0.44\textwidth]{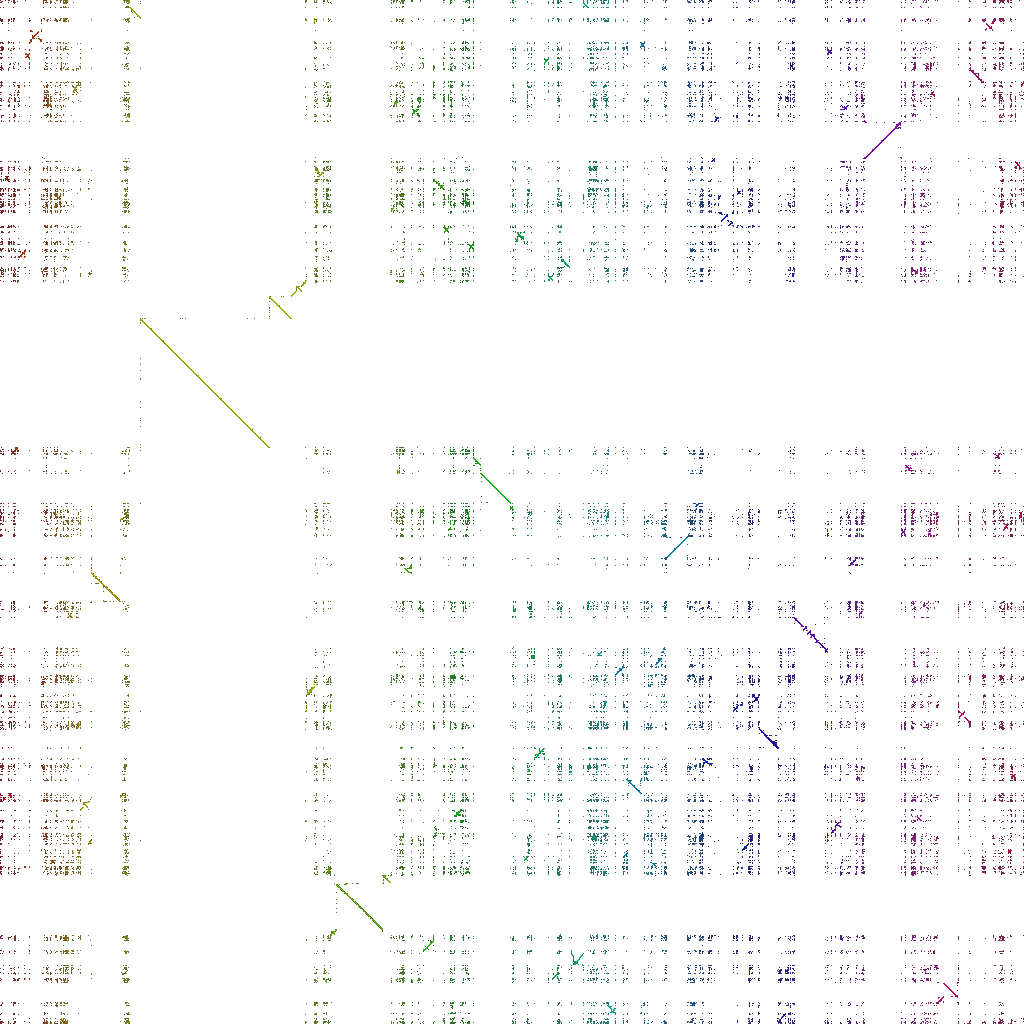}
\includegraphics[width=0.44\textwidth]{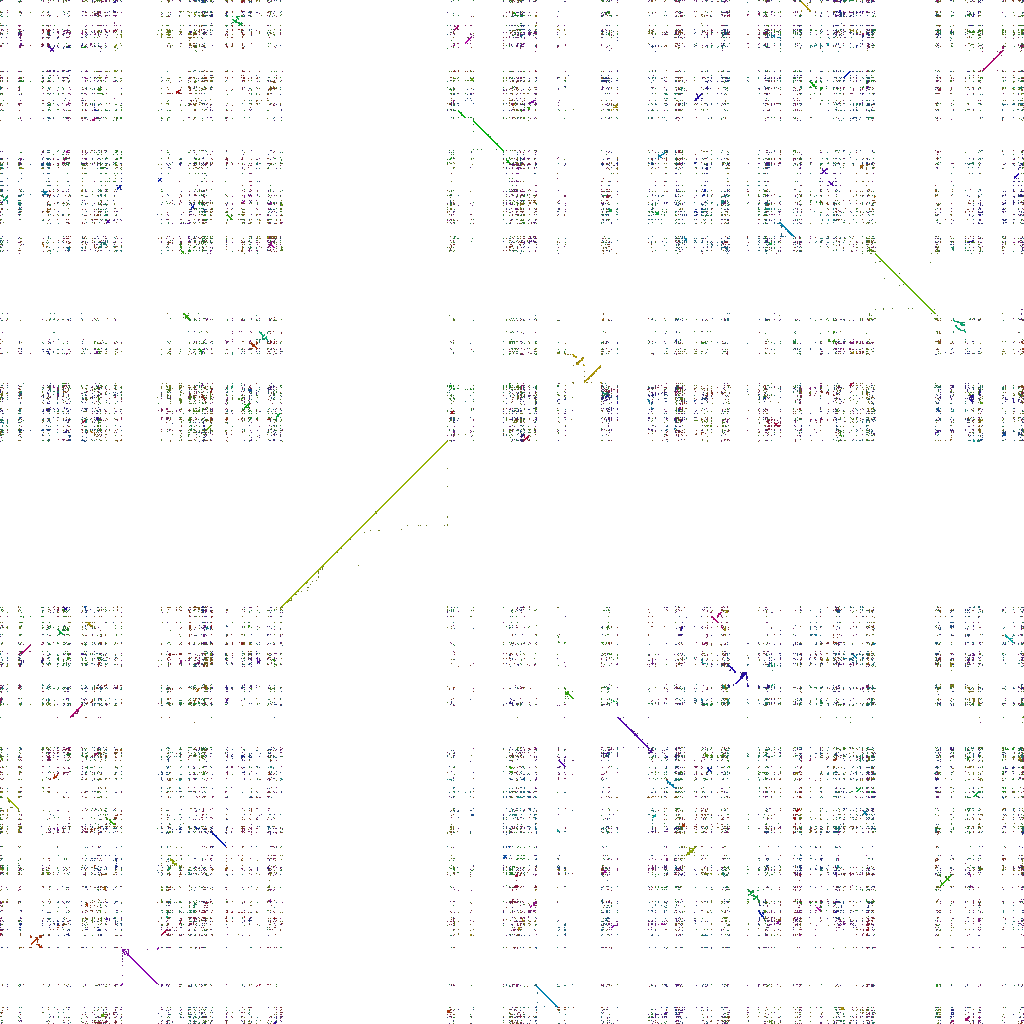}
\caption{A configuration in phase $A$ with $T = 20$ ($\kappa_0=5.0$, $\Delta=0.2$) in $\beta$ coordinates. The left-hand side chart is a projection on the $t - x$ plane, the right-hand side chart is a projection on the $x-y$ plane.}\label{fig:beta_A}
\end{figure}

\begin{figure}[H]
\centering
\includegraphics[width=0.31\textwidth]{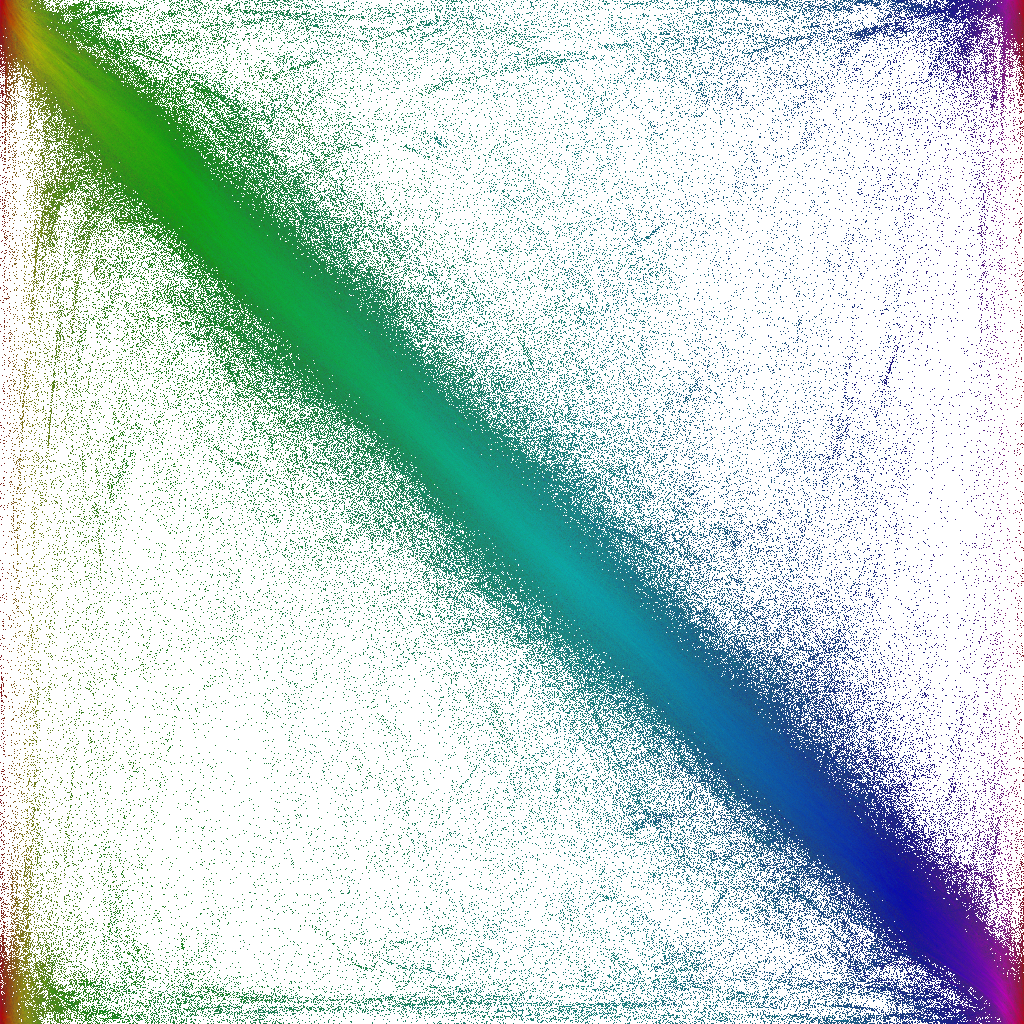}
\includegraphics[width=0.31\textwidth]{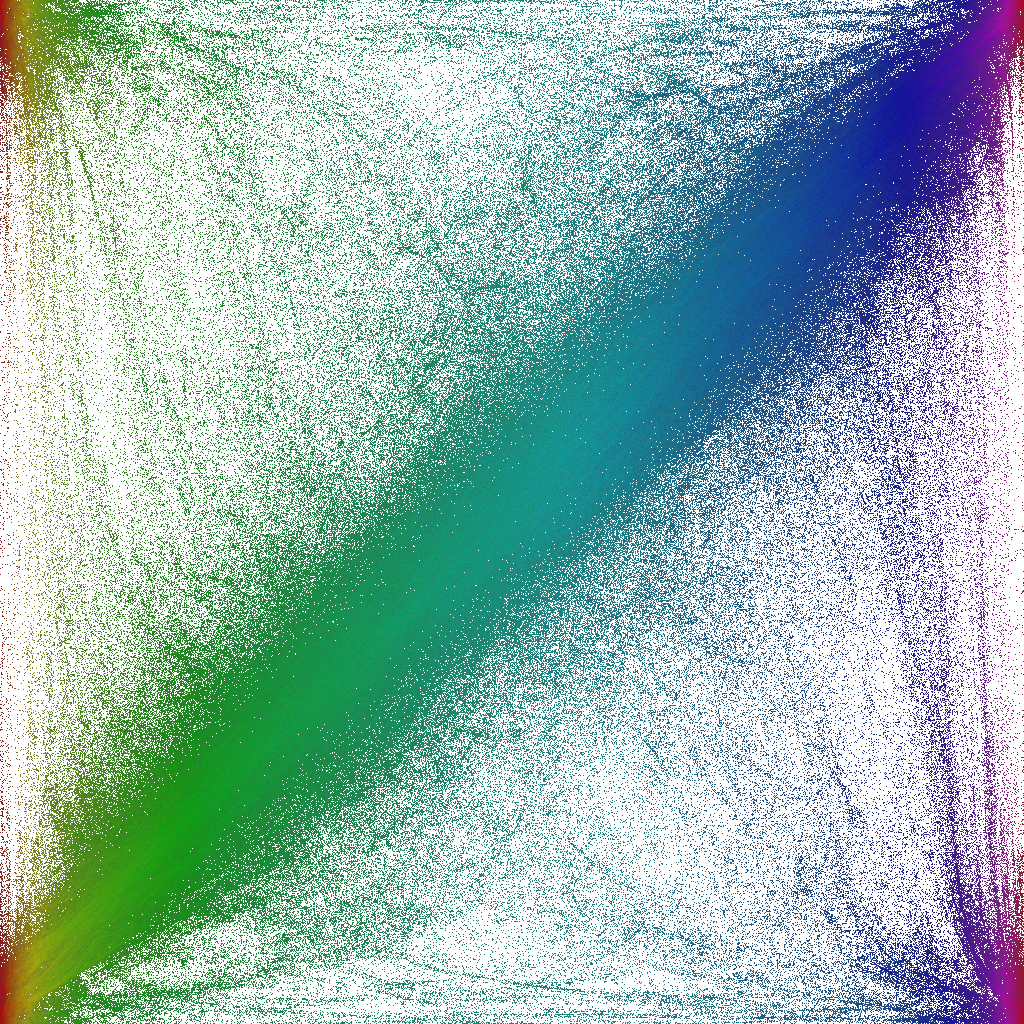}
\includegraphics[width=0.31\textwidth]{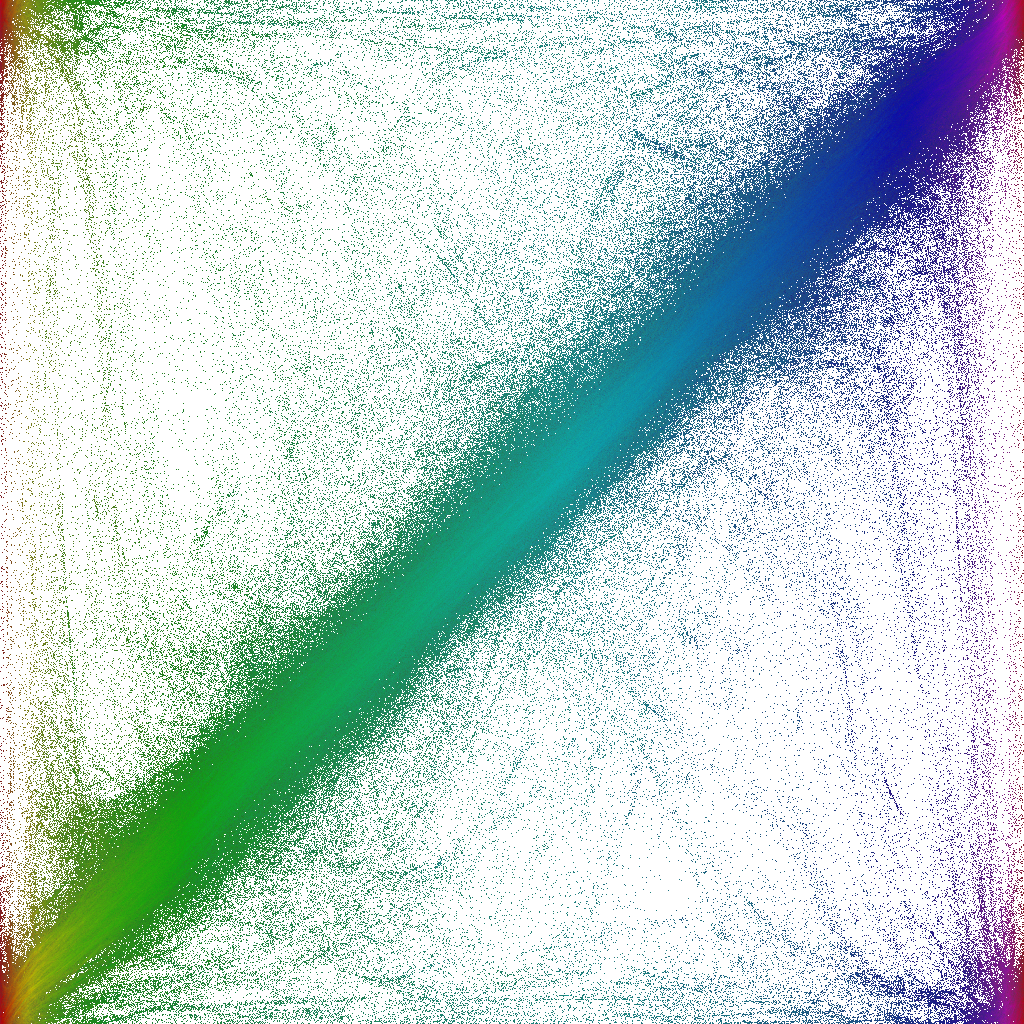}\\
\includegraphics[width=0.31\textwidth]{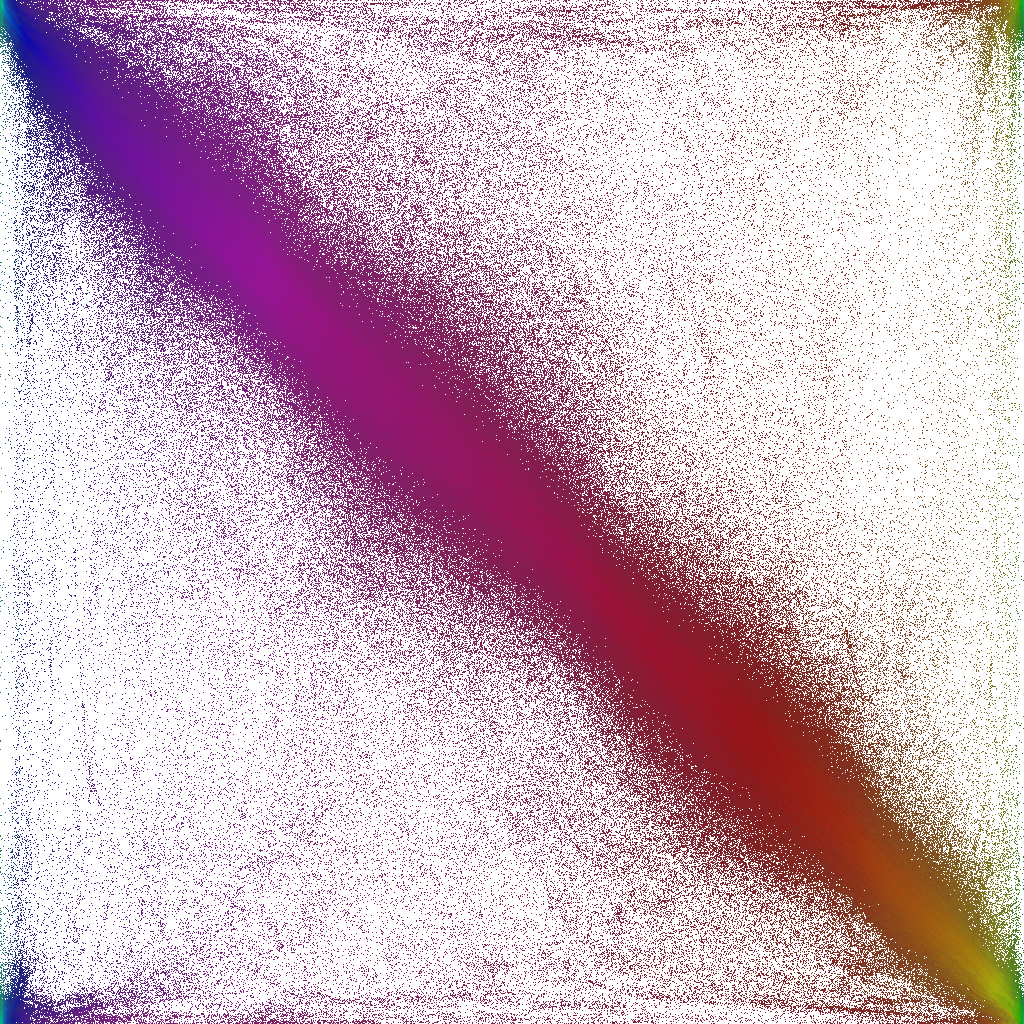}
\includegraphics[width=0.31\textwidth]{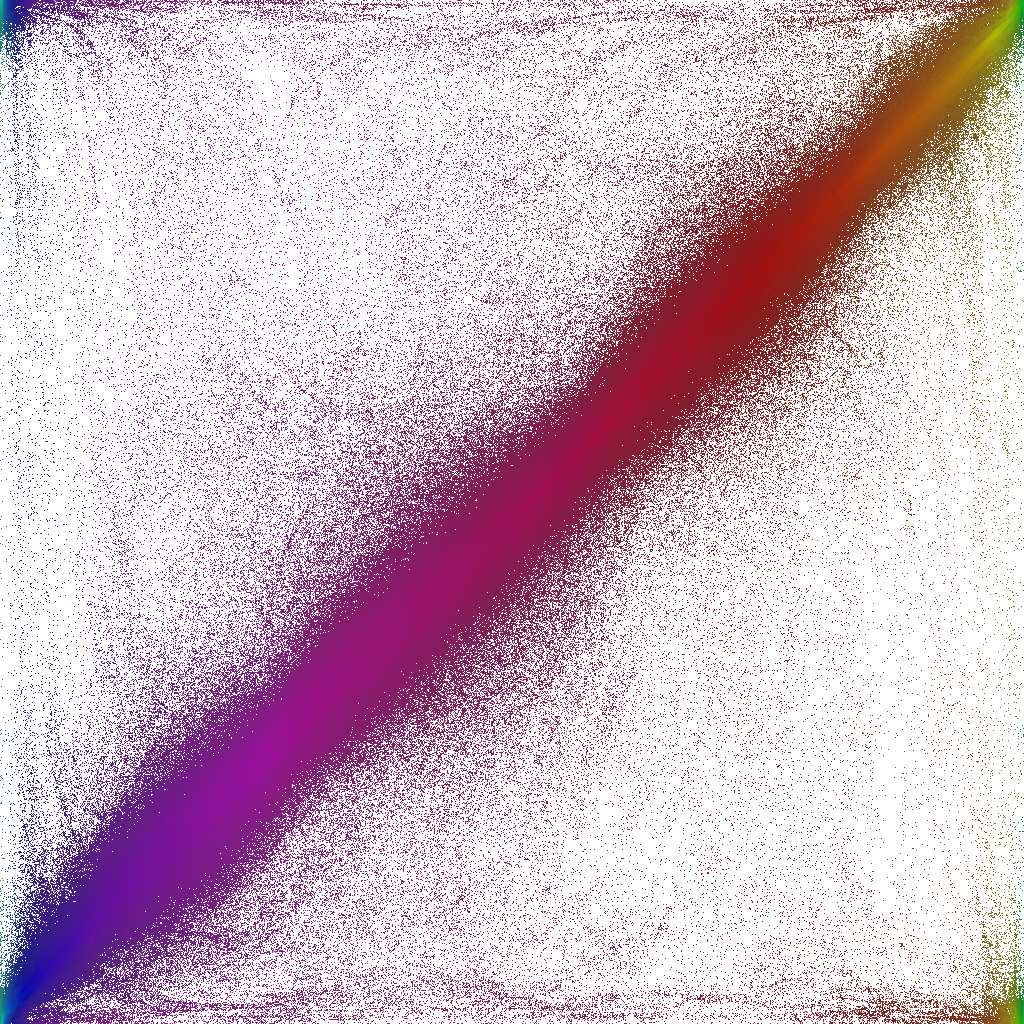}
\includegraphics[width=0.31\textwidth]{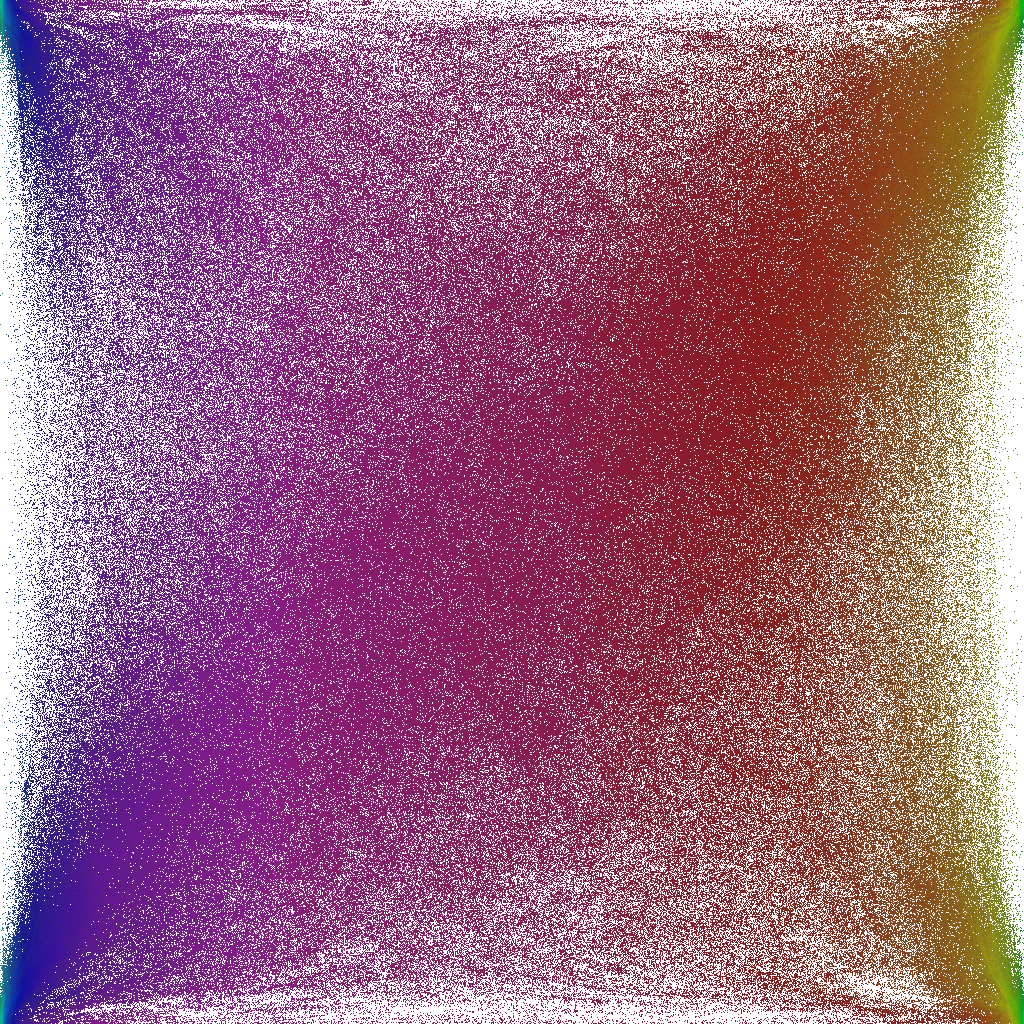}\\
\includegraphics[width=0.31\textwidth]{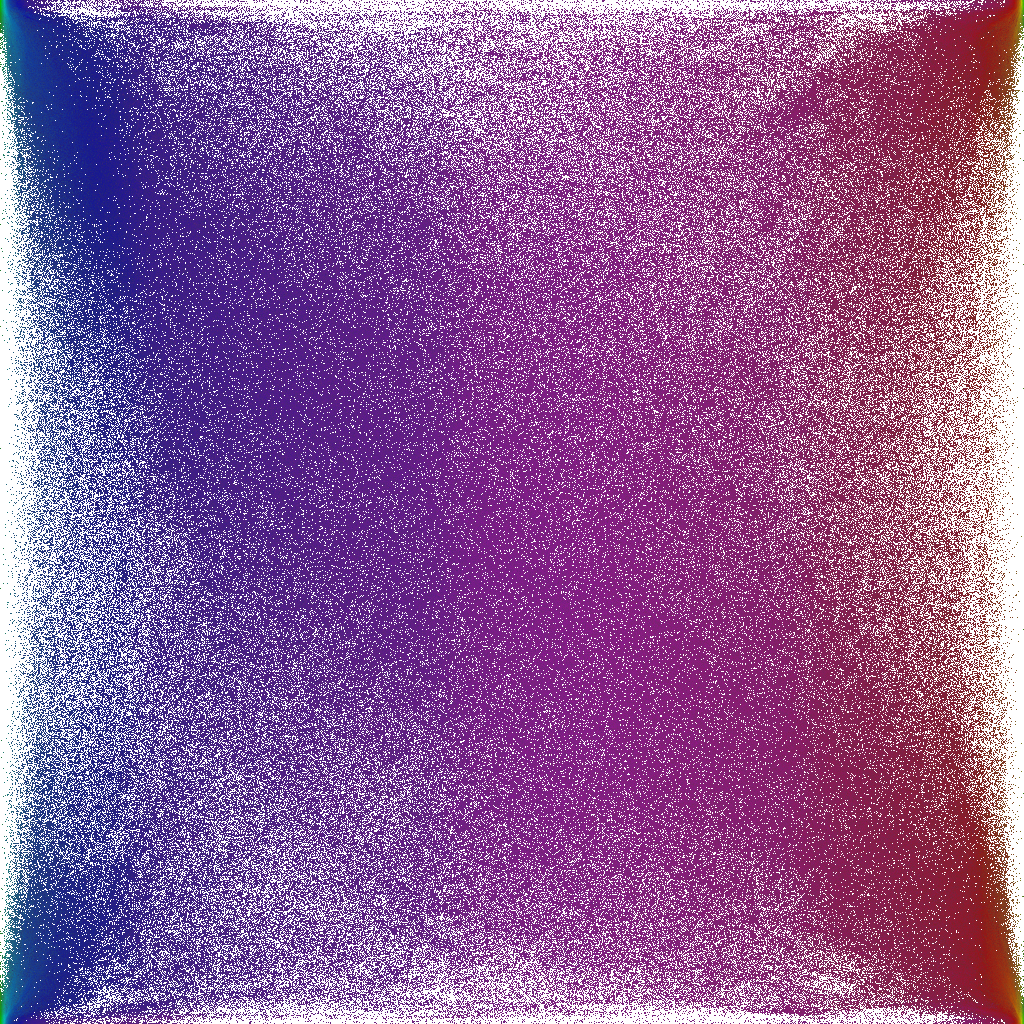}
\includegraphics[width=0.31\textwidth]{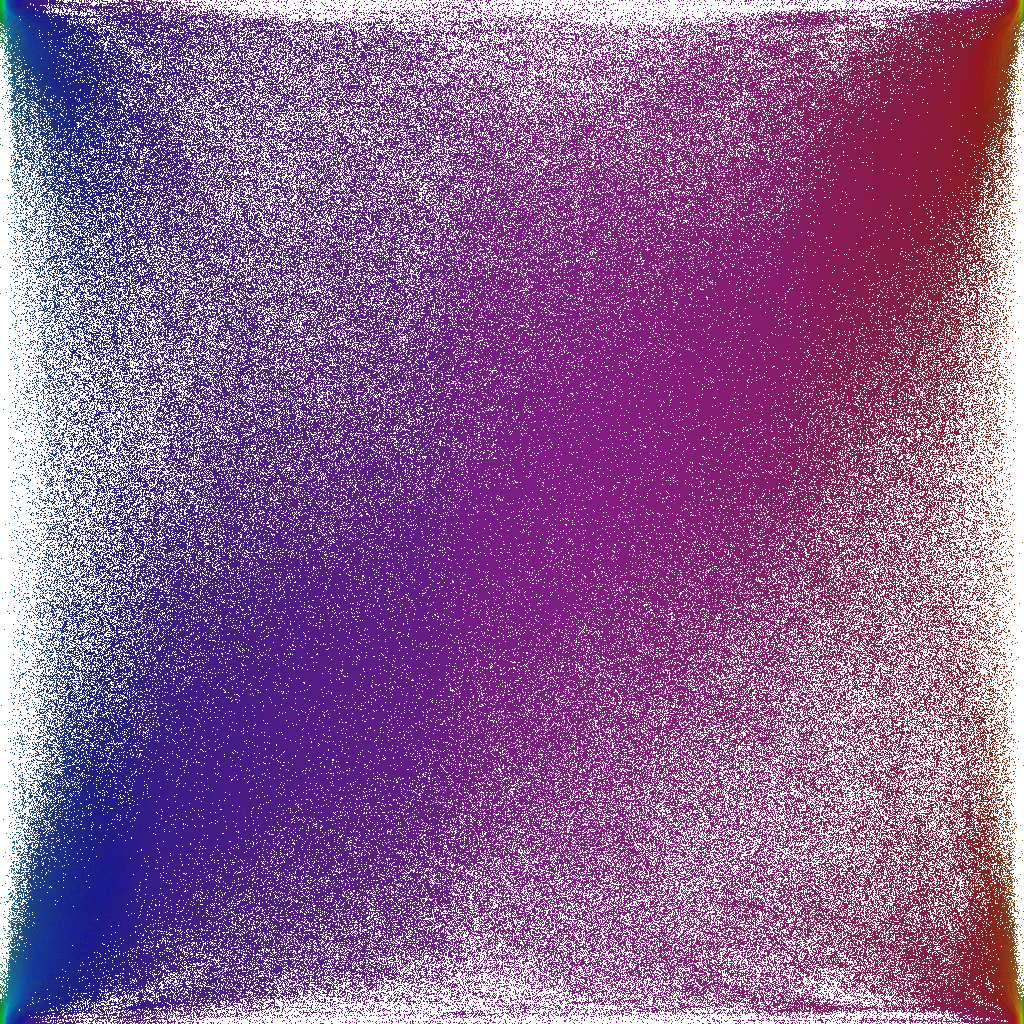}
\includegraphics[width=0.31\textwidth]{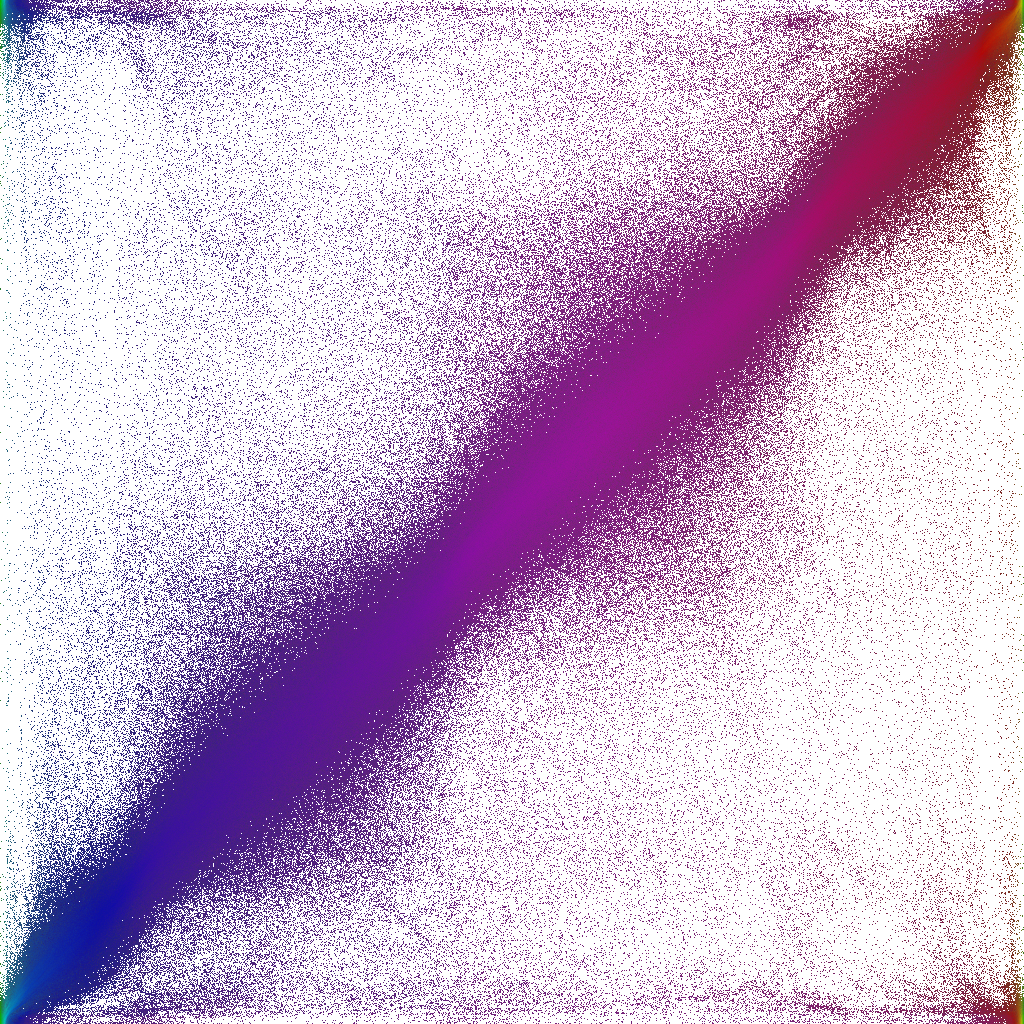}
\caption{Configurations in phase $C_b$ in $\beta$ coordinates for $T = 20$, $\Delta=0.2$ and $\kappa_0=2.5$ (top), $\kappa_0=2.0$ (middle), $\kappa_0=1.5$ (bottom). The left-hand side charts are projections on the $t - x$ plane, the middle charts on the $t-y$ plane and the right-hand side charts on the $t-z$ plane.}\label{fig:beta_Bif}
\end{figure}

\begin{figure}[H]
\centering
\includegraphics[width=0.31\textwidth]{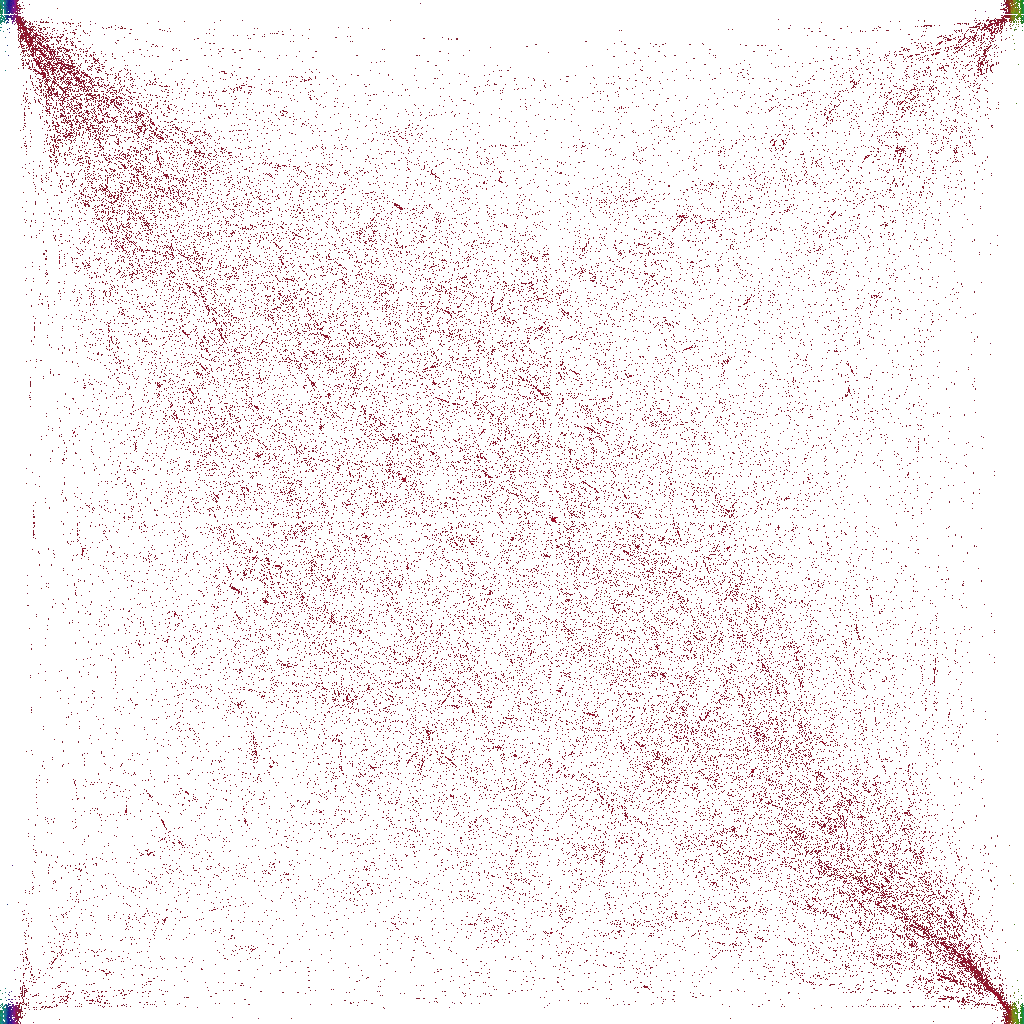}
\includegraphics[width=0.31\textwidth]{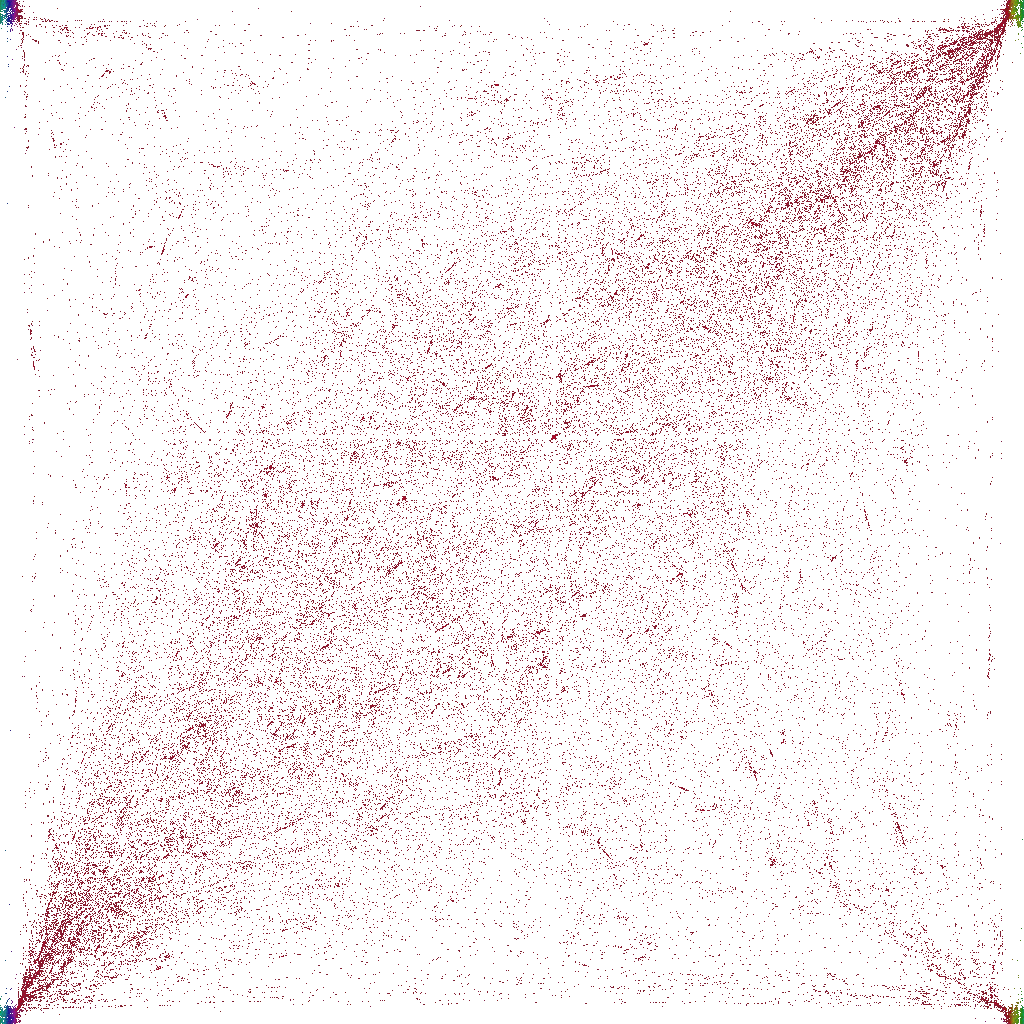}
\includegraphics[width=0.31\textwidth]{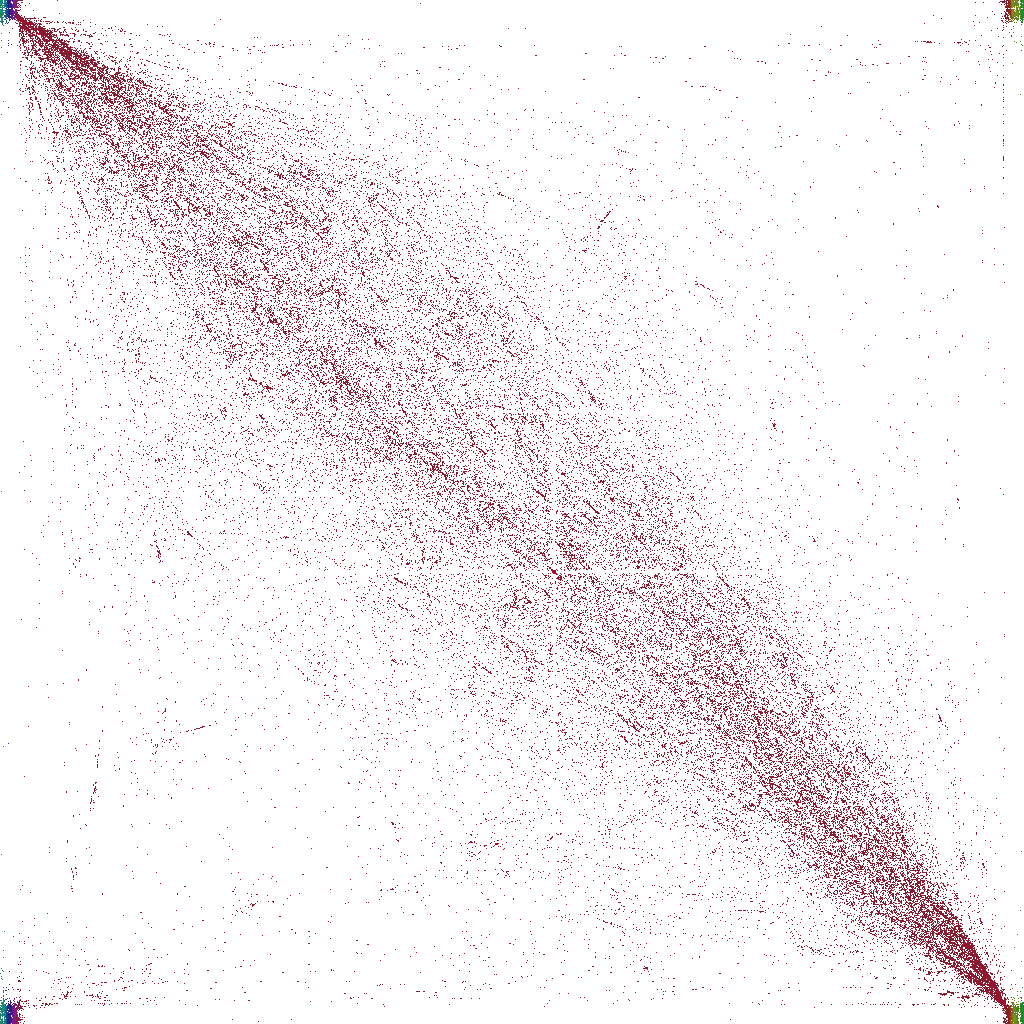}
\caption{A configuration in phase $B$ for $T = 4$ ($\kappa_0=4.4, \Delta=-0.7$) in $\beta$ coordinates. The left-hand side chart is a projection on the $t - x$ plane, the middle chart on the $t-y$ plane and the right-hand side chart on the $t-z$ plane.}\label{fig:beta_B}
\end{figure}

To visualize and analyze in detail the internal structure of geometric outgrowths, i.e., of the dense clouds of points in figures \ref{fig:cosmic_C} - \ref{fig:cosmic_A}, another parametrization might be more suitable. It can be introduced by first sorting all $ \tilde \ph $ field values so that
\beql{sorted}
0 \leq \tilde \ph_{i_1} \leq \tilde \ph_{i_2}\leq ... \leq \tilde \ph_{i_{N_4}} < 1,
\eeq
and then defining the map
\beql{BetaMap}
\tilde \ph \to \beta: \quad \beta_i = \frac{i}{N_4},
\eeq
where $i$ is the index (field position) in the sorted list \rf{sorted}. $\beta$ is by definition in the range. $[0,1]$. Since $\tilde \ph$ is a (discrete) harmonic function, $\beta$ monotonically interpolates between both sides of the elementary cell and thus can serve as a relational coordinate. It follows from the definition that the new $\beta$ coordinates will be stretched in the range where $\tilde \ph$ is dense and compressed where $\tilde \ph$ is sparse. As a result, the fractal geometric outgrowths get magnified relative to the ``central'' part of a triangulation; see figures \ref{fig:beta_C}~–~\ref{fig:beta_B}. Interestingly, the qualitative picture of generic triangulations does not change significantly in the semiclassical phase $C$, which suggests that the geometric outgrowths observed in this phase are small and shallow, as in figure~\ref{fig:beta_C}, where the voids and filaments structure is still visible in the $\beta$ coordinates. This is not the case in the other phases, as shown in figures \ref{fig:beta_A}~–~\ref{fig:beta_B}, where the new coordinates reveal much finer structures inside bigger and deeper outgrowths.

The new coordinates do not change qualitatively the results of the analysis of a phase $A$ configuration, where one still observes a number of separated and uncorrelated spacetime points giving rise to a quite homogeneous and isotropic geometry. The results observed in the bifurcation phase $C_b$ are more interesting, and they seem to change as one goes from the $C-C_b$ phase transition towards the $C_b-B$ phase transition; see figure~\ref{fig:beta_Bif} where we plot configurations for fixed $\Delta=0.2$ and various $\kappa_0=2.5$ (close to phase $C$), $\kappa_0=2.0$ (in the middle of phase $C_b$) and $\kappa_0=1.5$ (close to phase $B$). The top charts in figure~\ref{fig:beta_Bif} can be interpreted as a magnification of a single fractal outgrowth observed for $\kappa_0=2.5$ in figure~\ref{fig:cosmic_Bif} (bottom) in various directions, while middle and bottom charts are magnifications of similar outgrowths observed for $\kappa_0=2.0$ and $\kappa_0=1.5$, respectively. In each case, one clearly observes the time evolution of a very compact geometric object with no clear internal fine structure. For the configuration closest to phase $C$, the geometry is isotropic in all directions (top charts). This isotropy is broken as one approaches phase $B$ (middle and bottom charts). At the same time, the internal structure of the outgrowth becomes increasingly homogeneous, which manifests itself as a ``pillow-like'' picture.\footnote{We checked very carefully that the lack of any fine structures is not a result of finite numerical precision of the classical scalar field solution.}
It would be tempting to interpret such configurations as quantum spacetimes collapsing to a singularity, and in that case the observed anisotropy could be consistent with the BKL scenario. Finally, in phase $B$ the qualitative picture is quite similar, as shown in figure~\ref{fig:beta_B}, where no fine structure of the magnified outgrowth (i.e., the point in figure~\ref{fig:cosmic_B}) is observed, and the configuration looks quite  isotropic in all directions.

\section{Alternative spacetime foliations}\label{sec:foliations}

\begin{figure}[H]
\centering
\includegraphics[scale=1.2]{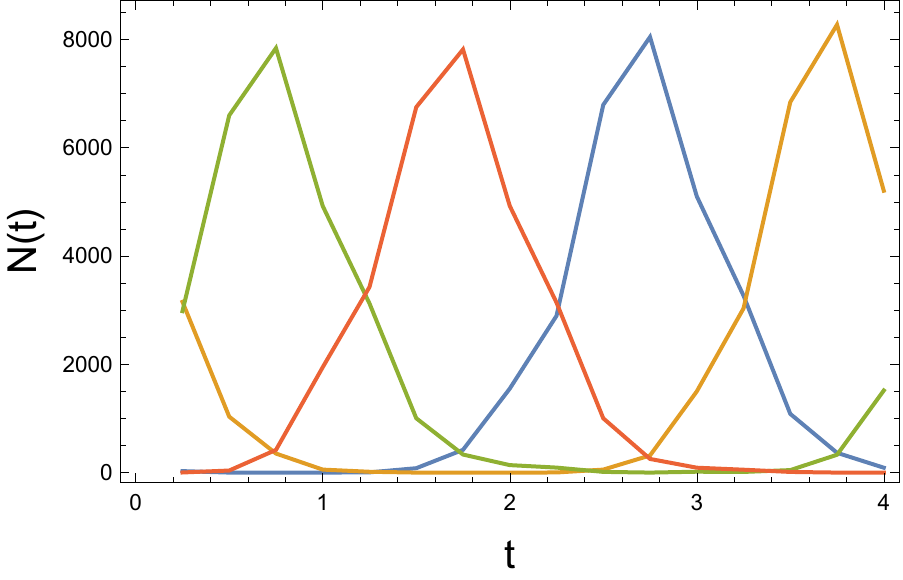}
\caption{Histograms of the original proper-time coordinate $t$ of simplices adjacent to the new boundary $H(\alpha_t)$ defined by the jump of the scalar field \rf{def2}. The data were measured for a generic triangulation in phase $C$. The time position of the simplices, and thus also the boundary, shifts with $\alpha_t$. In the histograms we used non-integer $t$ coordinates, depending on the simplex type, as explained in Section \ref{sub:maps}. 
}\label{fig:alphahist}
\end{figure}

As already mentioned in Sections \ref{sec:Intro} and \ref{sec:CDT}, CDT introduces a preferred spacetime foliation parametrized by the (lattice) proper-time coordinate $t$. As a result, the spatial slices (3D hypersurfaces built from tetrahedra in each integer time coordinate $t$) constitute a natural set of boundaries orthogonal to the time direction. The new idea introduced in Section \ref{sec:class} was to consider scalar field(s) with nontrivial jump(s) of magnitude $\delta=1$ on the boundaries in the time (or in spatial) direction(s). The scalar field solutions can then act as new time coordinates, with a natural choice of 
\begin{equation}
\tilde \ph_i^{(t)}(\alpha_t) = {\rm mod}(\bar\ph_i^{(t)}-\alpha_t,1),
\label{def2}
\end{equation}
where $\bar\ph_i^{(t)}$ is the classical solution of the scalar field with a jump on some of the time boundaries (spatial slices), and which can be viewed as a field taking values in $S^1$. The solution is parametrized by the real quantity $\alpha_t$ ($0~\leq~\alpha_t~<~1$). The field $\tilde \ph_i^{(t)}(\alpha_t)$ is by definition in the range $[0,1]$ and is periodic in $\alpha_t$ with period one. As already explained, one can consider an integer quantity $b(\tilde \ph_i^{(t)}(\alpha_t))$, defined in eq.~\rf{bi}, which measures the position of the jump of the scalar field \rf{def2}, i.e., the position of the new boundary $H(\alpha_t)$ orthogonal to the time direction. The nonzero (integer) values of $b(\tilde \ph_i^{(t)}(\alpha_t))$ indicate the number of new boundary faces (depending on $\alpha_t$) of a particular simplex. For a particular value of $\alpha_t$ there is a set of simplices for which $b(\tilde \ph_i^{(t)}(\alpha_t))>0$ and a set where $b(\tilde \ph_i^{(t)}(\alpha_t))<0$. These simplices lie on two opposite sides of the ($\alpha_t$-dependent) boundary. Note that in general $b(\tilde \ph_i^{(t)}(\alpha_t))$ and $b(\bar \ph_i^{(t)})$ are not the same, and thus the new 3D boundary $H(\alpha_t)$ is different than the original one, i.e., the spatial slice in $t$. The 3-volume (the number of tetrahedra) of the $H(\alpha_t)$ hypersurface is
\begin{equation}
V(\alpha_t) = \frac{1}{2}\sum_i |b(\tilde \ph_i^{(t)}(\alpha_t))|.
\label{volume}
\end{equation}
We can determine the vertices of the boundary tetrahedra by considering a simplex with $b(\tilde \ph_i^{(t)}(\alpha_t))>0$ and checking the neighboring simplices $j$ to find those for which $b(\tilde \ph_i^{(t)}(\alpha_t))<0$. Each such case defines a boundary face (tetrahedron). We repeat the same procedure for all simplices with $b(\tilde \ph_i^{(t)}(\alpha_t))>0$ to obtain a list of all boundary tetrahedra. Once the list is constructed, we check the neighborhood relations between the tetrahedra. Finally, we obtain a list of boundary tetrahedra where for each element the first 4 entries are the vertex labels of the tetrahedron, and the remaining 4 are the indices of tetrahedra opposite to the vertices (similar to the way we code 4D simplices in a CDT triangulation). The list is the analogue of a 3D foliation we used before to describe spatial slices, but now it is parametrized by $\alpha_t$. In all cases described here, the systems were periodic in time with the period $T=4$. The new 3D hypersurfaces $H(\alpha_t)$ shift with $\alpha_t$ as expected and are smeared along the original proper-time coordinate, as illustrated in figure~\ref{fig:alphahist}.

Obviously, in the toroidal spatial topology case examined here, 
a similar analysis can be performed also in the spatial directions.
One can introduce a set of four fields $\tilde \phi^\mu,\, \mu = x,y,z,t$ and the corresponding boundaries $H(\alpha_\mu)$ in the way already discussed, and then the hypersurfaces will be parametrized (shifted) by $\alpha_x$, $\alpha_y$, $\alpha_z$ and $\alpha_t$, respectively.

\subsection{The topology of the hypersurfaces $\bm{  H(\alpha)}$}\label{sub:topology}

\begin{figure}[H]
\centering
\includegraphics[scale=1.2]{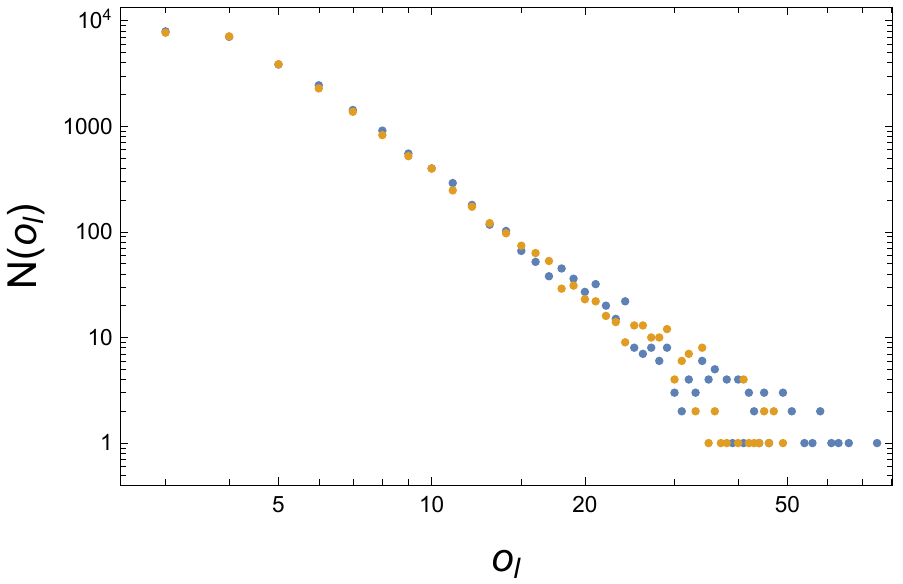}
\caption{Histograms of the order of links (related to 3D curvature) for $\alpha_t=0$ (blue) and $\alpha_t=0.5$ (orange) in a $C$ phase configuration ($\kappa_0 = 2.2,~ \Delta= 0.6 $ and $T=4$). }\label{fig:Olinks}
\end{figure}

\begin{figure}[H]
\centering
\includegraphics[scale=1.2]{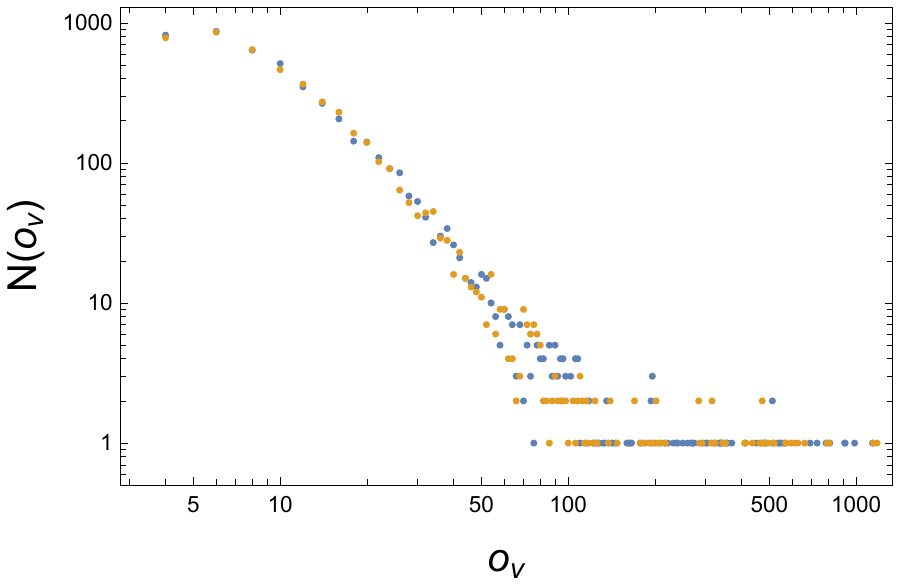}
\caption{Histograms of the order of vertices for $\alpha_t=0$ (blue) and $\alpha_t=0.5$ (orange) in a $C$ phase configuration
($\kappa_0 = 2.2,~ \Delta= 0.6 $ and $T=4$). }\label{fig:Overt}
\end{figure}

The first question to be asked is whether the 3D hypersurfaces obtained by the new foliation method outlined above are connected. This can easily be checked. We start from a random tetrahedron belonging to the hypersurface and move out measuring the volume distribution at the geodesic distance $r$ and, eventually, the total volume of the connected part of the hypersurface. We know the total volume $V(\alpha_t)$ defined by equation \rf{volume} and can check if all tetrahedra were visited. In all studied cases, they were all visited, and all hypersurfaces in the time direction (and similar hypersurfaces in all spatial directions) were fully connected. The studied cases were configurations from various CDT phases, and we checked the connectivity for many values of $\alpha$ in each spacetime direction. The conclusion is that in the case of CDT with the toroidal spatial topology\footnote{Here we consider systems with the toroidal spatial topology, so one can also define boundaries orthogonal to all three spatial directions, but one can study in the described way the scalar field coordinates and foliations in time direction for systems with a spherical or any other spatial topology.} the proposed method permits to define a set of connected 3D hypersurfaces in all spacetime directions. In each direction, these can be viewed as spacetime foliations, similar to those studied in a standard approach with the $t$ time foliation and 3D geometric states formed by tetrahedra. The second question is whether the 3D hypersurfaces satisfy the regular manifold conditions and thus preserve the 3D toroidal topology of the original spatial slices. This implies, for instance, that each triangle belonging to a hypersurface is a face of exactly two tetrahedra. In other words, each tetrahedron should have exactly 4 neighbors. We analyzed the neighborhood relations between tetrahedra belonging to the hypersurfaces and found that across a triangular face a tetrahedron could have 1, 3, 5 or a larger odd number of neighbors. This means that a triangle could belong, respectively, to 2, 4, 6 or more tetrahedra. Consequently, tetrahedra could have more than 4 neighbors. Their numbers are always even, and we found cases where the number of neighbors was 14, but larger even values are not excluded. We checked hypersurfaces in the $C$ phase for $\alpha_t=0$ and $\alpha_t=0.5$. In both cases we measured
the Euler characteristic
\beql{Euler}
\chi = N_3-N_2+N_1-N_0,
\eeq
(here $N_0,~N_1,~N_2,~N_3$ are the numbers of vertices, links, faces and tetrahedra forming  a given hypersurface $H(\alpha)$) which was large and negative (-208 and -142 respectively). In figure~\ref{fig:Olinks} we show distributions of the order of links in the two cases. We also checked the order of vertices. They range up to approximately 1200, see figure~\ref{fig:Overt}.

The conclusion at this point is that the new 3D foliation leaves $H(\alpha_t)$  {\it are not regular manifolds} and that multiple realizations of a sub-simplex with the same set of vertex labels do appear. However, the connectivity condition is still satisfied. In our Monte Carlo algorithm, we explicitly check the manifold (topology) conditions for the original time foliation into spatial slices. This is apparently not controlled by the Laplace solution of the classical scalar field. Looking also at the spatial directions, one may ask whether the original (locally minimal) boundaries used in our code are free of the topological defects described above. As already said, this is obviously true for the original time foliation, but checking the properties of spatial boundaries one finds that the algorithms we use produce geometric irregularities on boundaries of a similar nature as the $\it \alpha$-hypersurfaces. In the code we do not check if such irregularities appear, and indeed they may be produced. 

Finally, one may ask the question whether our interpretation of using $\alpha_t$ hypersurfaces (and similar hypersurfaces for $\alpha_x$, $\alpha_y$ and $\alpha_z$) as boundaries separating elementary cells is valid? What we mean is that irregularities of such hypersurfaces may lead to a situation where a part of a 4D elementary cell gets disconnected from the bulk by the irregular outgrowth on the hypersurface. We explicitly checked that such a situation never happens, i.e., each elementary cell is fully connected by 4D dual links, which do not cross the hypersurface. In the next subsection we will explain these observations.

\subsection{The hypersurfaces $ \bm{ H(\alpha)}$  evolved via  3D Pachner moves}\label{sub:evolution}

\begin{figure}[H]
\vspace{-1cm}
\centering
\includegraphics[scale=0.20]{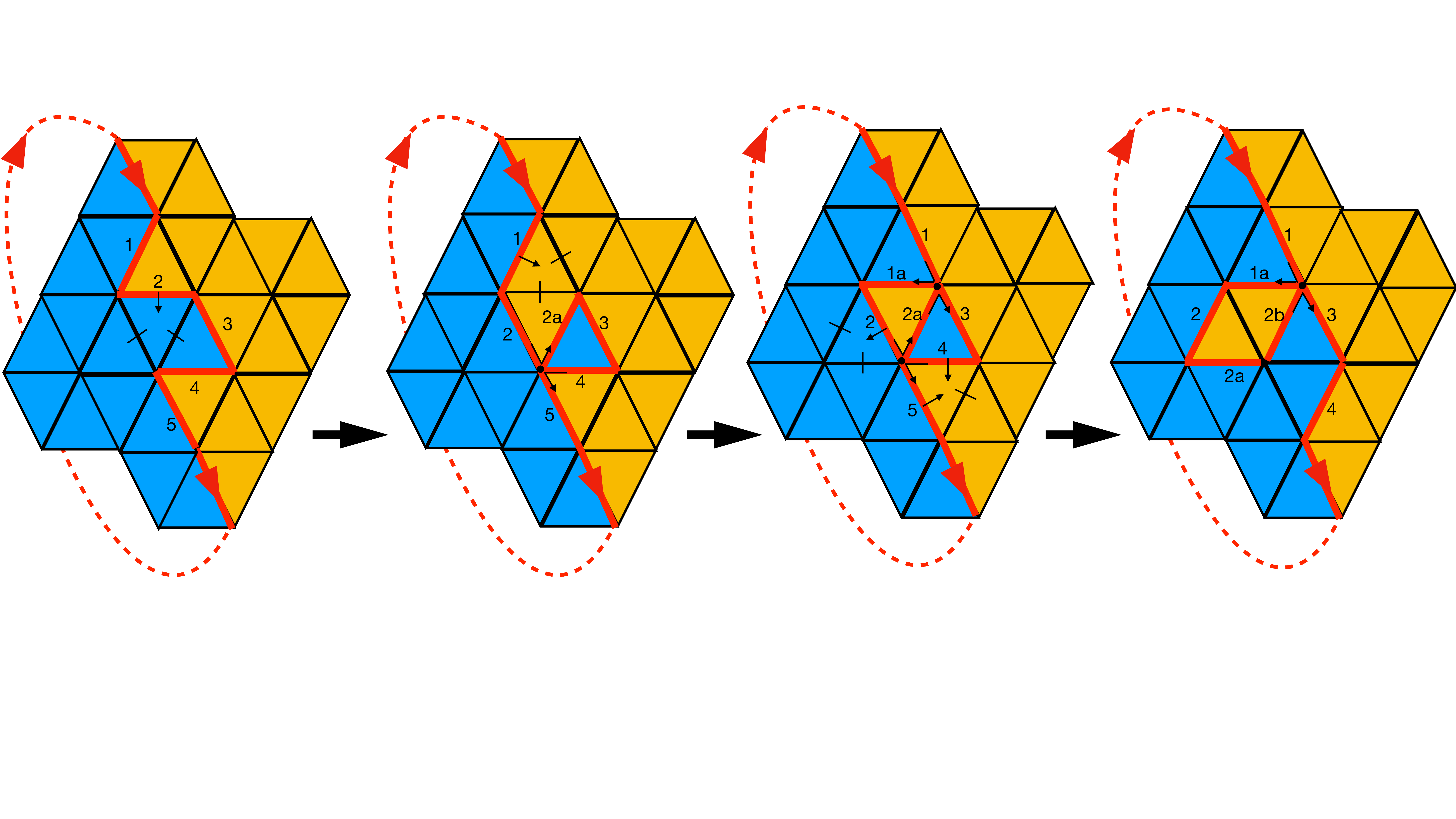}
\vspace{-2cm}
\caption{Shown is a part of a triangulation of a two-dimensional torus and a non-contractible boundary. First, we perform a Pachner move to transfer a blue  triangle to the other side of the boundary. We thereby create an outgrowth, as seen from the two-dimensional triangulation. The Euler characteristics $\chi$ decreases from 0 (the value for a closed curve) to -1, unless (as we would do if we viewed the Pachner move entirely from a one-dimensional point of view) we assign two vertices to the pinching point (or the intersection). In the next move, we create another outgrowth and another pinching point, and the Euler characteristics changes to -2. Finally, the last move removes an outgrowth, but there still remains one outgrowth and the 
Euler characteristics is -1.}
\label{figjan}
\end{figure}

Superficially, one may think that the variable $\alpha$ is continuous and that by varying it we get a 
continuous evolution of the 3-hypersurface $H(\alpha)$ defined by the jump of the classical scalar field solution \rf{def2}. On a discretized manifold this is however not the case. Suppose we analyze the hypersurface $H(\alpha)$ obtained for a particular value of $\alpha$ in one of the four directions, and the range of values for the field for this $\alpha$ is $\epsilon \leq \tilde\ph_i(\alpha) <1$, where $\epsilon>0$ is the minimal value of the field distribution observed at some (single) simplex $i_{min}$. If then we change $\alpha$ to $\alpha+\Delta \alpha$, where $\Delta \alpha< \epsilon$, it is clear that
\begin{equation}
b(\tilde\ph_i({\alpha+\Delta \alpha})) = b(\tilde\ph_i({\alpha})),
\end{equation}
and, consequently, the two hypersurfaces $H(\alpha)$ and $H(\alpha+\Delta \alpha)$ are identical.
Only if $\Delta \alpha$ becomes a little larger than $\epsilon$, the value of $b(\tilde\ph_{i_{min}}({\alpha+\Delta \alpha}))$ changes, and the simplex $i_{min}$ is moved to the other side of the boundary. The two hypersurfaces differ only by the position of this single simplex. Let us analyze what it means for the hypersurface $H(\alpha)$. The effect can be viewed as performing one of the so-called 3D Pachner moves on the hypersurface. Let us here recall that for triangulations in $d$ dimensions the Pachner moves are local changes described as follows: consider $n$ d-dimensional simplices in the triangulations, $n=1,\ldots,d+1$, which are glued together in such a way that they form a part of the boundary of a $d+1$-dimensional simplex. The (closed) boundary of the $d+1$-dimensional simplex has $d+2-n$ other $d$-dimensional simplices, which are also glued together. These two sets of $d$-dimensional simplices share a boundary consisting of $d-1$-dimensional simplices. Thus, one can replace the $n$ $d$-dimensional simplices in the original triangulation with the other $d+2-n$ simplices from the boundary of the $d+1$-dimensional simplex. There are $d+1$ types of such moves, one for each $n$. It is clear that this is precisely the situation we have in our case. We are given a hypersurface $H(\alpha)$, i.e., a three-dimensional triangulation. The way we change it is by ``moving'' a four-dimensional simplex that contains a certain number of three-simplices of the hypersurface to the other side of the boundary. In other words, we declare that the original three-simplices which belonged both to the hypersurface $H(\alpha)$ and to the given four-simplex do not belong to the hypersurface $H(\alpha+\Delta\alpha)$; instead, it is the other three-simplices of the four-simplex that belong to the new hypersurface $H(\alpha+\Delta\alpha)$: we have moved the four-simplex to the other side of the (new) hypersurface. There are only two problems with this: the Pachner moves can lead to degenerate triangulations (but with the same topology), and they may not lead to a three-dimensional manifold as viewed from the perspective of the embedding space of a given four-dimensional triangulation, as is the situation here. The situation is generic and occurs in any dimension $d$ and the reason is very simple: when performing the Pachner moves, new indices are assigned to the new vertices which were not part of the original  $d$-dimensional simplicial complex. However, if the vertices are already part of a given $d+1$-dimensional triangulation, and have some labels there, which we do keep, there is a chance that while performing the Pachner move we meet a vertex with the same label several times. This results in a situation where the $d$-dimensional triangulation may have self-intersections when viewed from the $d+1$-dimensional triangulation perspective, while from the point of view of Pachner moves in $d$-dimensions, the self-intersection vertices would have gotten different indices with 
no reference to an embedding space. This is precisely what we have observed, and we have illustrated the situation in the simplest of all cases, namely $d=1$,
in figure~\ref{figjan}.\\

Consequently, one can conclude that:
\begin{itemize}
\item The evolution of a hypersurface $H(\alpha)$  {is not continuous} in $\alpha$ but can be viewed as a discrete series of modifications of a boundary hypersurface. In each step, one or more simplices of the manifold are moved to
the other side of the boundary. This happens only for a discrete set of values of $\alpha$, which is an effect of the finite system size and of the discreteness of geometry. 
\item Each shift of the boundary $H(\alpha)$  can be viewed as a result of performing a number of 3D Pachner moves of the boundary.
\item $H(\alpha)$ hypersurfaces, viewed as embedded in a 4D CDT manifold, will in general not be 3D manifolds, but they are \emph{almost} manifolds in some sense, since a suitable additional labelling can turn them into 3D manifolds with the topology of 3-torus. A lower-dimensional analogy is a crumpled piece of paper smeared with glue, which causes the folding points to stick together.
\item Our algorithm to modify a (locally minimal) boundary in the 4D setup can also be interpreted in this setting. 
\end{itemize}
To summarize, the interpretation of the change of the hypersurfaces $H(\alpha)$ with $\alpha$ as a sequence of Pachner moves explains the properties 
of the surfaces that we have observed in Subsection \ref{sub:topology} above.\footnote{It should be noted that in the EDT simulations one usually uses the Pachner moves in a more restricted way, requiring that the moves should only create new triangulations where simplices are uniquely defined by their vertices. That will in general not be the case in an unrestricted use of the Pachner moves. However, even with their unrestricted use the underlying topology of the triangulation is not changed. The spurious change in topology we observe comes entirely from the embedding, as explained above.}

\subsection{The spatial volume distribution of the $\bm{H(\alpha)}$-hypersurfaces}

\begin{figure}[H]
\centering
\includegraphics[scale=0.95]{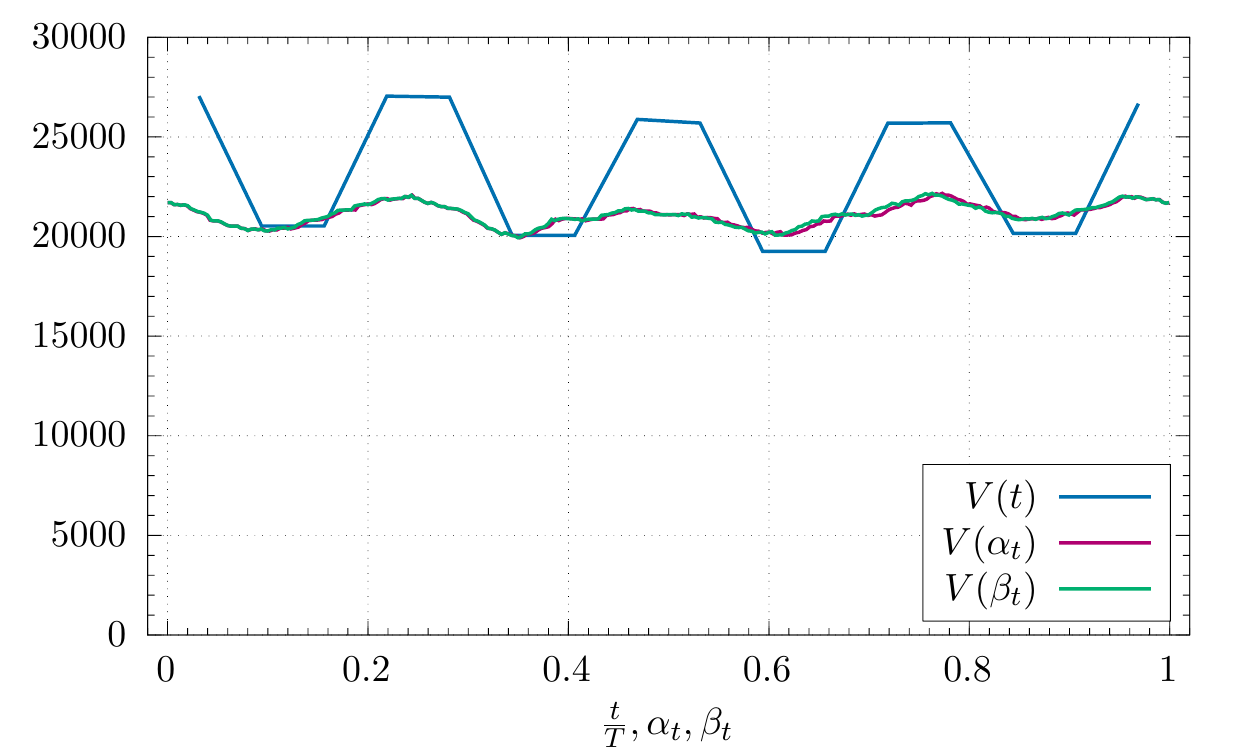}
\caption{The $\alpha_t$ and $\beta_t$ profiles in a single configuration in phase $C$ $(\kappa_0=2.2,~\Delta=0.6)$ with $T=4$, and the corresponding $t$-profile.
The $t$-profile was shifted to match the time values corresponding to the maxima of the $\alpha_t$ profile (see figure \ref{fig:alphahist}).
}\label{fig:Cprofile}
\end{figure}

\begin{figure}[H]
\centering
\includegraphics[scale=0.95]{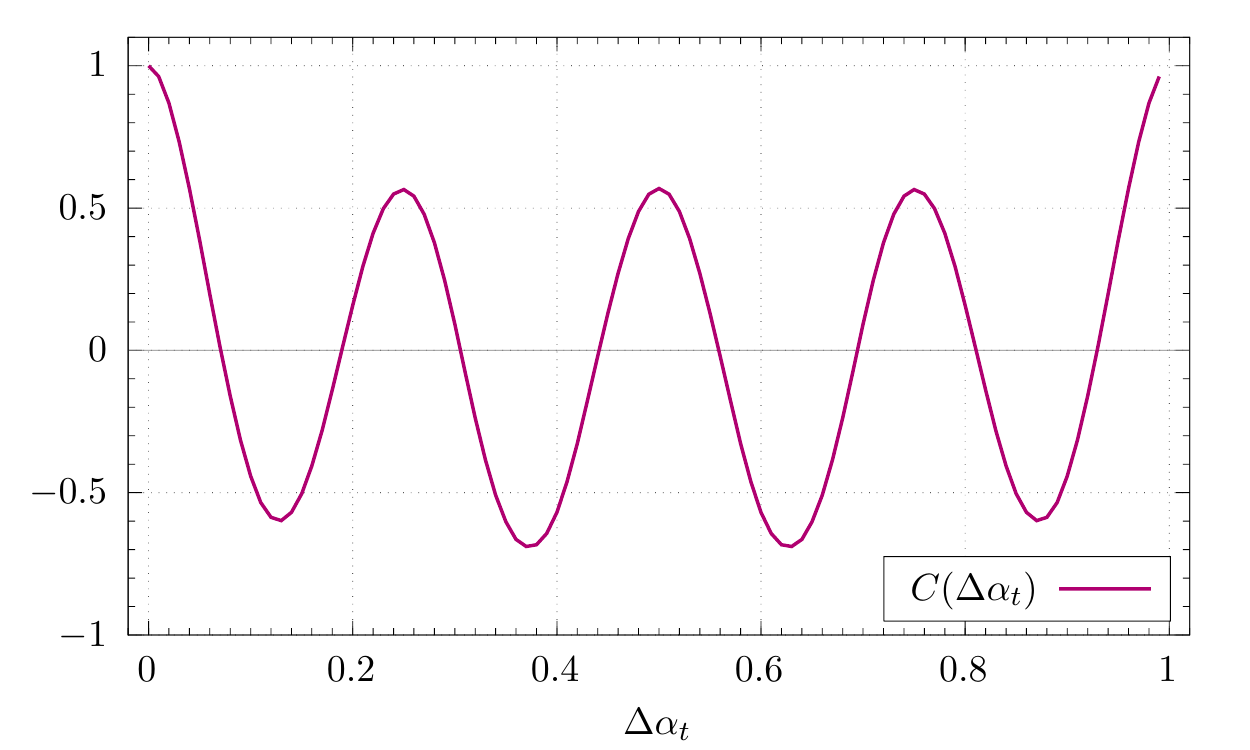}
\caption{Covariance of the $\alpha_t$-profile as a function of $\Delta\alpha_t$ (normalized by $C(0)=1$) in a single configuration in phase $C$ $(\kappa_0=2.2, ~\Delta=0.6)$ with $T=4$.}
\label{fig:cov}
\end{figure}

Varying $\alpha_t$ in the range between $0$ and $1$, for each configuration, one can measure the distribution of $V(\alpha_t)$, defined by eq.~(\ref{volume}), called here the {\it $\alpha_t$-profile}.
Below we illustrate the shape of $\alpha_t$-profiles for generic configurations in different CDT phases, starting with the semiclassical phase $C$, see figure~\ref{fig:Cprofile}. Values of $\alpha$ in each plot were taken in 100 steps of .01 (so $\alpha_\mu^i \equiv ({i-1})/{100}, i = 1,\dots, 100$). All measured systems were single configurations with the proper-time coordinate period $T=4$. In the plots we also show the volume profiles in the original proper-time coordinate (rescaled to fit the $[0,1]$ range), the $t$-profiles. We use generalized $t$ coordinates, in which we assign integer $t$ to each $(4,1)$ simplex and non-integer time coordinates $t +\frac{1}{4}$, $t +\frac{1}{2}$ and $t +\frac{3}{4}$ to the $(3, 2)$, $(2, 3)$ and $(1, 4)$ simplices, respectively, as discussed in Section~\ref{sub:maps}. The original proper-time foliation ($t$-profile) volume structure is also apparent in the new $\alpha_t$-profile function.
One can also measure the covariance function
\begin{equation}\label{covariance}
C(\Delta\alpha_\mu) = \frac{1}{\cal N}\sum_{i} (V(\alpha_\mu^i)-\bar{V})(V({\rm mod}(\alpha_\mu^i+\Delta\alpha_\mu\, ,1))-\bar{V}).
\end{equation}
$C(\Delta\alpha_t)$, normalized to be 1 at $\Delta\alpha_t=0$, for a single configuration in the $C$ phase is plotted in figure~\ref{fig:cov}. In this plot, the four layers are even more visible. Remember that the steps of $\alpha_t$ are $.01$, and one has all possible layers ($(4,1)$, $(3,2)$, $(2,3)$ and $(1,4)$).
 
\begin{figure}[H]
\centering
\includegraphics[scale=0.95]{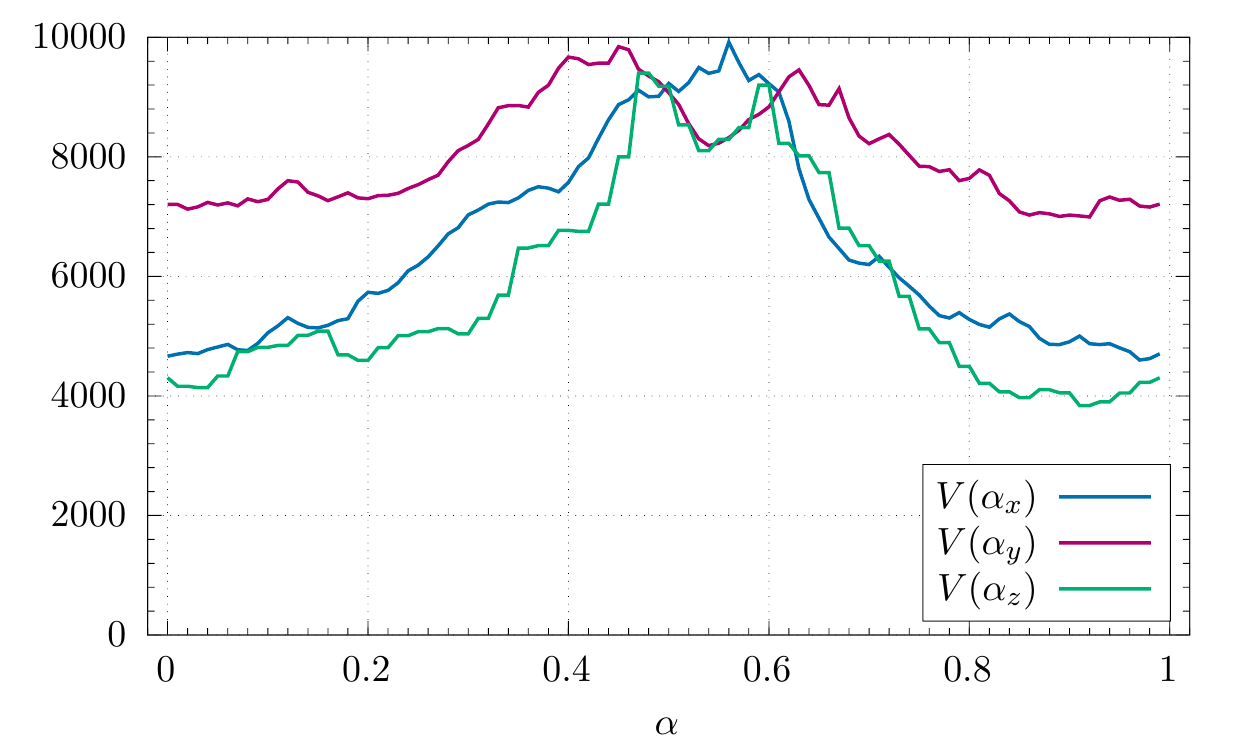}
\caption{The $\alpha_x,~\alpha_y$ and $\alpha_z$-profiles
in a configuration in phase $C$.
}\label{fig:CXprofile}
\end{figure}

\begin{figure}[H]
\centering
\includegraphics[scale=0.95]{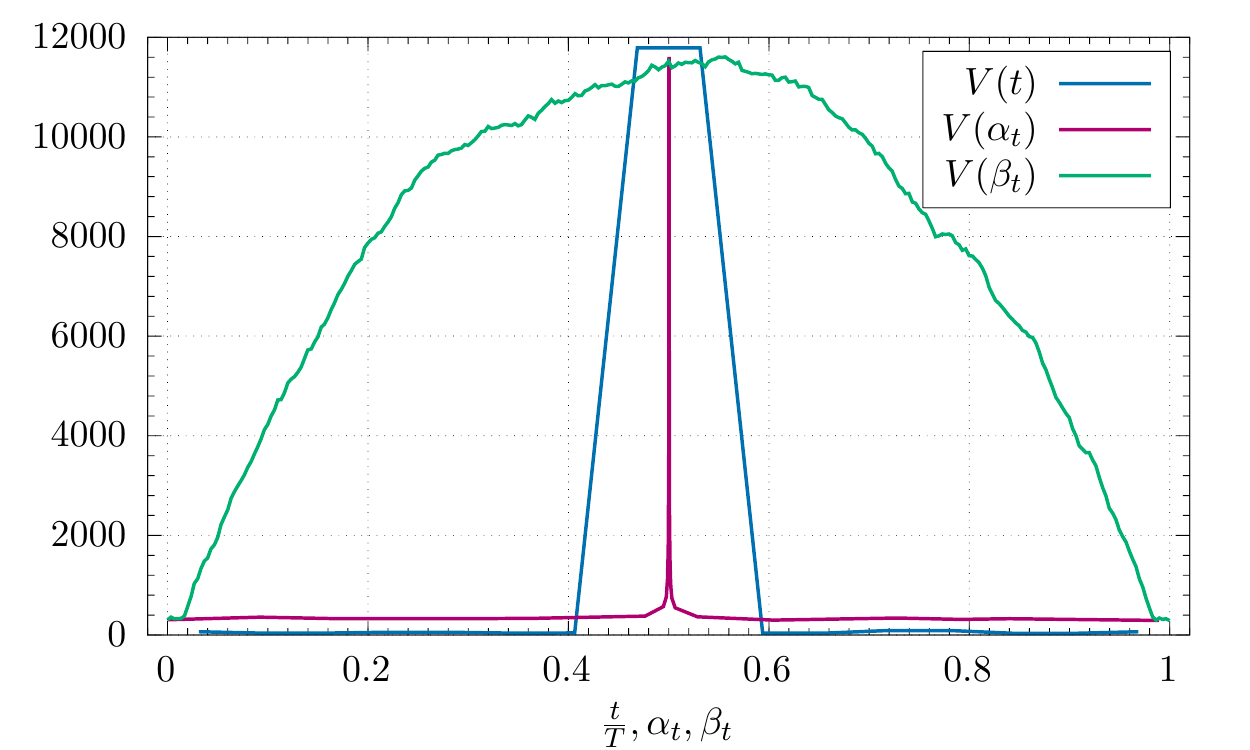}
\caption{The $\alpha_t$ and $\beta_t$ profiles in a single configuration in phase $B$ $(\kappa_0=4.4,~\Delta=-0.07)$ with $T=4$, and the corresponding $t$-profile.
Both profiles were shifted to place the maxima in the center of the plot.
The $t$-profile was additionally scaled by a factor $0.15$.
Superficially, the $t$-profile looks wider than the $\alpha_t$-profile, but this simply results from a low ``resolution'' of the $t$-profile which takes only $4\times T=16$  values in the time direction.
}\label{fig:Bprofile}
\end{figure}

\begin{figure}[H]
\centering
\includegraphics[scale=0.95]{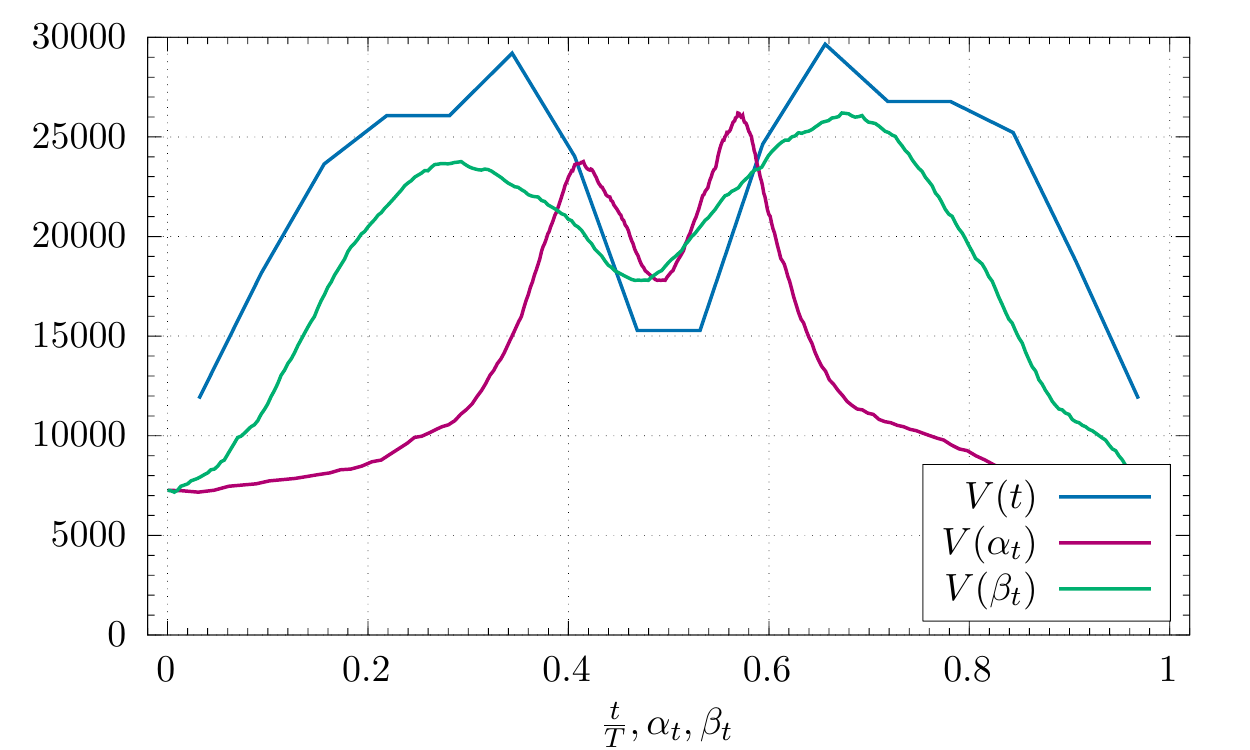}
\caption{The $\alpha_t$ and $\beta_t$ profiles in a single configuration in phase $C_b$ $(\kappa_0=2.0,~\Delta=0.1)$ with $T=4$, and the corresponding $t$-profile.}
\label{fig:CBprofile}
\end{figure}

For the toroidal CDT, the $\alpha$ volume and covariance functions can also be measured in all spatial directions. For illustration, in figure~\ref{fig:CXprofile} we show (volume) $\alpha$-profiles in the three spatial directions for the same configuration in phase $C$. The profiles can be averaged over many measured configurations, which may eventually lead to the reconstruction of the effective CDT action, now not only in time (as it was done for the original $t$ coordinate) but also in the spatial directions.
As can be seen from the volume $\alpha$-profile functions, the spatial distributions are concentrated around a certain value of $\alpha_\mu^i$, and consequently the covariance functions in the spatial directions look different than the one in the time direction.
\begin{figure}[H]
\centering
\includegraphics[scale=0.95]{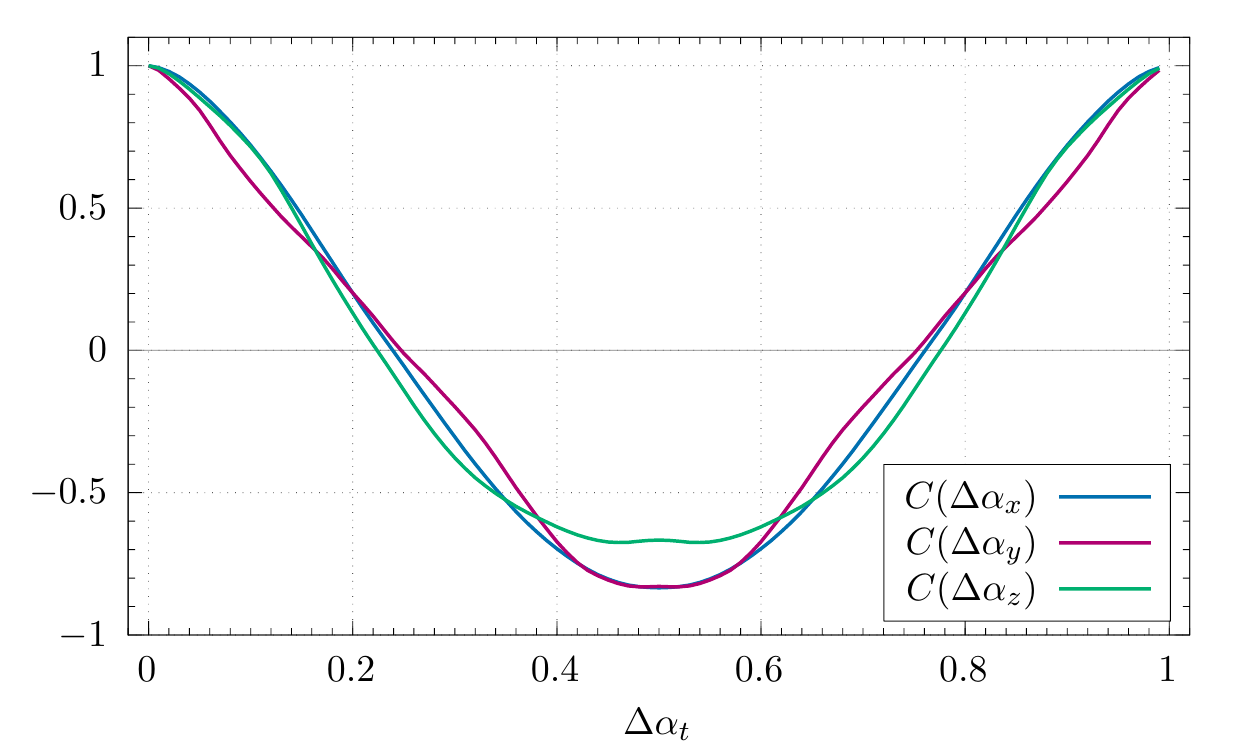}
\caption{Covariance functions in the $x,~y,~z$ directions
in a configuration in phase $C$, normalized by $C(0)=1$.}
\label{fig:covxyz}
\end{figure}
Similar plots for the $\alpha_t$-profiles in phases $B$ and $C_b$ are shown in figures \ref{fig:Bprofile}~-~\ref{fig:CBprofile}. One can see the appearance of time compactification in the $B$ phase and the typical saw-like volume structure in the $C_b$ phase, although in this case the $\alpha_t$-profile seems distorted compared to the $t$-profile. We will return to this in the next subsection.

\subsection{The spatial volume distributions in the $\beta$-parametrization}

\begin{figure}[H]
\centering
\includegraphics[scale=0.95]{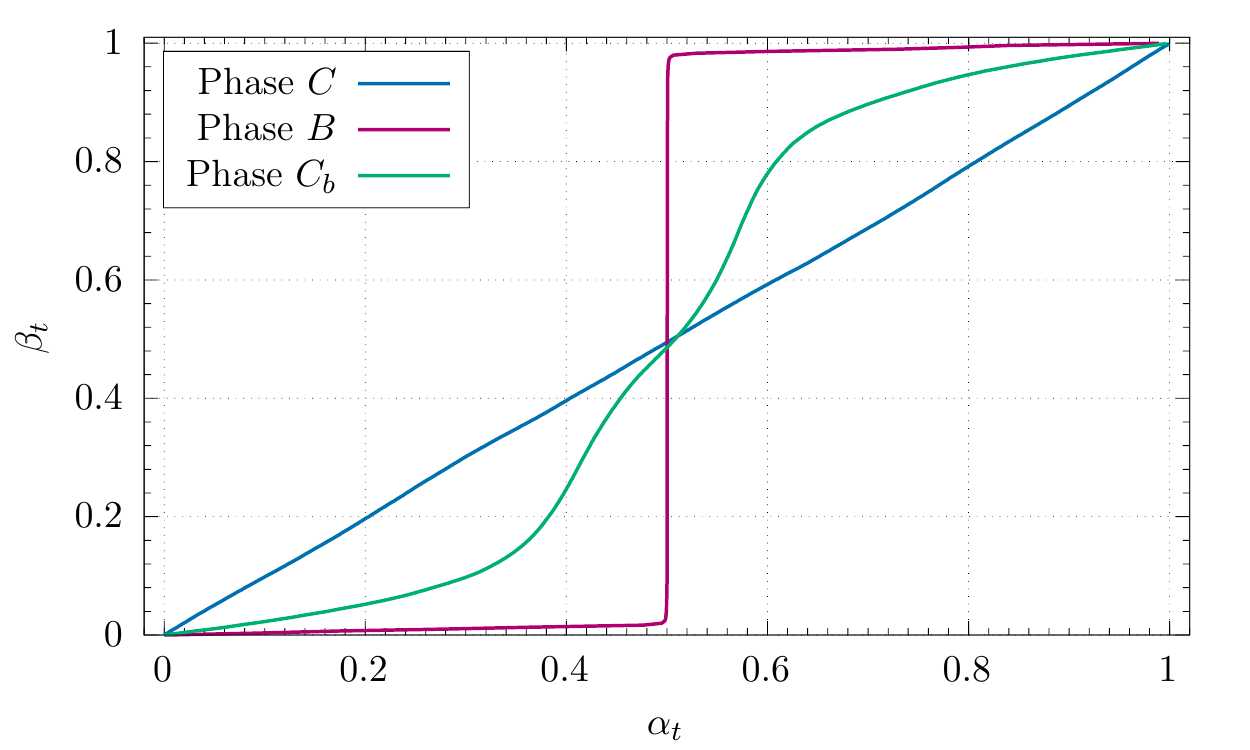}
\caption{$\beta_t$ as a function of $\alpha_t$ in a configuration in phases $C$, $B$ and $C_b$.
}\label{fig:alphabeta}
\end{figure}

By means of eqs.~\rf{sorted} and \rf{BetaMap} in section \ref{sub:maps} we introduced the $\beta$-coordinates, which, as we will now argue, are useful for measuring distances between the different foliation leaves $H(\alpha)$. Let us consider  the evolution of a boundary between $\alpha=0$ and $\alpha=1$. One can see that for increasing $\alpha$, gradually all the simplices in the manifold are moved from one side of the boundary to the other. It is tempting to define a distance between two boundaries at different values of $\alpha$ as the number of transfers of simplices necessary to evolve the boundary $\alpha$ into the boundary $\alpha'$. For each $\alpha$ we may define {$\beta(\alpha)$} as the number of transfers between the $\alpha=0$ boundary (where $\beta=0$) and the $\alpha$ boundary, normalized by the total number of simplices $N_4$. Note that this is exactly equivalent to the definition of $\beta$ used in Section \ref{sub:maps} (eq.~\rf{BetaMap}) if we set $\beta(\alpha) = \beta_i$, where $i$ is the index (field position) in the sorted list \rf{sorted} of a simplex that joins the ${H(\alpha})$ hypersurface at a given step of the boundary evolution. The new parameter $\beta$ is again in the range $0\leq \beta <1$ and can easily be measured for any configuration in each direction. In figure~\ref{fig:alphabeta} we show $\beta_t$ as a function of $\alpha_t$ (the index denotes again the time direction) in a configuration in phase $C$. One can see that the two definitions coincide in this case, and in practice $\beta_t \approx \alpha_t$. Consequently, the $\beta_t$-profile is almost identical to the $\alpha_t$-profile, as shown in figure~\ref{fig:Cprofile}. This is different in other phases. A plot of $\beta_t$ as a function of $\alpha_t$ in a configuration in phase $B$ is shown in figure~\ref{fig:alphabeta}. In this case, the whole change in $\beta_t$ is concentrated in a very narrow neighborhood of $\alpha_t \approx 0.5$, for which value we observe a blob in the $\alpha_t$-profile (conf. figure~\ref{fig:Bprofile}). As a result, almost all boundary transfers happen in this neighborhood, and the distribution of $V(\beta_t)$ is completely different than that of $V(\alpha_t)$. The difference is conspicuous in figure~\ref{fig:Bprofile}, where the narrow peak in the $\alpha_t$-profile is greatly expanded in the new $\beta_t$ parametrization. In the $C_b$ phase, the relation between $\alpha_t$ and $\beta_t$ is different yet again non-trivial, as exemplified in figures \ref{fig:alphabeta} and \ref{fig:CBprofile}. Both the peaks of the $\alpha_t$-profile are squeezed in a part of the $[0,1)$ range, leaving the rest of the profile much flatter, whereas the $\beta_t$-profile is much more regular.
    
\section{Dynamical scalar fields}\label{sec:dynamical}
The simplest quantum matter which can be added to the quantum geometry of CDT is a scalar field. Models of this type were studied in EDT and CDT, mostly for the spherical spatial topology but recently also for the toroidal spatial topology. For such models, the lattice regularized path integral of quantum gravity \rf{PI} includes also an integral over scalar fields $\ph$: 
\beql{PInew}
 {\mathcal Z}_{CDT} = \sum_\mathcal{T}\int \mathcal{D}[\ph] \rme^{- \left(S_R[\mathcal{T}] + S_M[\{\ph\},\cT] \right)}.
\eeq
The dynamical scalar field $\ph$ was
in all cases located in the simplices, and the following action of a massless field was considered:
\begin{equation}
 S_M[\{\ph\},\cT] = \frac{1}{2} \sum_{i \leftrightarrow j}(\ph_i-\ph_j)^2= \sum_{i,j} \ph_iL_{ij}\ph_j,
 \label{action}
\end{equation}
where, in the 4D case, the discrete Laplacian is given by 
\begin{equation}
 L_{ij}= 5~ \delta_{ij} - A_{ij},
 \label{lapl2}
\end{equation}
with $A_{ij}=1,0$ being the symmetric adjacency matrix on the dual lattice; see Section~\ref{sec:class} for a discussion. 
The Gaussian form of the field means that in principle the
field can be integrated out using the flat measure 
\begin{equation}
\mathcal{D}[\ph] = \prod_i{}' \frac{\rmd \ph_i}{\sqrt\pi},
 \label{measure}
\end{equation}
contributing to the geometric action $S_R[\cT] \to S_R[\cT]+ S_\mathrm{quant}^\mathrm{eff}[\cT]$ a term 
\begin{equation}
 S_\mathrm{quant}^\mathrm{eff}[\cT] = \frac{1}{2}\log \det({\bL}'(\cT)),
\end{equation}
where ${\bL}'(\cT)$ is the Laplacian matrix $\bL(\cT)$ in the subspace orthogonal to the constant zero-mode of $\bL$. In the measure 
we also eliminate the integration over the zero mode (hence the ``prime'' index in equation \rf{measure}). The dependence on geometry sits in the dependence of ${\bL}'(\cT)$ on the adjacency matrix $\bA$ defined for a given triangulation $\cT$, which is modified by geometric moves. The dynamical field $\ph$ can be rescaled $\ph\to \lambda \ph$, but this rescaling can be eliminated by the change of measure and in effect included in the redefinition of the cosmological constant.

To summarize the results of our earlier research: the inclusion of an interaction of geometry with the massless scalar field(s) did not change the geometric properties observed without such fields, at most shifting values of the coupling constants by finite numbers \cite{scalar}. Including a potential (like a mass term) suppresses field fluctuations
but also does not lead to a visible change of the geometric phase structure.
We also tried to increase the number of scalar fields, considering
several copies of the field 
\begin{equation}
 S_M[\{\ph\},\cT] \to \sum_\mu S_M[\{\ph^\mu\},\cT].
\end{equation}
The effect was the same as with a single scalar field. We conclude that the dependence of the determinant $S_\mathrm{quant}^\mathrm{eff}[\cT]$ on $\cT$ is weak and, in practice, we can treat it as a constant.

\subsection{Jumps}

\begin{figure}[H]
\centering
\includegraphics[scale=0.2]{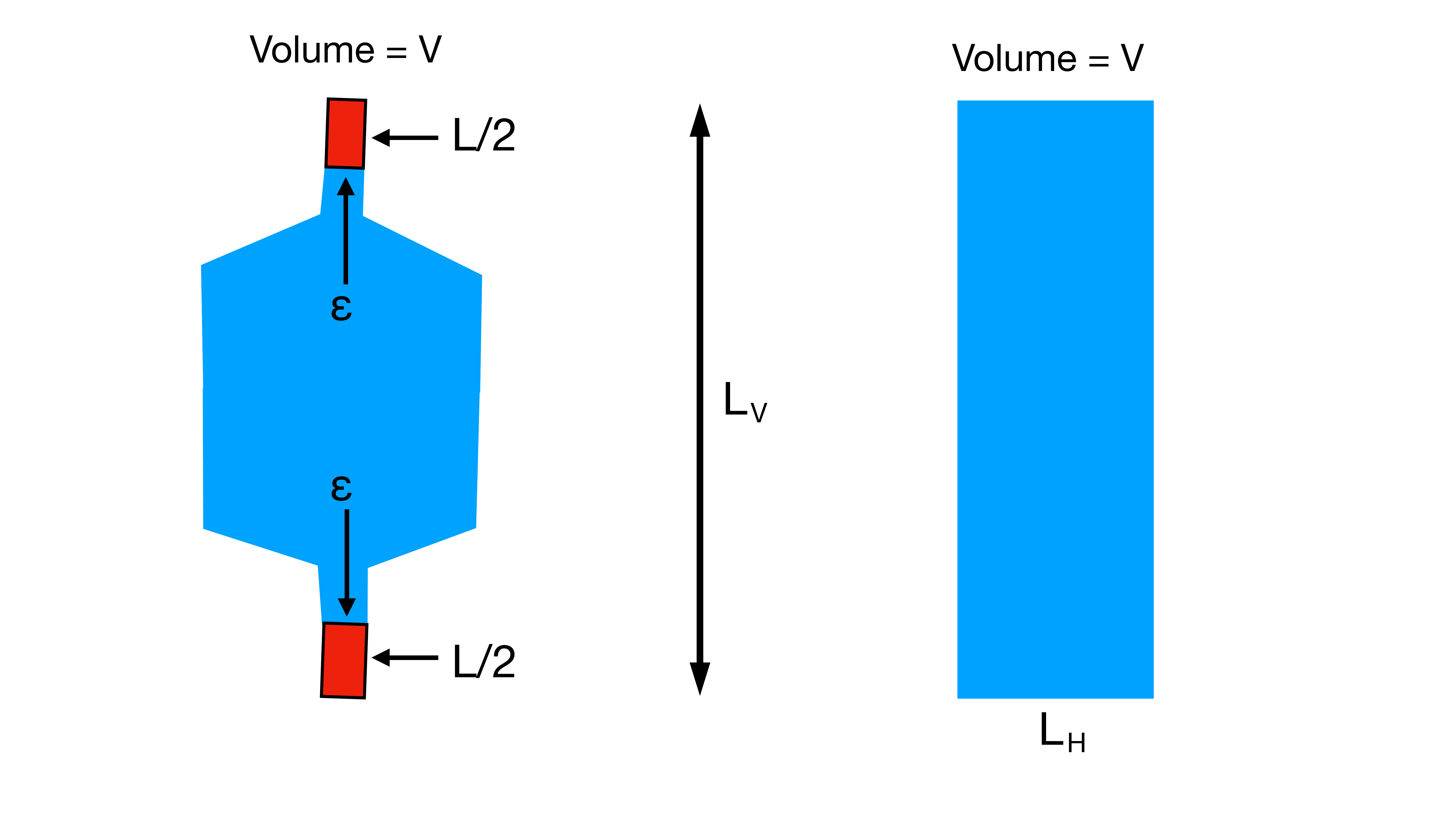}
\vspace{-1cm}
\caption{Left: a pinched torus with the opposite sides identified. Going from the bottom to the top, $\phi$ increases from 0 to $\delta$; specifically, in the lower red part it changes from 0 to $\delta/2$, in the blue region it stays constant and equal to $\delta/2$, and in the upper red part it changes from $\delta/2$ to $\delta$. $\phi$ is constant in the horizontal direction. The volume of the red region, the only region of the field change, is $L\cdot \epsilon$. Right: a torus where $\phi$ is constant in the horizontal direction and uniformly increases from 0 to $\delta$ from bottom to top. The two tori are assumed to have the same vertical length $L_V$ and the same volume $V$ (which for the right figure can be written as  $V=L_V L_H$).}\label{fig-pinch}
\end{figure}

The new aspect introduced in \cite{LetterQuantum} and studied in detail here is based on two major generalizations of the CDT model:
\begin{itemize}
 \item The spatial topology was chosen to be toroidal $T^3$. Effectively the topology is toroidal $T^4$ since we also assume periodicity in the time direction. The system can be treated as infinite, with the elementary cell repeated periodically in four directions. 
 \item The scalar field was defined as taking values on a circle of  circumference $\delta$ rather than in $\mathbb{R}$ and forced to wind around  the circle when moving around a non-contractible loop in one of the directions on $T^4$. {This can alternatively be viewed as a field taking values in $\mathbb{R}$ with a jump of magnitude $\delta$ when crossing the (unphysical) boundary of an elementary cell; see Section \ref{sec:class} for details.}
\end{itemize}
The latter modification thoroughly changes the dynamics of the geometry-matter interaction. Previously, for the $\mathbb{R}$-valued scalar field without jumps imposed, the constant field configuration (i.e., the classical solution) resulted in the absolute minimum (zero) of the matter action. Now, this solution with a zero winding number is excluded, yet there is a way of rearranging the geometry that makes the action decrease virtually to zero. For an illustration in the simple case of a two-dimensional torus see figure~\ref{fig-pinch}. 
The argument is independent of the number of dimensions as long as at least one direction is periodic. 

The left-hand side picture shows a torus with volume $V$ and vertical length $L_V$, which is pinched to a cylinder of circumference $\epsilon$ and length $L$. The scalar field winds once around a circle of circumference $\delta$ when we move around a non-contractible loop in the vertical direction, or, equivalently, the field jumps by $\delta$ when passing a boundary between the lower and the upper edge of the picture (the opposite sides of the picture are identified). We consider a specific field configuration, where the field $\phi$ changes uniformly from 0 to $\delta/2$ over a distance $L/2$ in the lower red part, stays constant and equal to $\delta/2$ in the blue part and changes from $\delta/2$ to $\delta$ in the upper red part. The region where the field changes is joined smoothly to the region where it is constant. The total matter action of this field configuration is
\beql{jan50}
S_{M}[\{\phi\},\delta, \cT_L] = \Big( \frac{\delta}{L}\Big)^2 L \, \epsilon  
= \delta^2 \frac{\epsilon}{L}, 
\eeq
and the minimal action for a classical field configuration for this geometry is even lower.\footnote{Note that the field configuration used in \rf{jan50}, even if smoothly joining the regions where $\phi$ changes and where $\phi$ is constant, will in general fail to satisfy Laplace's equation, i.e., it will not have the minimum value of the action \rf{action}. We only use it to show that by changing geometry the actual solution to Laplace's equation with winding number 1 can be made arbitrarily small. On the other hand, the solution $\phi$ used in \rf{jan51} is the minimum for the given geometry since it has winding number 1 and satisfies Laplace's equation.} Clearly, this value can be made arbitrarily small when $\epsilon \to 0$, and this is even more true in higher dimensions. The right-hand side picture in figure~\ref{fig-pinch} also shows a torus with volume $V$ and vertical length $L_V$. For this geometry, the action is minimal for a field changing uniformly from 0 to $\delta$ when we move from bottom to top, and is equal to
\beql{jan51}
S_{M}[\{\phi\},\delta,\cT_R] = \Big( \frac{\delta}{L_V}\Big)^2 L_V L_H  = \delta^2 
\frac{V}{L_V^2},\qquad V= L_H L_V,
\eeq
which is bounded from below when $V$ and $L_V$ are fixed. Let us discuss the consequence of this for the full quantum theory. We consider the action of a single scalar field, 
\begin{equation}
 S_M[\{\ph\}, \delta, \cT] = \frac{1}{2} \sum_{i \leftrightarrow j}(\ph_i - \ph_j - \delta B_{ij})^2= \sum_{i,j} \ph_i L_{ij}\ph_j 
 -2 \delta \sum_i \ph_i b_i  + \delta^2 \cdot V .
 \label{actionb}
\end{equation}
Here $B_{ij}=\pm 1,$ when the boundary face $i\to j$ is crossed in the positive (negative) direction, and $B_{ij}=0$ otherwise; $b_i = \sum_j B_{ij}$ and $V = \frac{1}{2} \sum_{i,j} B^2_{ij}$. Note that now the size of the jump $\delta$ fixes the scale of the field $\ph$. The action (\ref{actionb}) is still Gaussian but with a linear term. Like before, the field $\ph$ can be integrated out. We use the standard method to eliminate the term linear in $\ph$ by a shift. We decompose the field into the classical part $\bar \ph_i$ and the quantum part $\xi_i$:
\beql{decompose}
\ph_i=\bar \ph_i + \xi_i
\eeq
Since both $\ph_i$ and $\bar{\ph}_i$ have winding number 1, the fluctuation field $\xi_i$ is a scalar field with winding number 0, like an ordinary scalar field taking values in $\mathbb{R}$. We modify the integration measure
\begin{equation}
\mathcal{D}[\ph] = \mathcal{D}[\xi]
\end{equation}
and rewrite the action (\ref{actionb}) as
\begin{align}
 S_M[\{\ph\}, \delta, \cT] &= \sum_{i,j} \xi_i L_{ij}\xi_j + \sum_{i,j} \bar\ph_i L_{ij}\bar\ph_j - 2 \delta \sum_i \bar\ph_i b_i + \delta^2 \cdot V \nonumber\\
 &= \sum_{i,j} \xi_i L_{ij}\xi_j + S_M[\{\bar\ph\},\delta,\cT].\label{newaction}
\end{align}
After integrating out the quantum field, we see that now the field with a jump contributes to the geometric action 
\beql{newterm}
  \tilde{S}_\mathrm{quant}^\mathrm{eff}[\cT, \delta]= S_\mathrm{quant}^\mathrm{eff}[\cT]+\Delta S^\mathrm{eff}[\cT,\delta], \quad \Delta S^\mathrm{eff}[\cT,\delta] =
  S_M[\{\bar{\phi}\},\delta,{\cal T}].
\eeq
The extra correction term $ \Delta S^\mathrm{eff}[\cT,\delta]$ is nothing else than the scalar field action \rf{actionb} evaluated at the classical solution $\bar\ph$. It can be written in many equivalent ways, e.g.,
\begin{eqnarray}
\Delta S^\mathrm{eff}[\cT,\delta]&=&- \delta \sum_i \bar\ph_i b_i + \delta^2 \cdot V \nonumber\\
    &=& - \delta^2 \sum_{i,j}\left(b_i \tilde{L}^{-1}_{ij} b_j - \frac{B_{ij}^2}{2}\right)\nonumber\\
    &=& - \frac{1}{2}\sum_{i,j} \delta B_{ij}(\bar\ph_i - \bar\ph_j - \delta B_{ij}),
\label{classaction} 
\end{eqnarray}
where we used the fact that the classical field $\bar \ph$ satisfies
\begin{equation}
 \sum_j L_{ij}\bar\ph_j = \delta \cdot b_i,\quad \bar\ph_i = \delta \sum_j \tilde{L}^{-1}_{ij} b_j.
 \label{classical}
\end{equation}
It is worth mentioning that, according to \rf{classaction}, the action $S_M[\{\bar{\phi}\},\delta,{\cal T}]$ of the classical solution $\bar{\phi}$  can be written entirely in terms of the values of $\bar{\phi}_i$ next to the boundary with the jump, despite the fact that the action itself is independent of the precise location of the boundary. The purely quantum contribution $S_\mathrm{quant}^\mathrm{eff}[\cT]$ is thus exactly the same as for the case with no jump ($\delta=0$) and the (purely classical) correction $\Delta S^\mathrm{eff}[\cT,\delta]=S_M[\{\bar{\phi}\},\delta,{\cal T}]$  is quadratic in the jump size $\delta$. We now have the following situation: for a given geometry, i.e., a given triangulation ${\cal T}$, the contribution from the quantum fluctuations  of the scalar field is the same whether the scalar field takes value in $\mathbb{R}$ (and thus just fluctuates around 0) or in a circle $S^1$ of circumference $\delta$ (and fluctuates around the classical solution $\bar{\phi}_i$ with winding number 1). However, in the latter case the minimum of the classical action $S_M[\{\bar\ph\},\delta,\cT]$ depends in a crucial way on the triangulation $\cT$.  Triangulations that are pinched as shown in figure~\ref{fig-pinch} have the smallest matter action but, in general, the geometric Regge (Einstein-Hilbert) part of the action is larger for them than for non-pinched triangulations. Thus, there is a competition between matter and the geometric action. In the case of a scalar field winding around the time direction, this can easily be illustrated using a simple minisuperspace approximation. We refer to Appendix 3 for details. The conclusion is that for a small jump magnitude $\delta< \delta_c$, the geometric part of the action prevails, and generic triangulations in the path integral are quite similar to the ones that dominate when no matter field with a jump is present. However, for a large jump magnitude $\delta>\delta_c$, the total (geometric+matter) action is the lowest for pinched triangulations, and the system fluctuates around them. Thus, we have a picture where for small $\delta<\delta_c$, the effect of the scalar field is small, and we can say that the scalar field couples to and follows the geometry. However, when $\delta > \delta_c$, the scalar field pinches the geometry to a spatial volume which
is small or maybe even zero, and (almost) all changes of $\phi$ take place in this region of very small volume. Thus, $\phi$ basically splits a spacetime with a non-trivial winding number in the time direction into two parts: one (of cutoff size) with a nonzero winding number and one (dominating)
with a zero winding number. Therefore, for $\delta=\delta_c$ we should observe a new type of a phase transition caused entirely by the scalar field, a phase transition in which the effective spacetime topology can change from toroidal to a simply connected one. This analysis is of course based on a very simple minisuperspace action (see Appendix 3), which might be a good description in the time direction but not necessarily in the spatial directions, where there is no minisuperspace approximation. Therefore we now turn to numerical Monte Carlo simulations. In Section \ref{sec:dynamical_1} we discuss the case of an $S^1$ scalar field in the time direction in  CDT  with the $T^3$ spatial topology.\footnote{We stated above that in this situation the phenomenon of pinching should be independent of the spatial topology. This is presumably true. However, we might fail to discern it if the spatial topology is $S^3$ and the system is in the semiclassical phase $C$. The reason is that in this case we generally already have a geometric pinching, in fact a whole ``stalk'' of cut-off size width, even without  a scalar field. In that situation there will be no problem for the scalar field to produce a jump of $\delta$ in the stalk, and there should not be any real difference in the effect of a scalar field with values in $\mathbb{R}$ and a scalar field with values in $S^1$ and a non-trivial winding number.} Then, in Section \ref{sec:dynamical_2} we investigate the case of three scalar fields winding around spatial directions.

\subsection{Results for a single scalar field with a jump in the time direction}\label{sec:dynamical_1} 

\begin{figure}[H]
\centering
\includegraphics[scale=0.7]{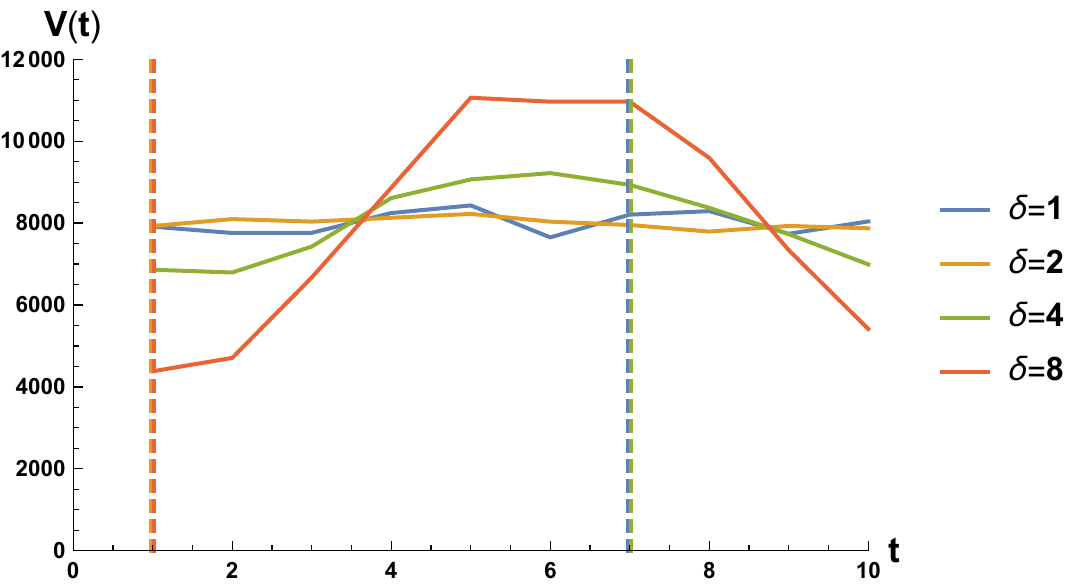}
\includegraphics[scale=0.7]{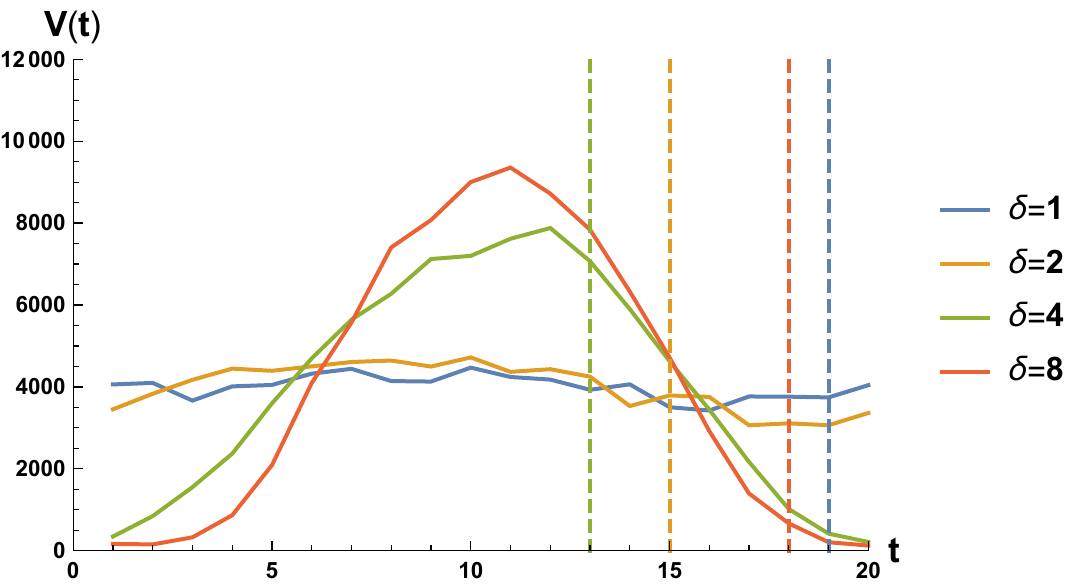}
\caption{Spatial volume $t$-profiles in single generic configurations inside the semiclassical phase $C$ ($\kappa_0=2.2, \Delta=0.6$) for $T=10$ (left) and $T=20$ (right) with scalar field jump magnitudes $\delta=1,2,4,8$. For each configuration the position of a jump of the scalar field is denoted by a dashed vertical line.}\label{fig:VolJumpInTime}
\end{figure}

Below we present the results obtained for one dynamical scalar field with a jump of magnitude $\delta$, or, in other words, a scalar field taking values in a circle of circumference $\delta$  in the time direction. All measured systems were toroidal CDT configurations inside the semiclassical $C$ phase region ($\kappa_0=2.2$, $\Delta=0.6$), and the Monte Carlo simulations were performed for the lattice volume $N_{4,1}= 160\rmk$ and the proper-time periods $T=10$ and $T=20$. In the Monte Carlo code, the jump was effectuated on the crossing between the $t=T$ and the $t=1$ (periodic) proper-time coordinate, i.e., between the field values inside the $(1,4)$ simplices (with 1 vertex in $t=T$ and 4 vertices in $t=1$) and the $(4,1)$ simplices (with 4 vertices in $t=1$ and 1 vertex in $t=2$), so that the time-boundary was the spatial slice in the layer $t=1$.\footnote{As already discussed, the formulation is independent of the boundary position, and thus one could as well use any other spatial layer or a more complicated boundary in time direction.} Spatial volume $t$-profiles (in the original $t$ coordinate: $V(t)=$ number of tetrahedra in a spatial slice $t$) for single generic configurations with several different jump magnitudes $\delta=1,2,4,8$ are presented in figure~\ref{fig:VolJumpInTime}. To facilitate the comparison, the profiles measured for various $\delta$ were shifted in the (periodic) proper-time axis so that the maxima are placed at the centers of the charts. 

\begin{figure}[H]
\centering
\includegraphics[scale=0.7]{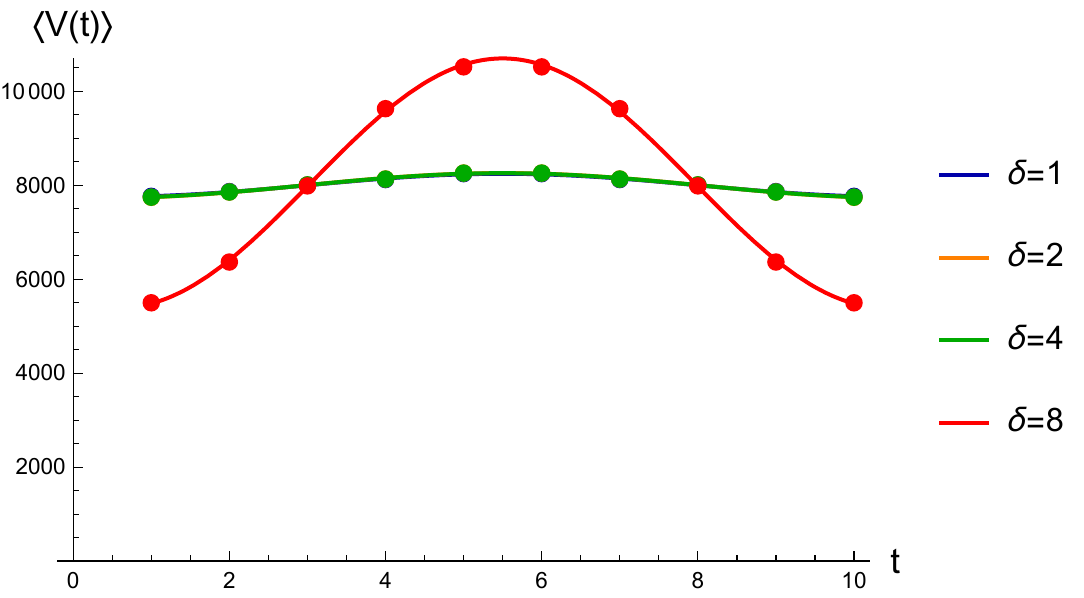}
\includegraphics[scale=0.7]{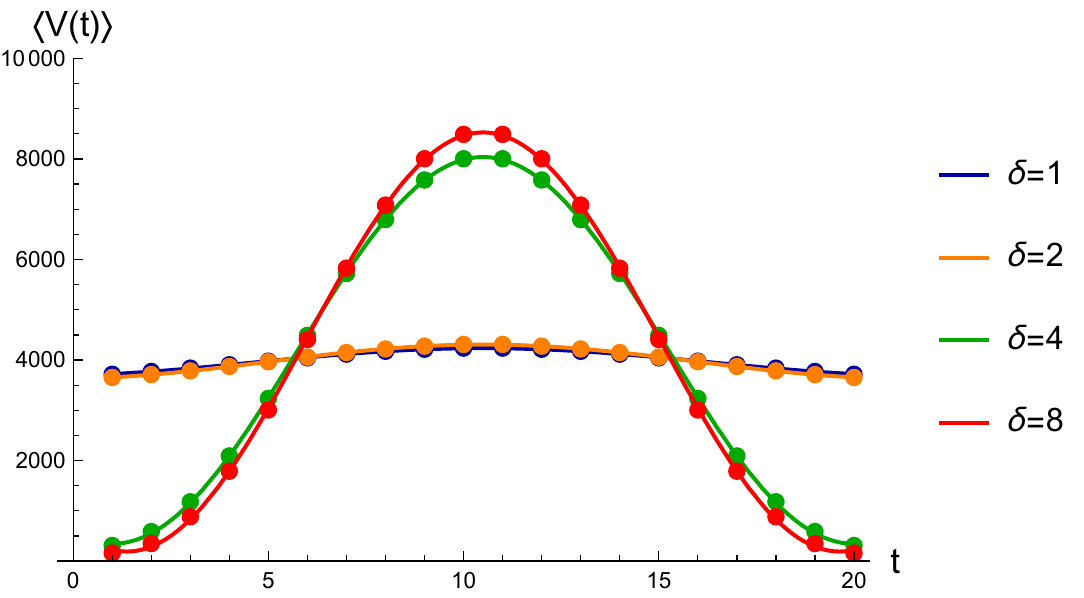}
\caption{Spatial volume $t$-profiles averaged over many MC configurations inside the semiclassical phase $C$ ($\kappa_0=2.2, \Delta=0.6$) for $T=10$ (left) and $T=20$ (right) with scalar field jump magnitudes $\delta=1,2,4,8$. Error bars for mesured data points were estimated using single-elimination (binned) jacknife procedure (for most points they are below the resolution of the plots). Solid lines are fits of the function: $c+a \cos(b(t-t_0))$.
In the left plot, the curves for $\delta=1,2,4$ overlap within the picture resolution.}

\label{fig:VolJumpInTimeAvg}
\end{figure}

For small jump magnitudes ($\delta = 1, 2$) one observes flat volume profiles characteristic for toroidal CDT in the pure gravity case (i.e., without the scalar field), while for large jump magnitudes ($\delta = 4,8$) the volume profiles are completely changed, showing the blob-like configurations (somewhat similar to the left-hand side picture in figure~\ref{fig-pinch}). 
The pinching becomes more pronounced for larger $T$. In view of the discussion in the last subsection, this is very understandable. With the same four-volume $V$, it is a larger deformation of the geometry to perform a pinching of $V(t)$ to small values if $T$ is small and thus the minimal value of $V(t)$ is larger. figure~\ref{fig:VolJumpInTime} also provides a clear illustration of the fact that the precise location of the hypersurface where the scalar field jumps has no effect on the interaction between the scalar field and the geometry. In the figure we have shown the location of the jump in the numerical code, and it is clearly unrelated to the position of the region where the geometry is pinched by the scalar field, even though when looking at eq.~\rf{classaction} (as already mentioned there) one could be misled to think that all physics of the classical scalar field is related to the location of the jump. 

Figure \ref{fig:VolJumpInTimeAvg} presents the volume profiles averaged over many Monte Carlo configurations. In order to get rid of the time-translation symmetry (the center of volume of each configuration can perform a random walk around the periodic time axis), the center of volume  of each individual $t$-profile  was  shifted to a universal position $t_0=T/2+0.5$. Because of this shifting, one can observe  artificial small "blobs" for small jump magnitudes ($\delta=1,2$). Nevertheless, it is easily seen that the phase transition takes place above $\delta=4$ for $T=10$ and above $\delta=2$ for $T=20$, respectively. Figure \ref{fig:VolJumpInTimeAvg} also contains fits of the cosine relation resulting from the minisuperspace model discussed in Appendix~3. It is remarkable that despite our computer generated data are based on the full non-perturbative model including all microscopic degrees of freedom, the averaged profiles (obtained after integrating out all degrees of freedom but the scale factor) are so well explained by the simple minisuperspace approximation, where the scale factor (time dependence) is the only dynamical variable.

\subsection{Results for three scalar fields with one or more jumps in spatial directions}\label{sec:dynamical_2}

This subsection presents the results obtained for dynamical scalar fields with jumps in spatial directions. In each case, the system contained three scalar fields, and we could adjust the jump magnitudes $\delta_1, \delta_2,\delta_3$. In the Monte Carlo code, the jump of each scalar field was realized when crossing a 3D boundary orthogonal to one of three independent non-contractible loops winding around the toroidal spatial directions.  In practice, we measured systems where one, two, or all three fields had the same jump magnitude $\delta$, i.e., where: (1) $\delta_1=\delta, \delta_2=\delta_3=0$, (2) $\delta_1=\delta_2=\delta, \delta_3=0$ or (3)  $\delta_1=\delta_2=\delta_3=\delta$, for various choices of $\delta$. Therefore, one can view the systems as having  $n=1,2 \text{ or } 3$ scalar fields taking values on a circle of circumference $\delta$ and having winding number 1, and the remaining $3-n$ fields taking values in $\mathbb{R}$ (with no winding number imposed). The analyzed systems were all at the same point ($\kappa_0=2.2, \Delta=0.6$) in the semiclassical C phase, with the volume $N_{4,1}=160\rmk$ and the number of time slices $T=4$ {(in the end of this subsection we also present results for a larger system with $N_{4,1}=720\rmk$ and $T=20$, obtained at the point ($\kappa_0=4.0, \Delta=0.2$), also inside the C phase)}.

\begin{figure}[H]
\centering
\includegraphics[scale=0.6]{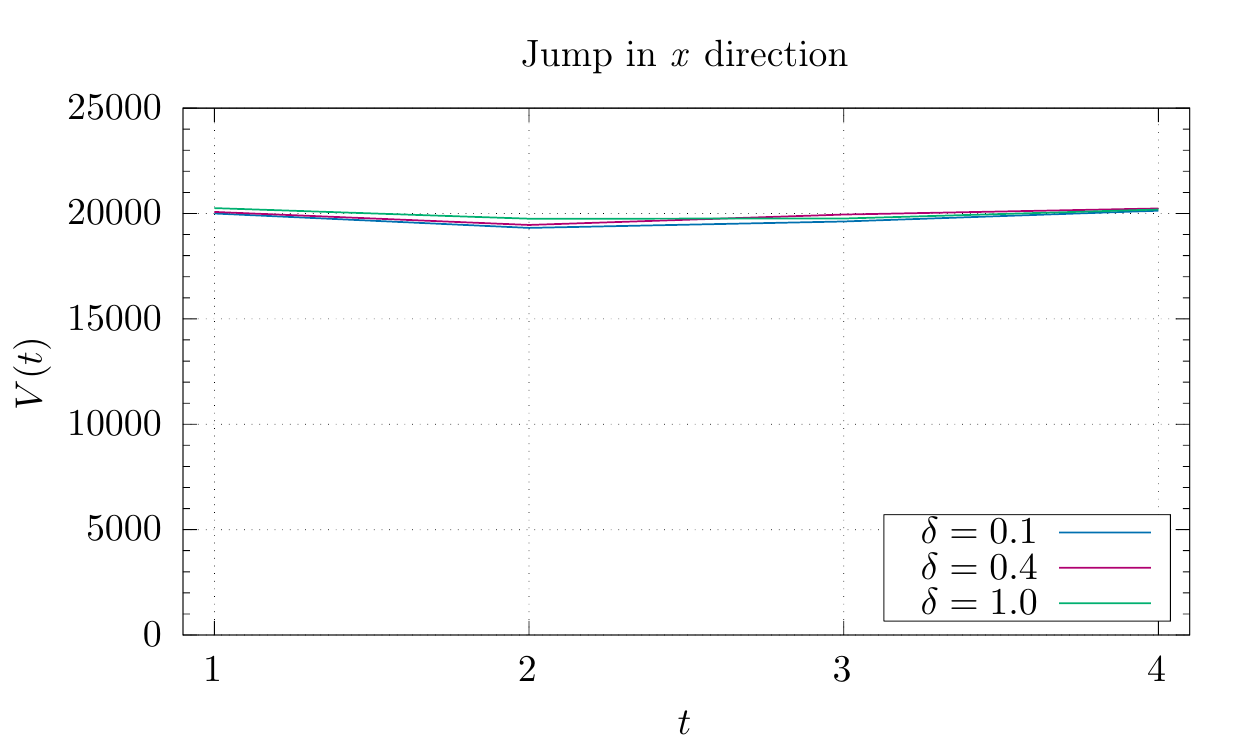}
\includegraphics[scale=0.6]{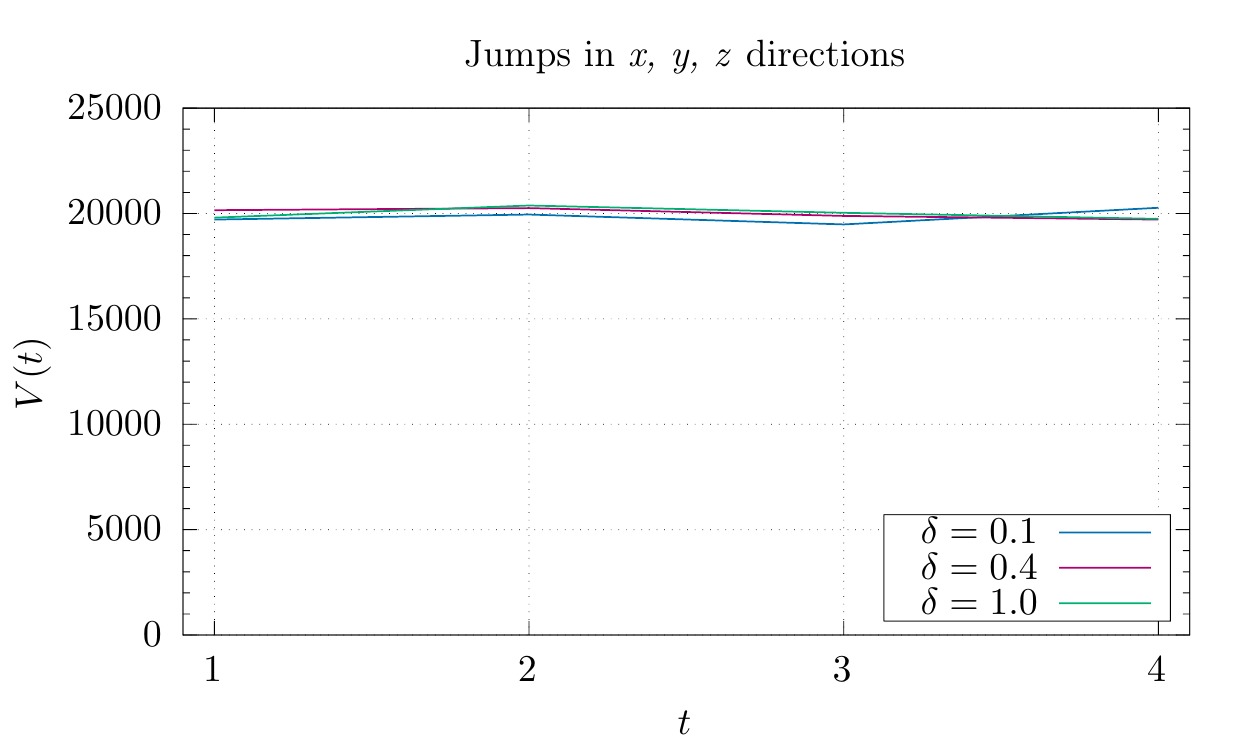}
\caption{Spatial volume $t$-profiles in single generic configurations inside the semiclassical phase $C$ ($\kappa_0=2.2, \Delta=0.6$) for $T=4$ and with dynamical scalar fields with jump in one spatial direction (left) and in three spatial directions (right).}
\label{fig:VolJumpInSpace}
\end{figure}

{For the sake of order}, we start our analysis with the spatial volume t-profiles for a single generic configuration observed for the cases when the field jumps in one or three spatial directions. In this case, as can be seen in figure~\ref{fig:VolJumpInSpace}, one does not observe the pinching effect in the volume profiles even for the largest measured scalar field jump magnitude $\delta$, but this is most likely due to the very small extent of the periodic time axis (fixed at $T=4$), which prevents blob-like volume profiles from forming {(as we will show later, such non-trivial volume profiles can be observed for larger $T=20$)}.\footnote{A similar behavior was earlier observed in the spherical CDT pure gravity case, where the blob-like volume profile  resulting from a non-trivial minisuperspace effective potential term could be observed only for large enough $T$. For small $T$ the observed volume profile was flat, but one could still measure the same effective potential term as for large $T$.}

To extract more information about the (change in) geometric structure caused by the dynamical scalar field(s) with a certain (large) jump magnitude, one can repeat the analysis of Section~\ref{sec:class}, i.e., define coordinates given by the classical scalar field solutions in all spatial and time directions. To facilitate comparison with the results for the pure gravity case presented in Sections \ref{sec:class} and \ref{sec:foliations}, we rescaled the obtained solutions to the classical Laplace's equation \rf{eq:poisson} to get the standard jump magnitude ($\delta=1$) independently of the actual jump magnitude of the dynamical scalar field(s) $\delta$. This can be interpreted as introducing  {\it new independent classical fields} $\bar\ph^\mu(\delta=1)$ on top of the dynamical fields $\ph^\mu(\delta)$ or, alternatively, as computing the (rescaled) {\it expected value of the dynamical field(s)}

\beql{rescaledfield}
\langle \ph^\mu(\delta) \rangle \equiv \delta \cdot \bar\ph^\mu(\delta=1).
\eeq

\begin{figure}[H]
\centering
\includegraphics[scale=0.6]{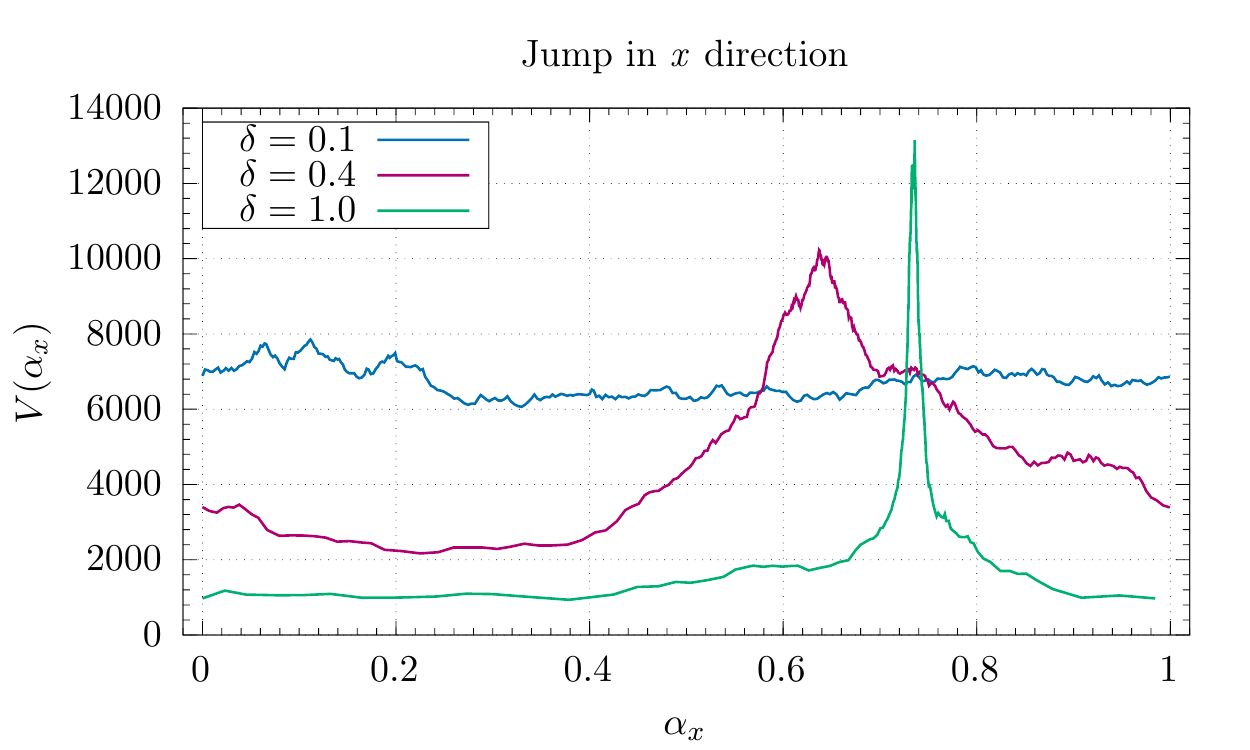}
\includegraphics[scale=0.6]{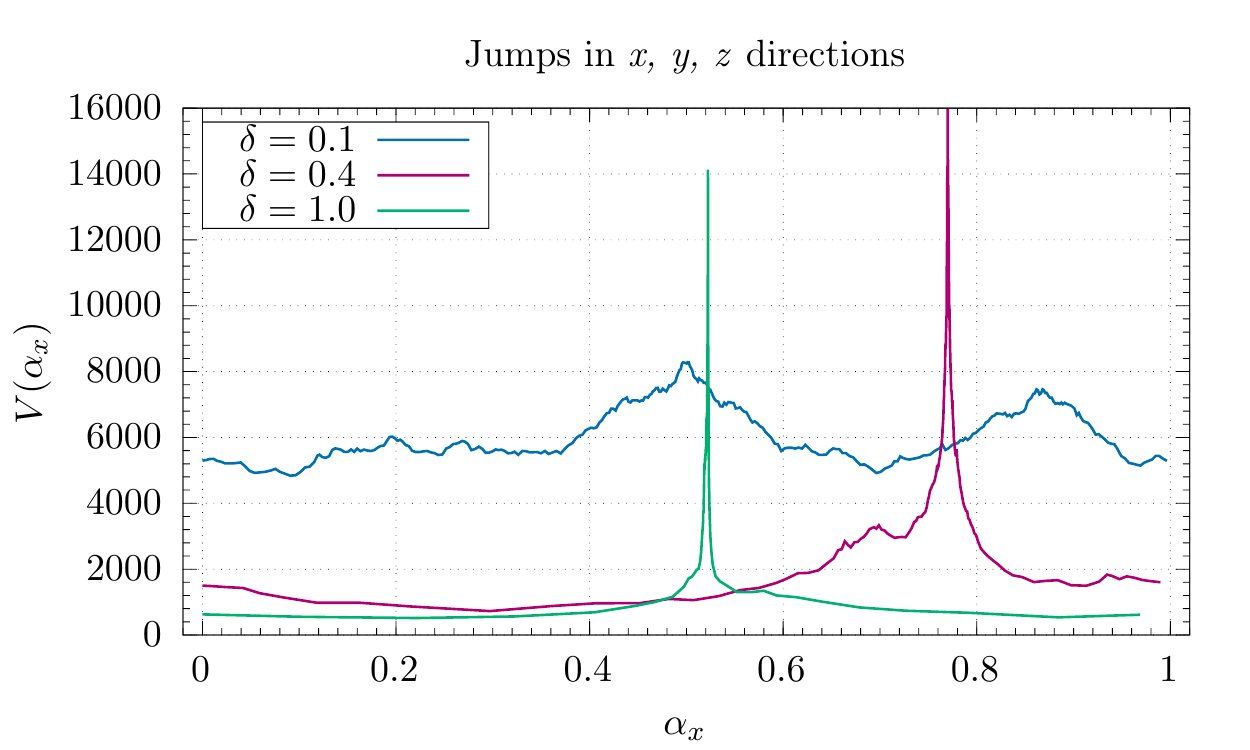}\\
\includegraphics[scale=0.6]{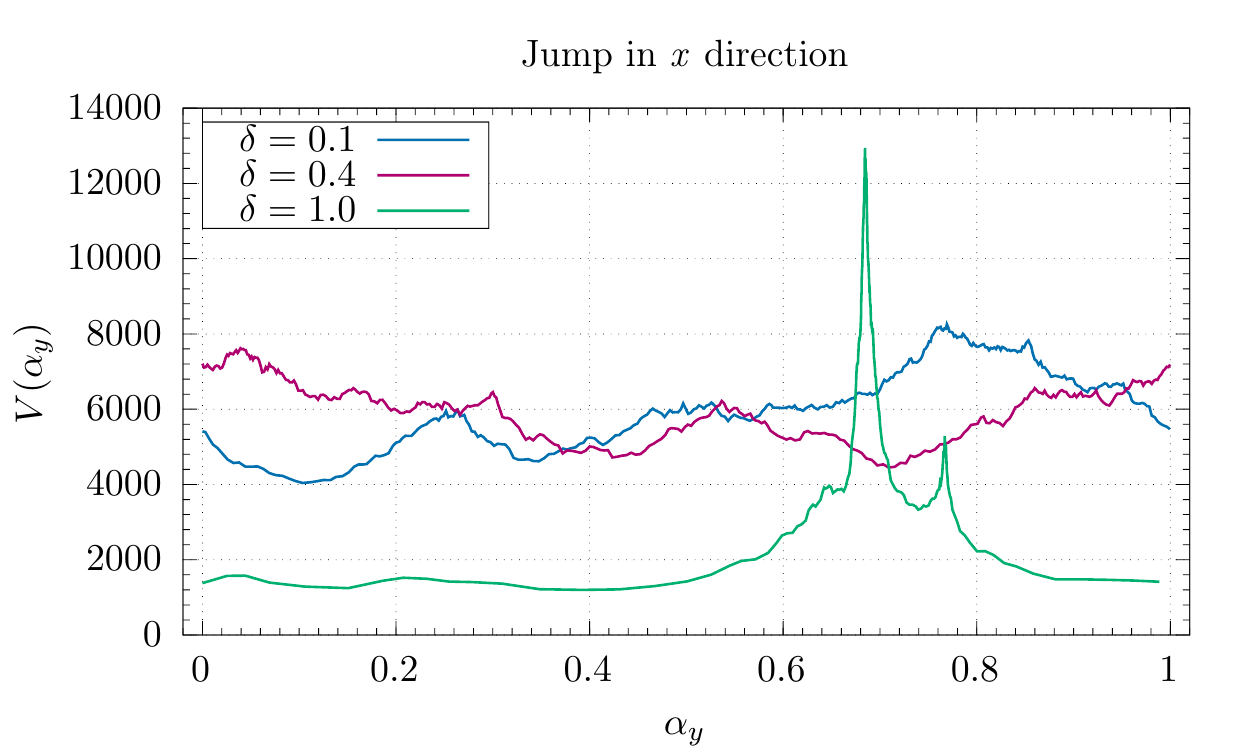}
\includegraphics[scale=0.6]{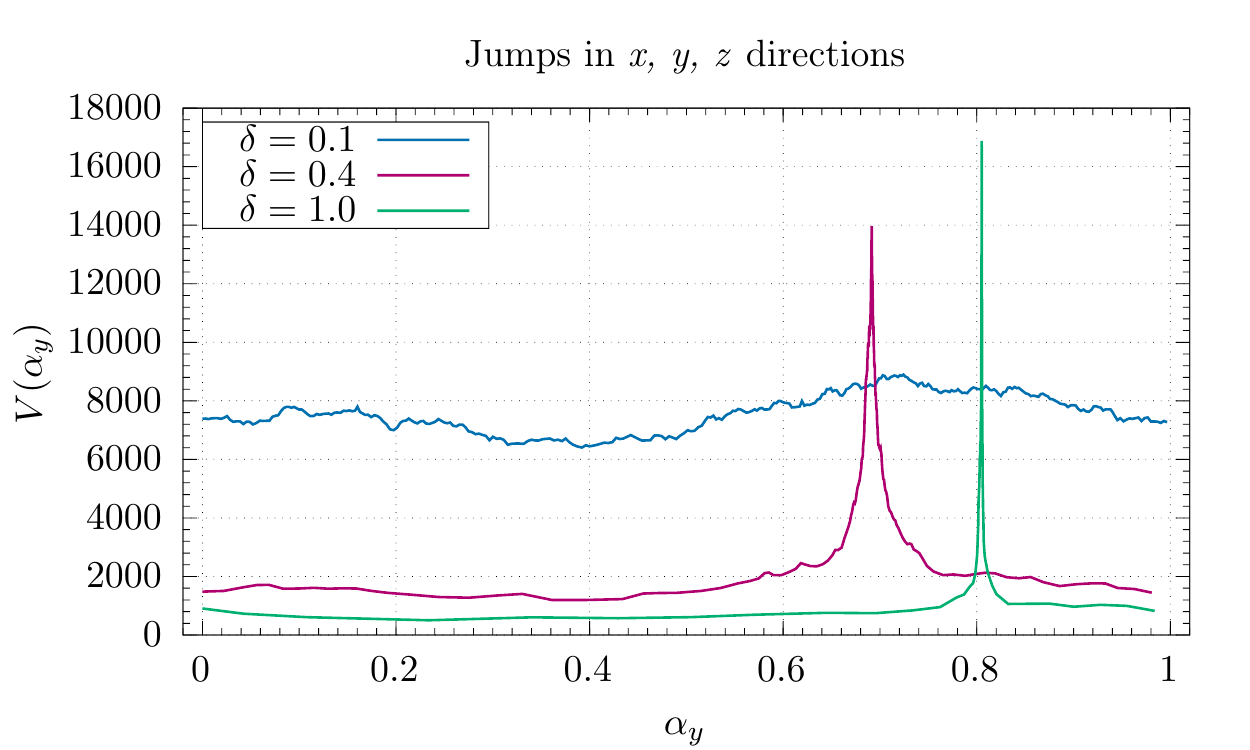}\\
\includegraphics[scale=0.6]{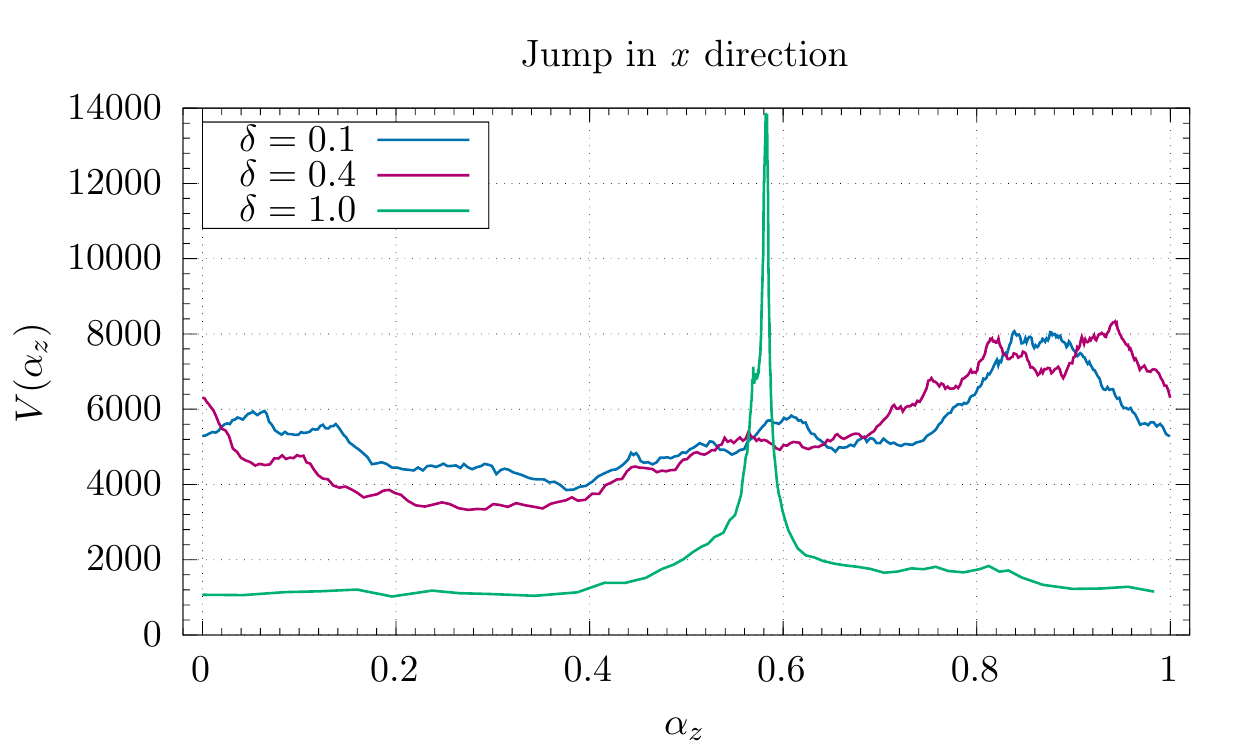}
\includegraphics[scale=0.6]{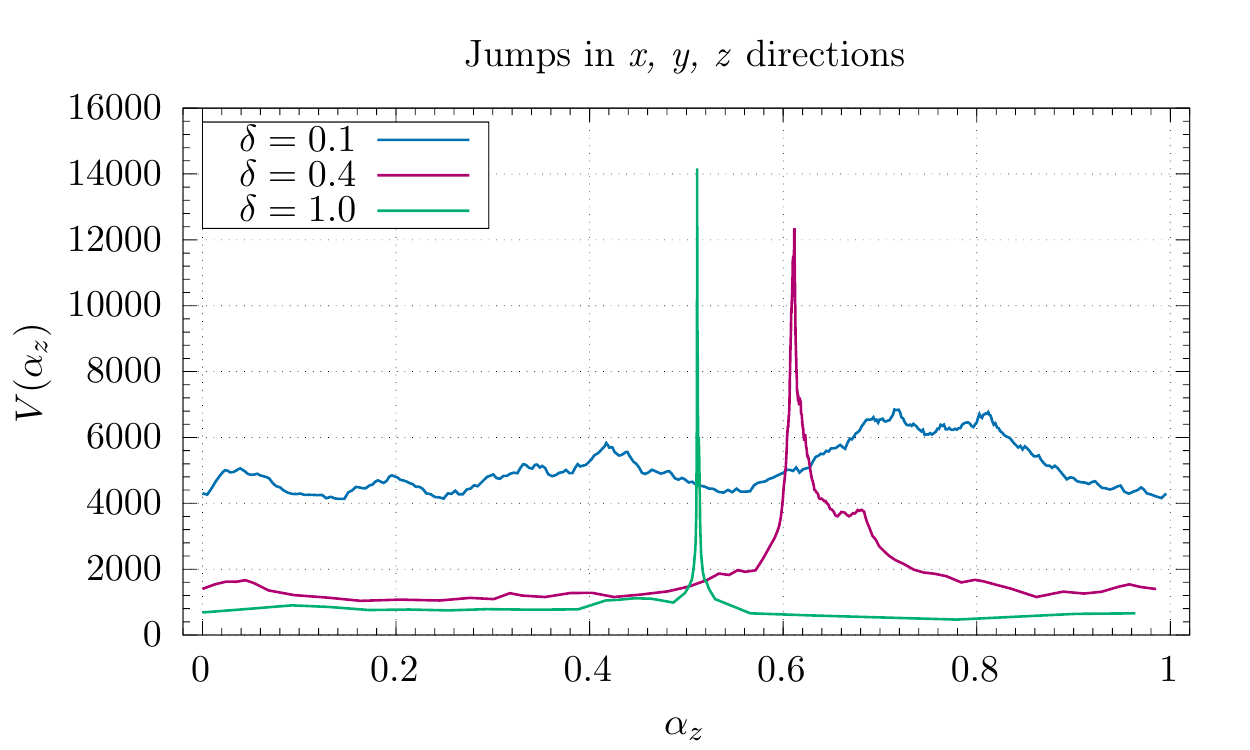}
\caption{$\alpha_x-$ (top), $\alpha_y-$ (middle) and $\alpha_z-$ (bottom) profiles in the $x,~y,~z$ directions in
single generic configurations inside the semiclassical phase $C$ ($\kappa_0=2.2, \Delta=0.6$) for $T=4$ and with dynamical scalar fields with jump in one spatial direction (left) and in three spatial directions (right).}
\label{fig:alphaJumpInSpace}
\end{figure}

This way one can, for example, measure the $\alpha$-profiles not only in time but also in the spatial directions (see Section \ref{sec:foliations} for discussion). The $\alpha$-profiles in spatial directions, presented in figure~\ref{fig:alphaJumpInSpace}, are visibly pinched for large 
jump magnitudes, and the effect depends on the number of fields with a jump. It is also readily seen that in the case where the jump of the field takes place only in one spatial direction, say $x$, the blob-like volume profiles in the (orthogonal) spatial directions $y$ and $z$ are also observed for a large value of the jump  ($\delta = 1.0$), as in the left-hand side plots of figure~\ref{fig:alphaJumpInSpace}. This is a strong evidence that the observed effect results from a genuine pinching of geometry caused by the scalar field(s) winding around a circle, as discussed above, the effect being clearly stronger for more numerous scalar fields with a jump (conf. the right-hand side plots in figure~\ref{fig:alphaJumpInSpace}).

\begin{figure}[H]
\centering
\includegraphics[width=0.45\textwidth]{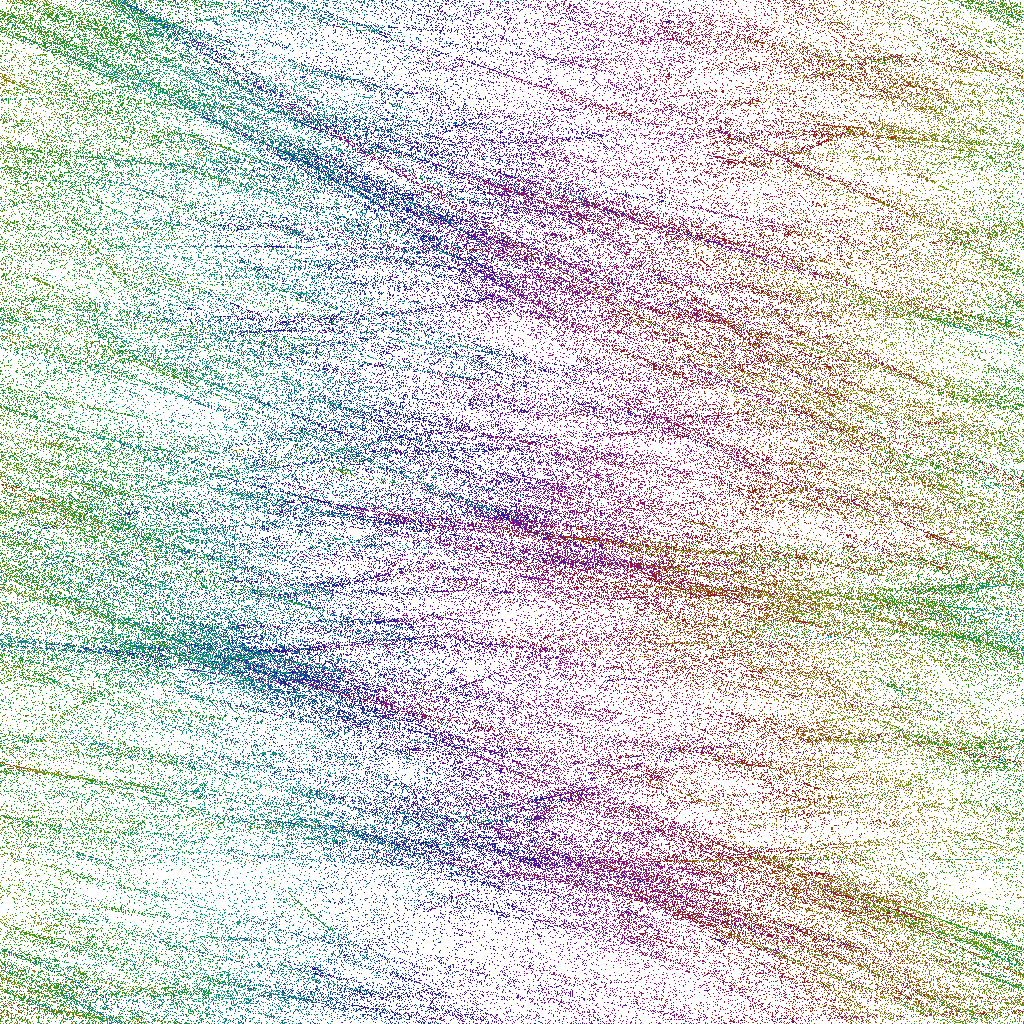}
\includegraphics[width=0.45\textwidth]{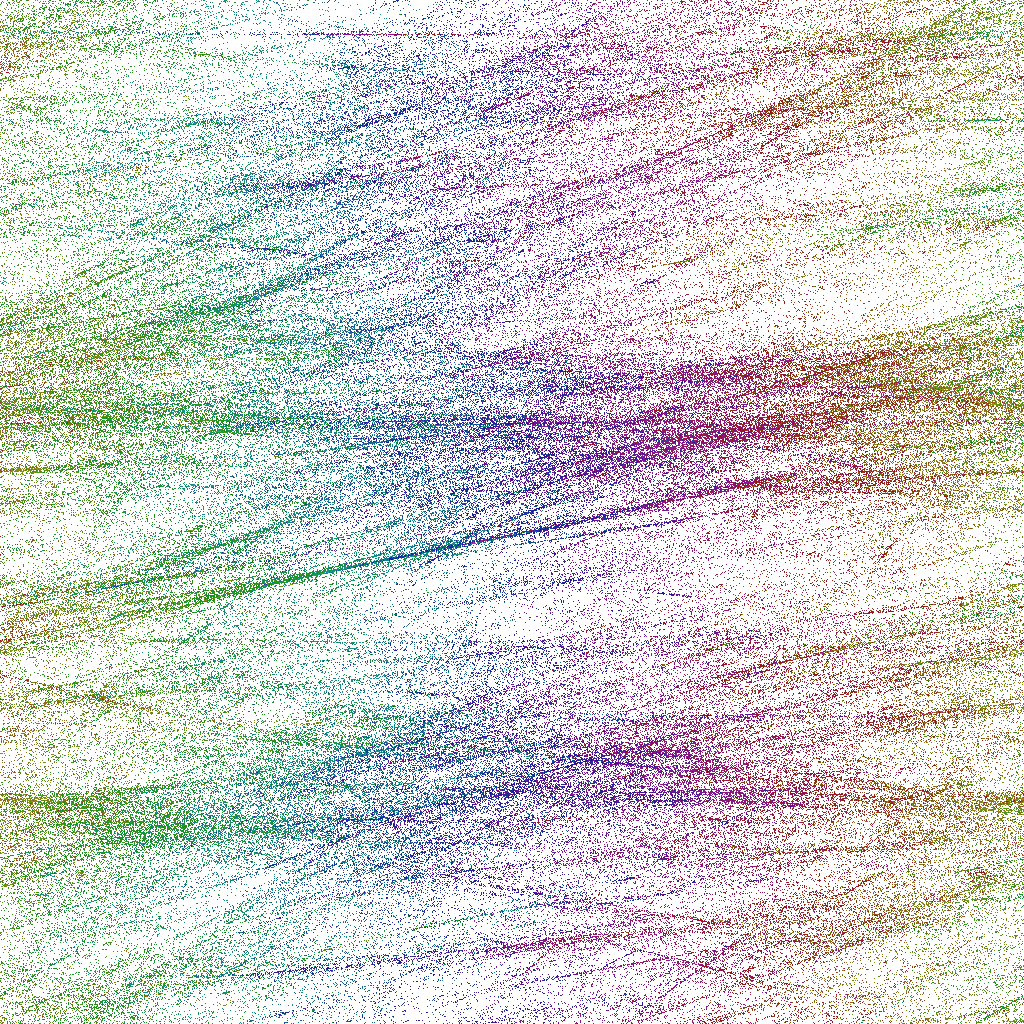} \\
\includegraphics[width=0.45\textwidth]{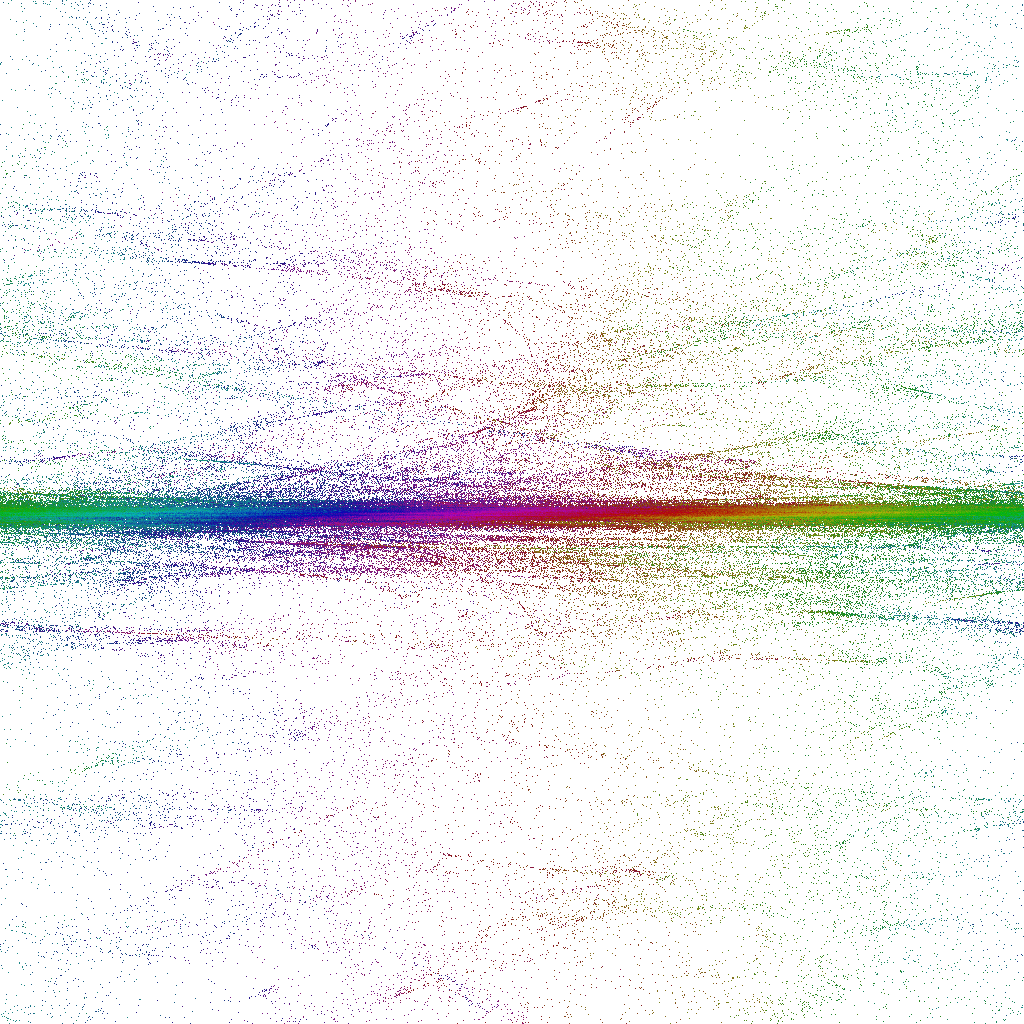}
\includegraphics[width=0.45\textwidth]{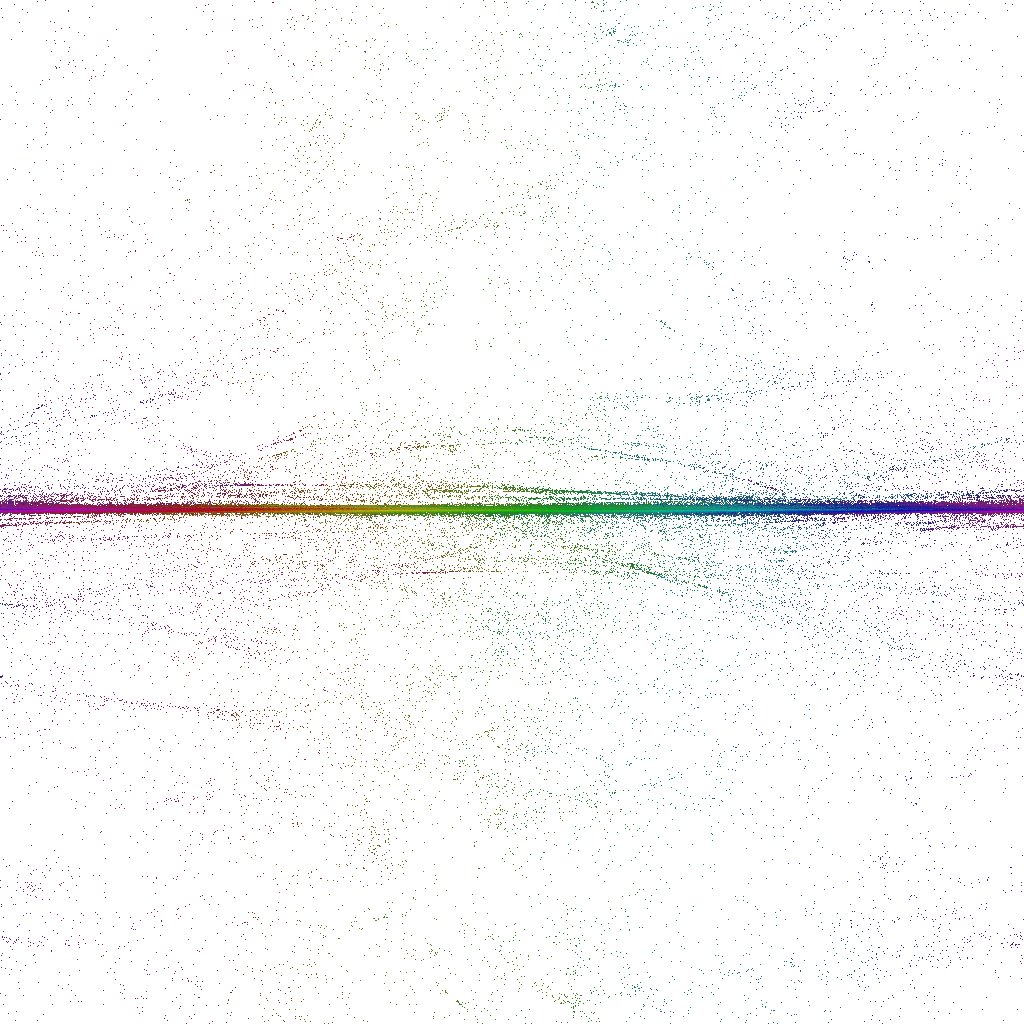}\\
\caption{Density maps in $\bar{\phi}$ coordinates (for the definition see Section \ref{sub:maps}) representing the effect of the spatial pinch in $t-x$ directions  for configurations in phase $C$ ($\kappa_0=2.2, \Delta=0.6$) with $T=4$. The left-hand side charts are for a single jump in $x$ direction and the right-hand side charts are for three jumps in all spatial directions.
 Top: configurations with a small jump magnitude ($\delta=0.1$). Bottom:  configurations with a large jump magnitude ($\delta=1.0$).
}\label{fig:VolJumpInSpaceTX}
\end{figure}

\begin{figure}[H]
\centering
\includegraphics[width=0.45\textwidth]{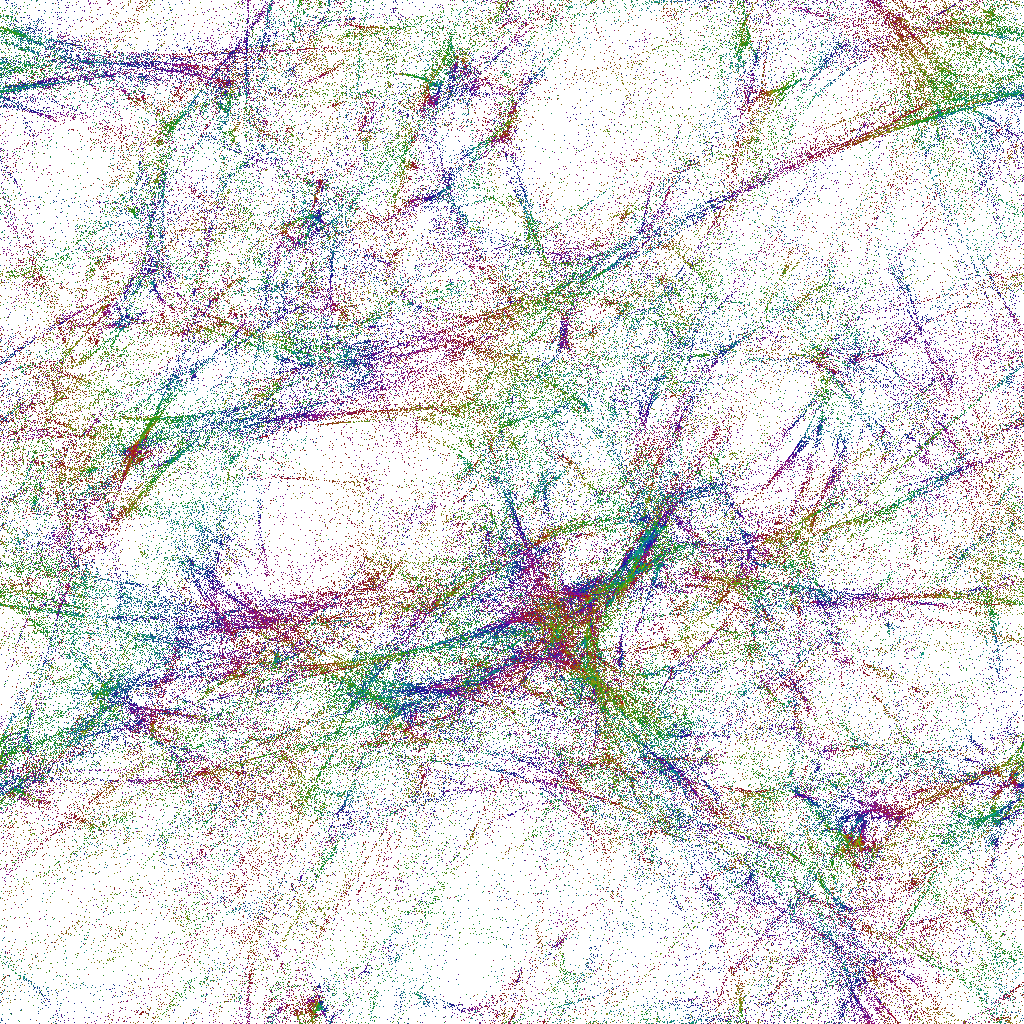}
\includegraphics[width=0.45\textwidth]{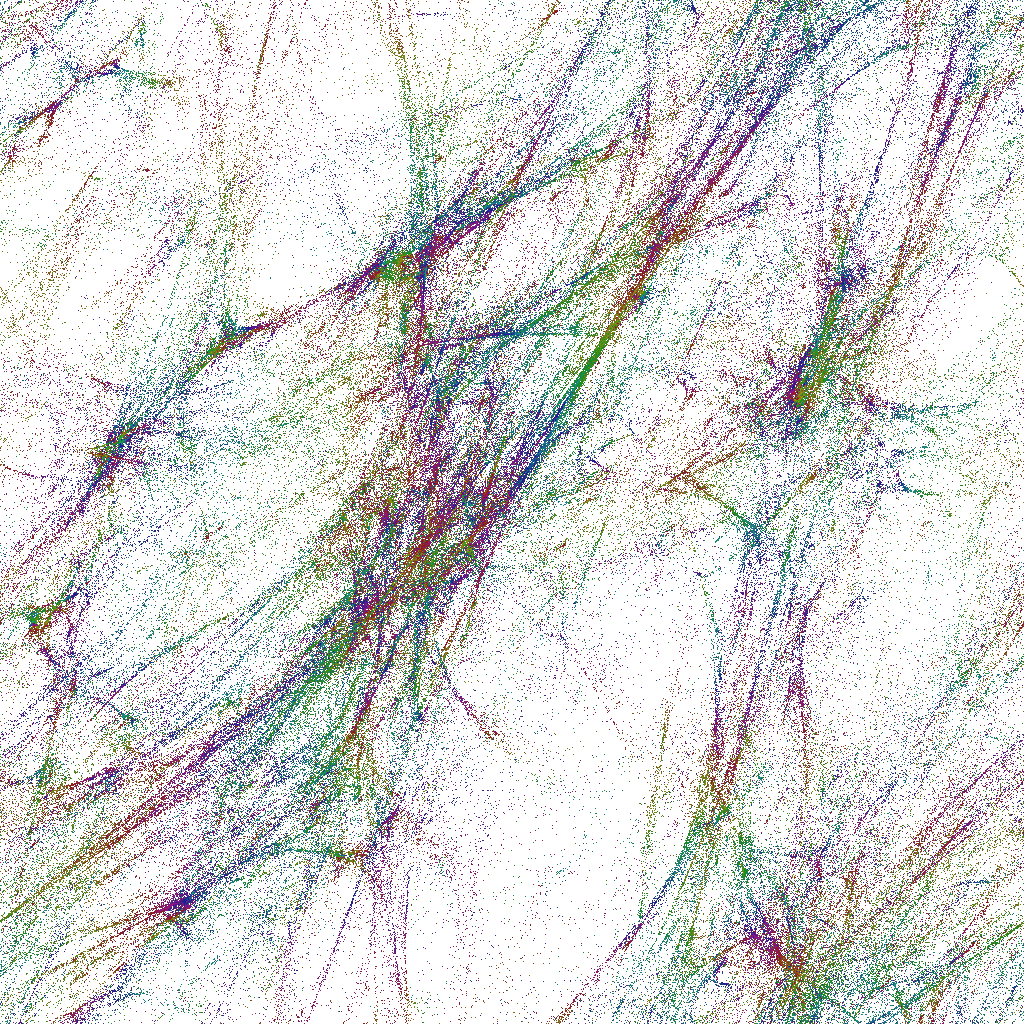} \\
\includegraphics[width=0.45\textwidth]{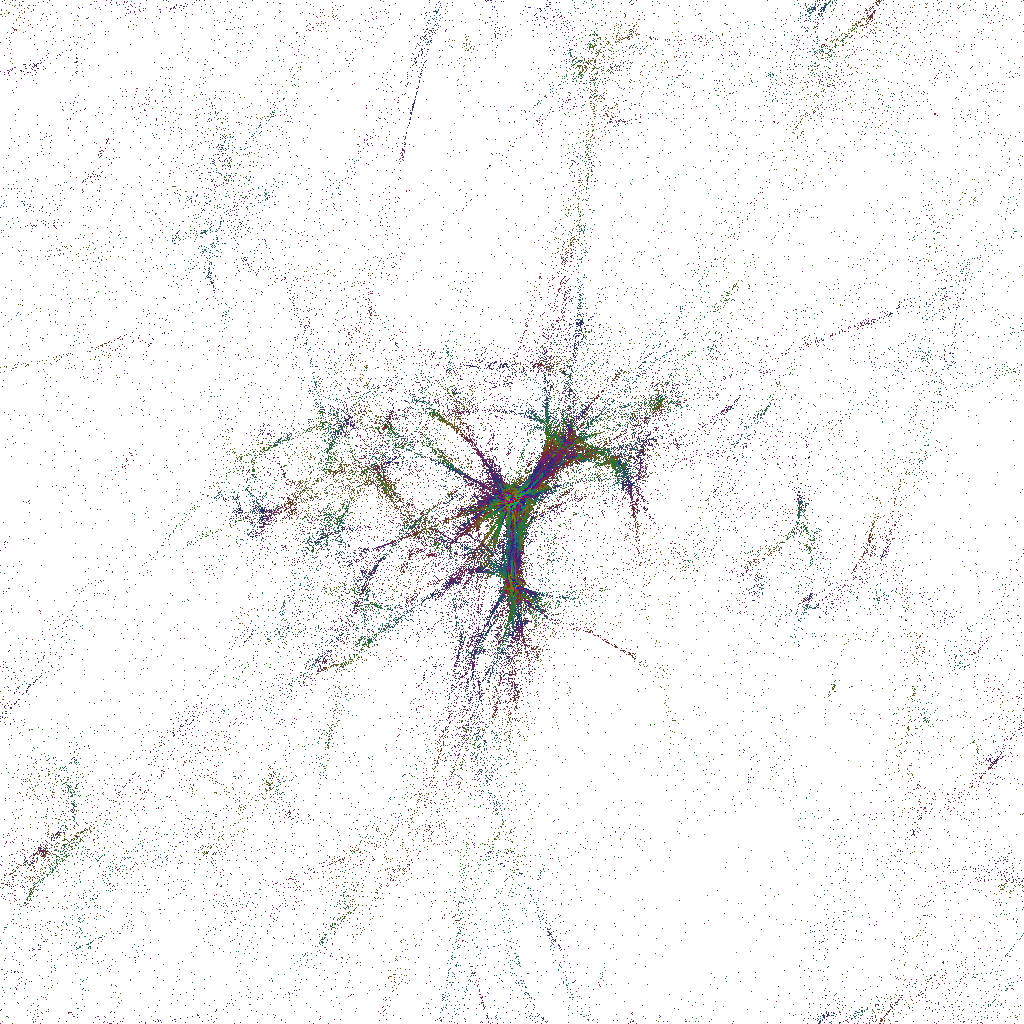}
\includegraphics[width=0.45\textwidth]{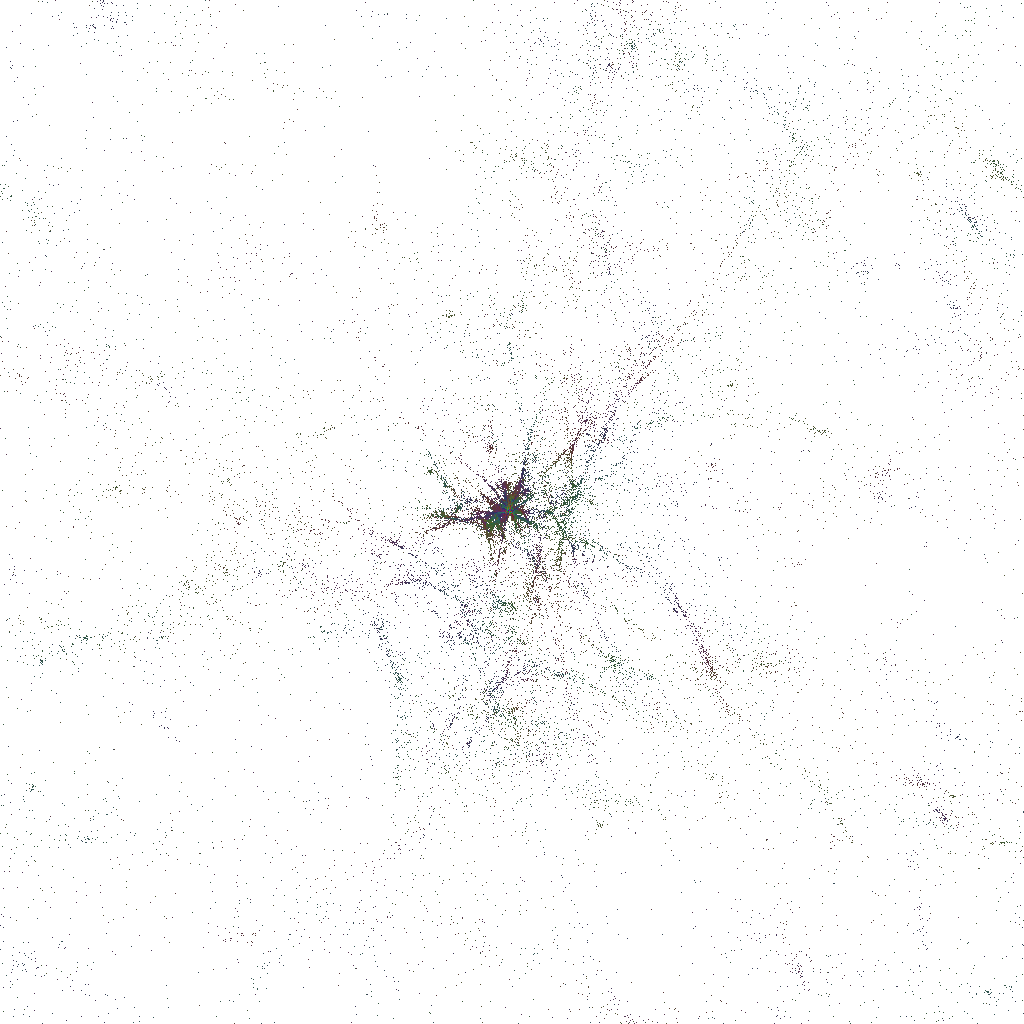}\\
\caption{
Density maps in $\tilde \ph$ coordinates (for the definition see Section \ref{sub:maps}) projected on the $x-y$ plane for configurations in phase $C$ ($\kappa_0=2.2, \Delta=0.6$) with $T=4$ and with dynamical scalar fields with jump in one spatial direction (left) and in three spatial directions (right). Top: configurations with a small jump magnitude ($\delta=0.1$). Bottom:  configurations with a large jump magnitude ($\delta=1.0$).}
\label{fig:jumps_C}
\end{figure}

Using the classical scalar field solutions as coordinates, one can also measure the density maps defined in Section \ref{sub:maps} and observe if and how they are affected by dynamical scalar fields. 
figure~\ref{fig:VolJumpInSpaceTX} presents the density maps projected on the $t-x$ plane, and 
figure~\ref{fig:jumps_C} shows the density maps projected on the $x-y$ plane. The system has three scalar fields with either one jump in the $x$ direction only (left-hand side charts) or three jumps 
in all three spatial directions (right-hand side charts). For small jump magnitudes (top plots), one observes in all directions the cosmic void and filament structures, which look qualitatively the same as in the pure gravity case (see figure~\ref{fig:cosmic_C} for comparison). For large jump magnitudes (bottom plots), the density maps qualitatively change as the geometry gets effectively compressed to a single outgrowth in all spatial directions (as already discussed, for $T=4$ the time direction is not compressed), the effect visibly increasing in strength with the number of scalar fields with a jump. These results are easily explicable by the pinching phenomenon discussed above.

\begin{figure}[H]
\centering
\includegraphics[height=5.5cm]{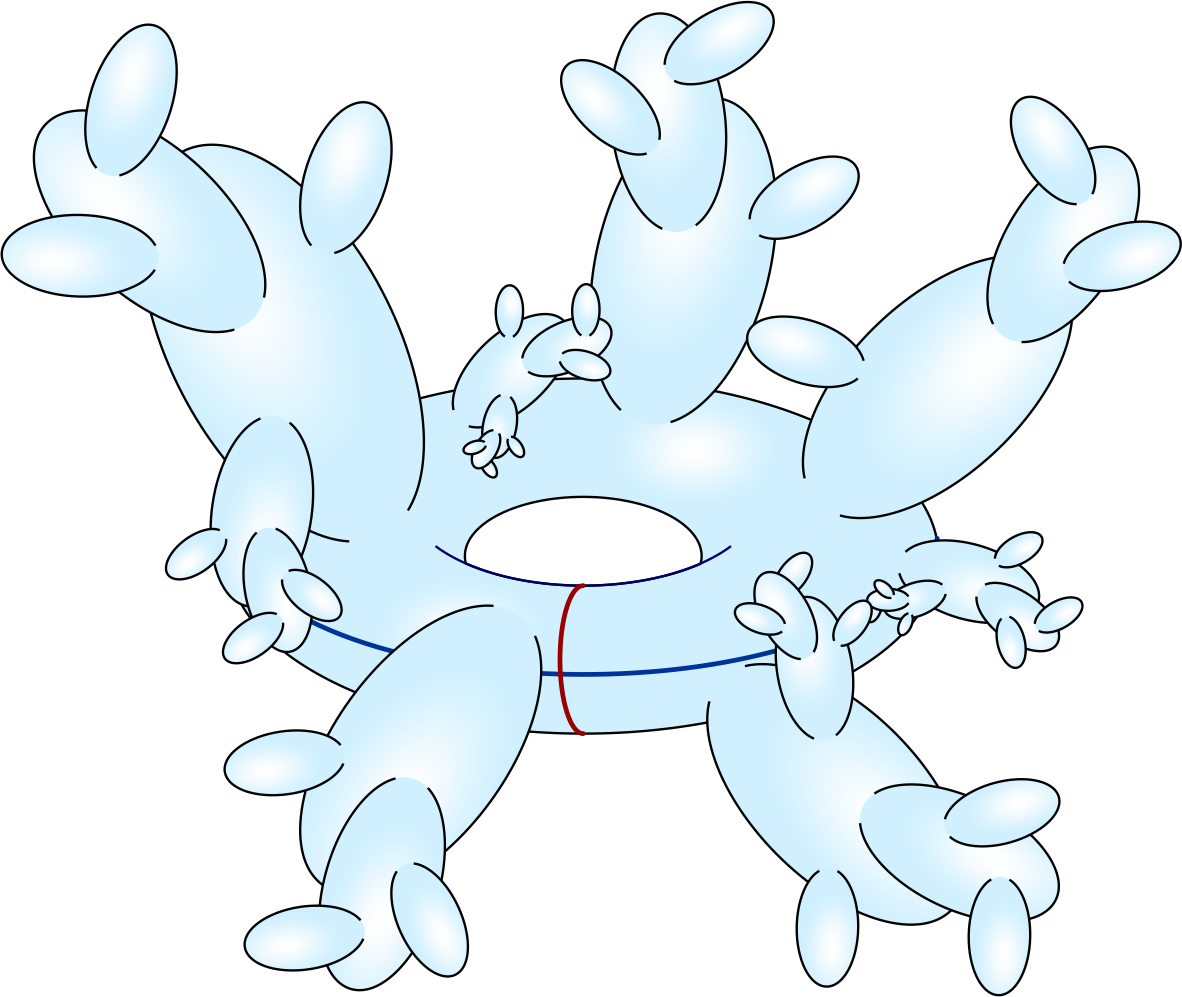}
\includegraphics[height=5.5cm]{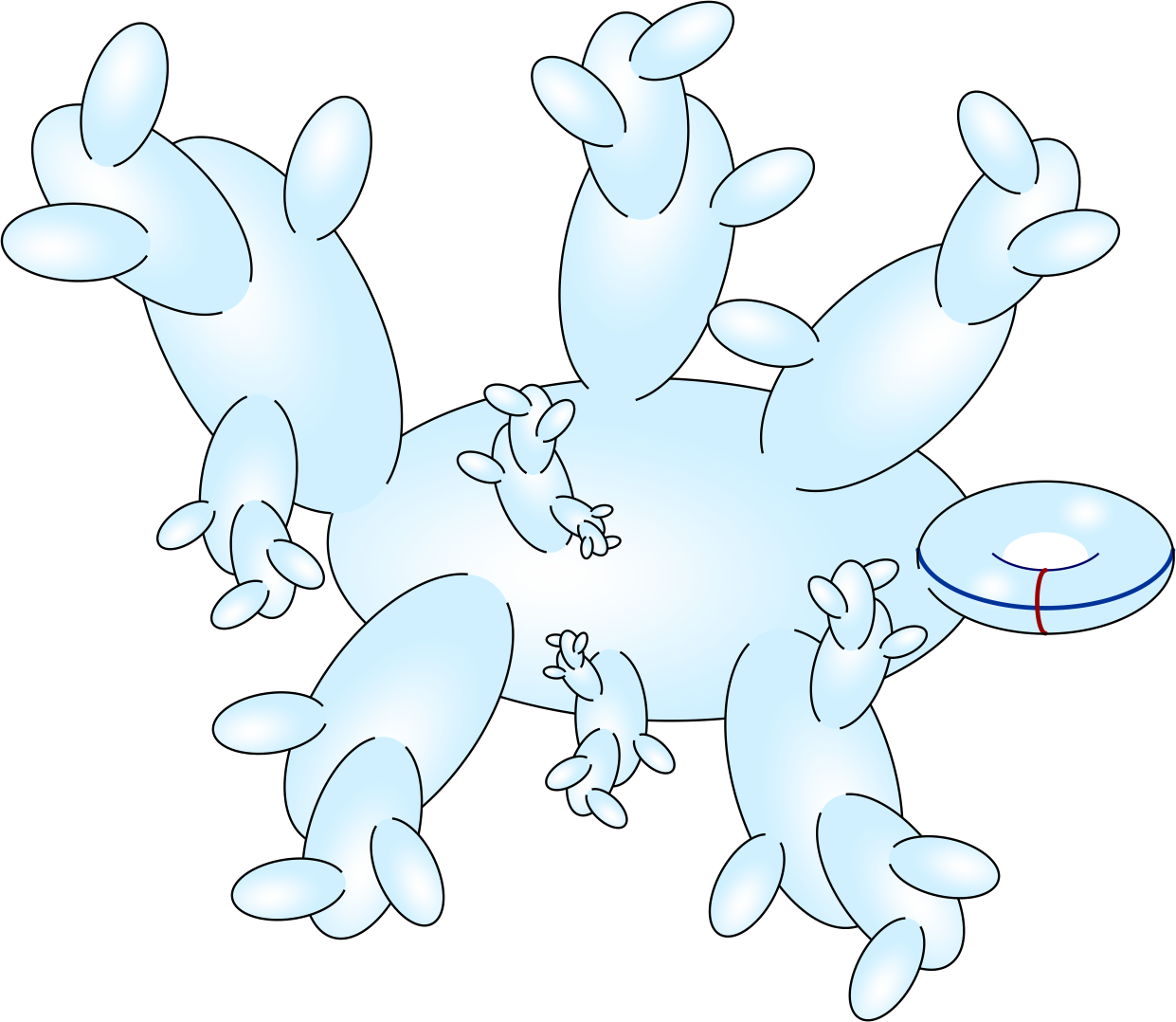}
\caption{Cartoon 2D pictures representing the generic features of CDT quantum geometries for the pure gravity case / a small jump magnitude (left) and for a large jump magnitude (right).}\label{fig:ballon}
\end{figure}

To illustrate this, let us analyze a simple 2D example, where a fractal geometry can be compared to a toroidal balloon with outgrowths, as shown in figure~\ref{fig:ballon}. 
For the pure gravity case (and for a small jump magnitude), the geometry typically looks like in the left plot with a large central part and a number of relatively small outgrowths. The scalar fields with large jump magnitudes compress the central part, where (almost) all change of the field occurs, and, because of the total volume constraint, transfer the volume into one of the outgrowths, where the field is much more uniform, leading to the picture on the right plot.

\begin{figure}[H]
\centering
\includegraphics[width=0.45\textwidth]{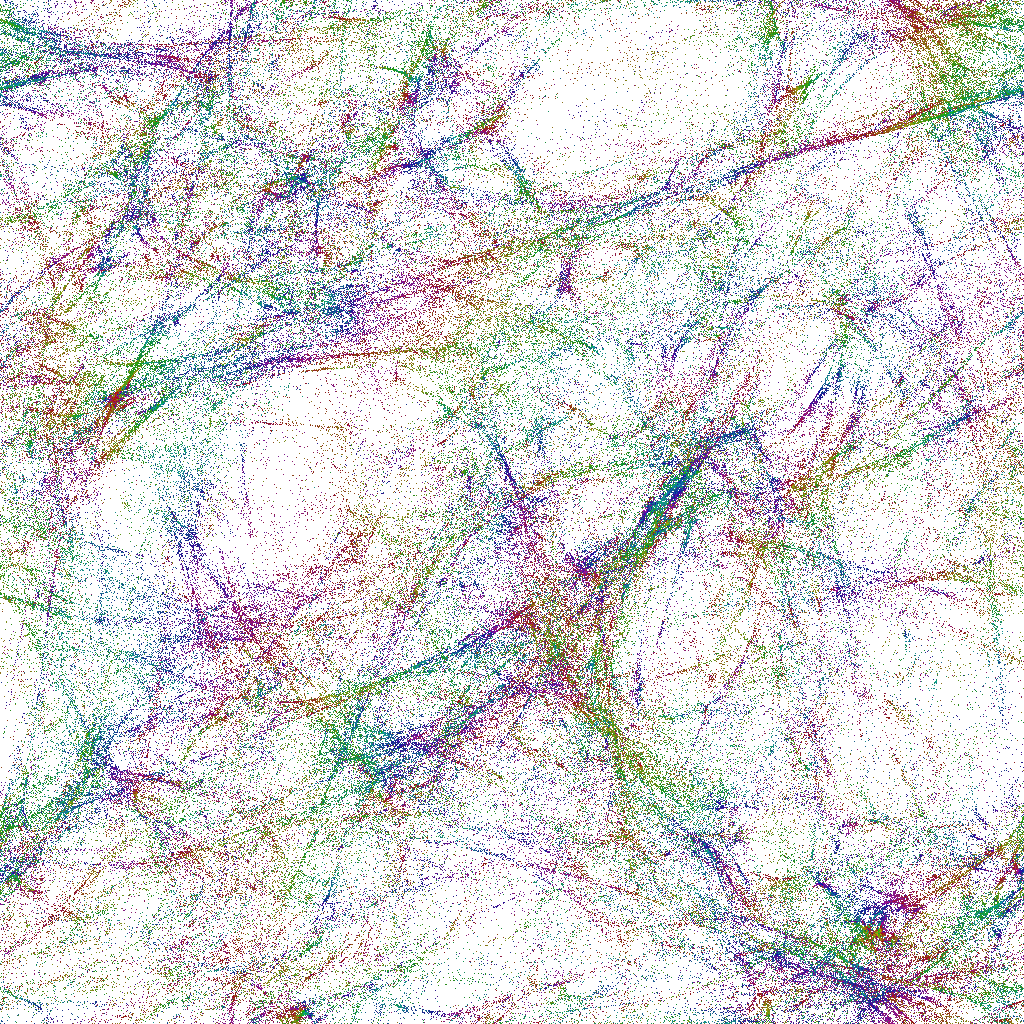}
\includegraphics[width=0.45\textwidth]{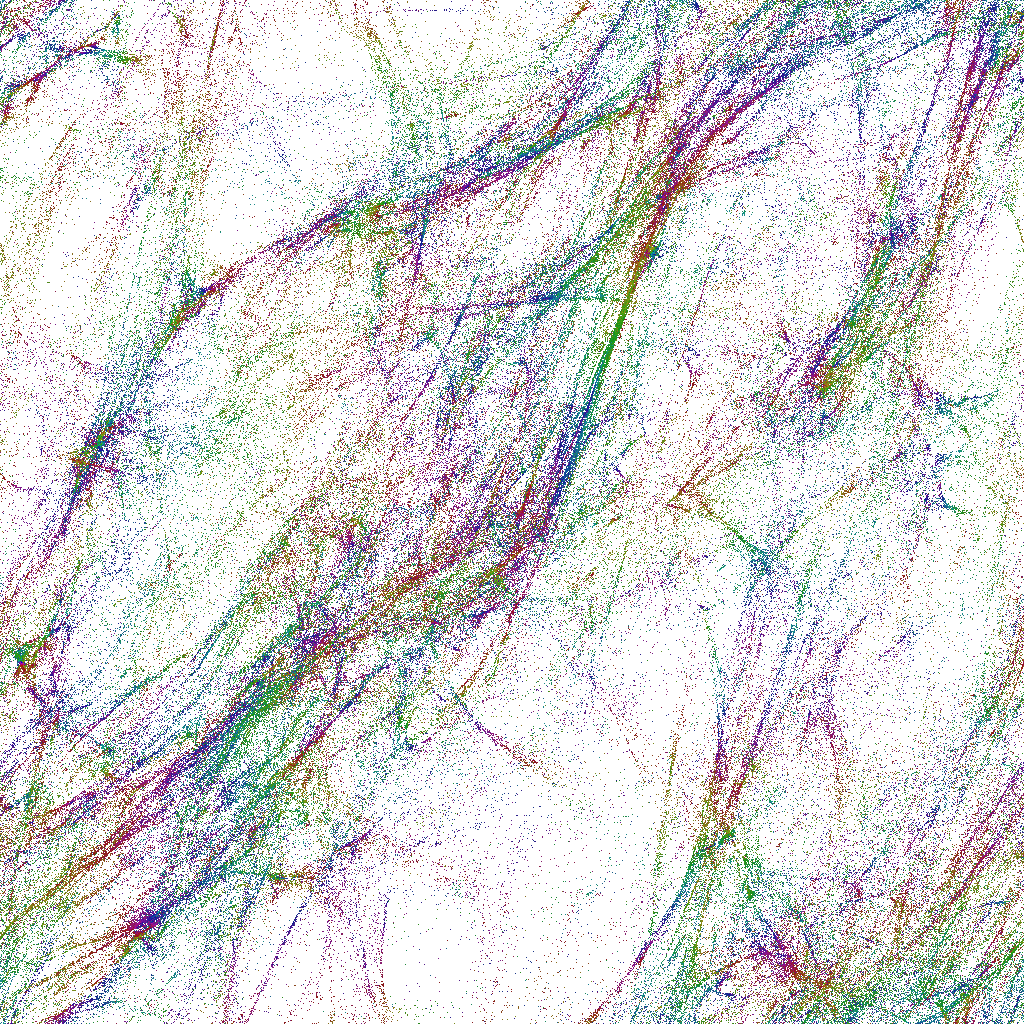} \\
\includegraphics[width=0.45\textwidth]{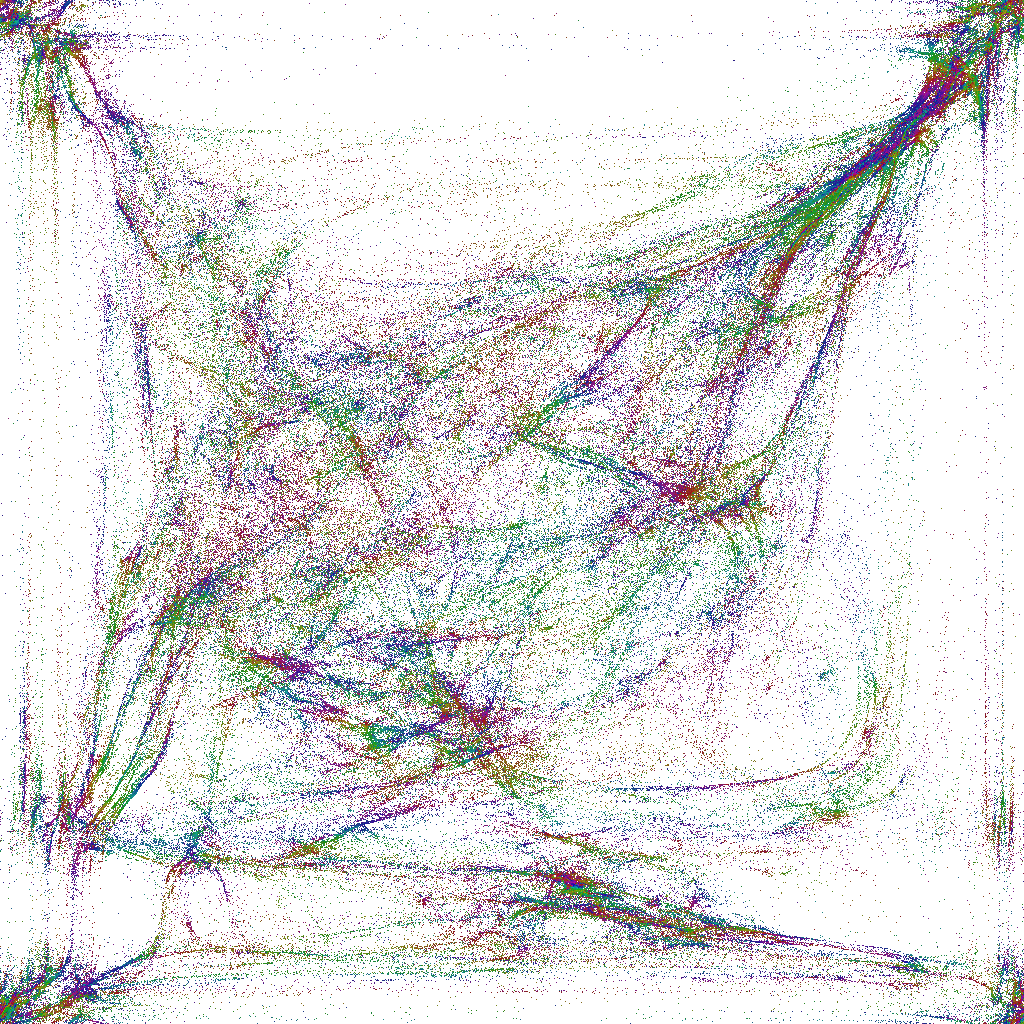}
\includegraphics[width=0.45\textwidth]{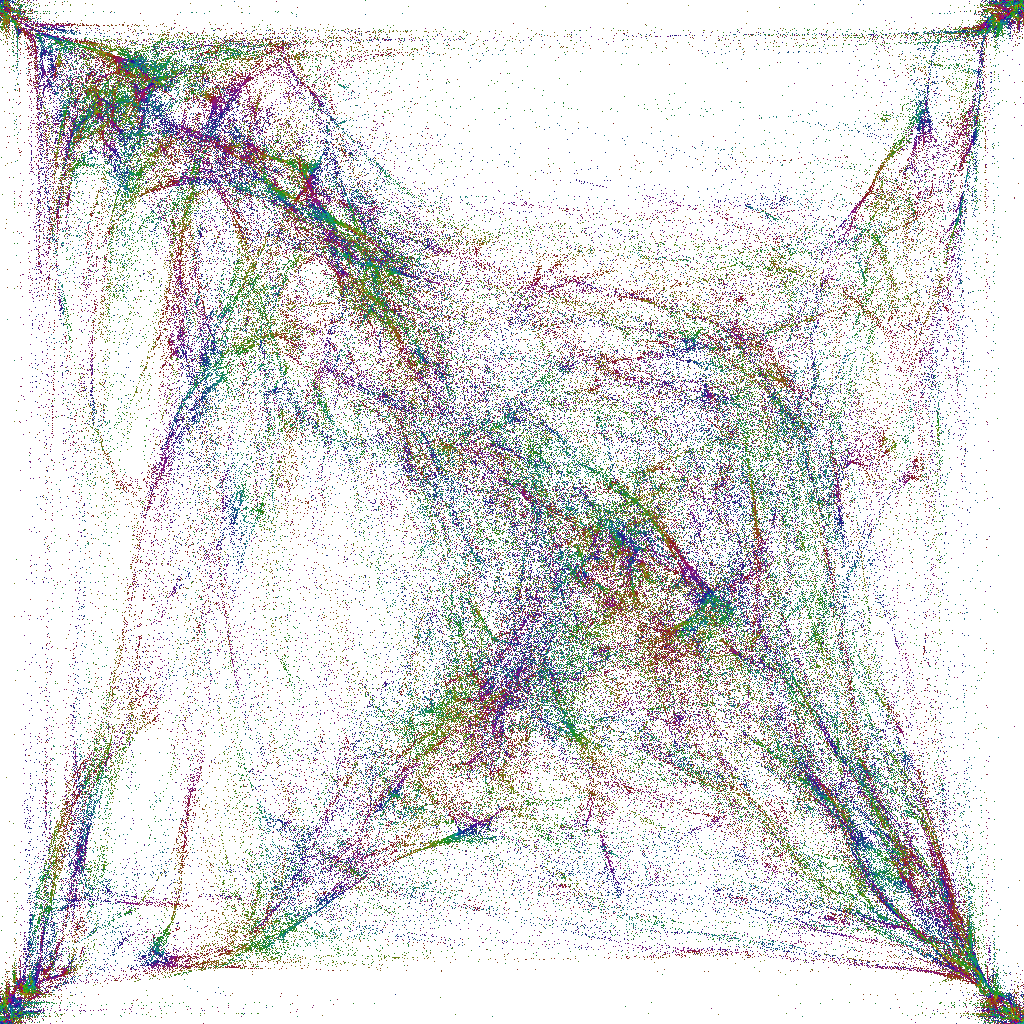}\\
\caption{
Density maps in $\beta$ coordinates (for definition see Section \ref{sub:maps}) projected on the $x-y$ plane for configurations in phase $C$ ($\kappa_0=2.2, \Delta=0.6$) with $T=4$ and with dynamical scalar fields with jump in one spatial direction (left) and in three spatial directions (right). Top: configurations with a small jump magnitude ($\delta=0.1$). Bottom:  configurations with a large jump magnitude ($\delta=1.0$).}\label{fig:Betajumps_C}
\end{figure}

One could na\"ively think that as an effect of the geometry pinching caused by the dynamical scalar fields with (large) jumps, one would obtain a compactified geometry similar to the geometry of the bifurcation phase $C_b$ or (for even larger jump magnitudes) to a collapsed geometry of the $B$ phase.
Interestingly, this is not the case. {As can be seen in figure \ref{fig:VolJumpInSpaceTX}, for
suffiently large jump size the spherical outgrowth spreads over time, }and the fine structure of the semiclassical phase $C$  geometry survives the pinching effect as is illustrated in figure~\ref{fig:Betajumps_C}, where we show the density maps in $x-y$ directions, now in the $\beta$-coordinates introduced in Section \ref{sub:maps}. In these coordinates, the field condensations get stretched and, as a consequence, the geometric outgrowths, i.e.,  the dense regions in figure~\ref{fig:jumps_C}, get magnified. One can clearly see the very nontrivial internal structure of the outgrowths, again with the {\it cosmic voids and filaments} characteristic for the phase $C$ region. Thus, the internal geometry of the large outgrowths created by the pinching effect of the dynamical scalar fields with jump(s) is now completely different than the (almost) homogeneous geometry of the large outgrowths observed in phases $C_b$ and $B$ (see figures \ref{fig:beta_Bif} and \ref{fig:beta_B} for the pure gravity case). 

\begin{figure}[ht]
\centering
\includegraphics[scale=0.95]{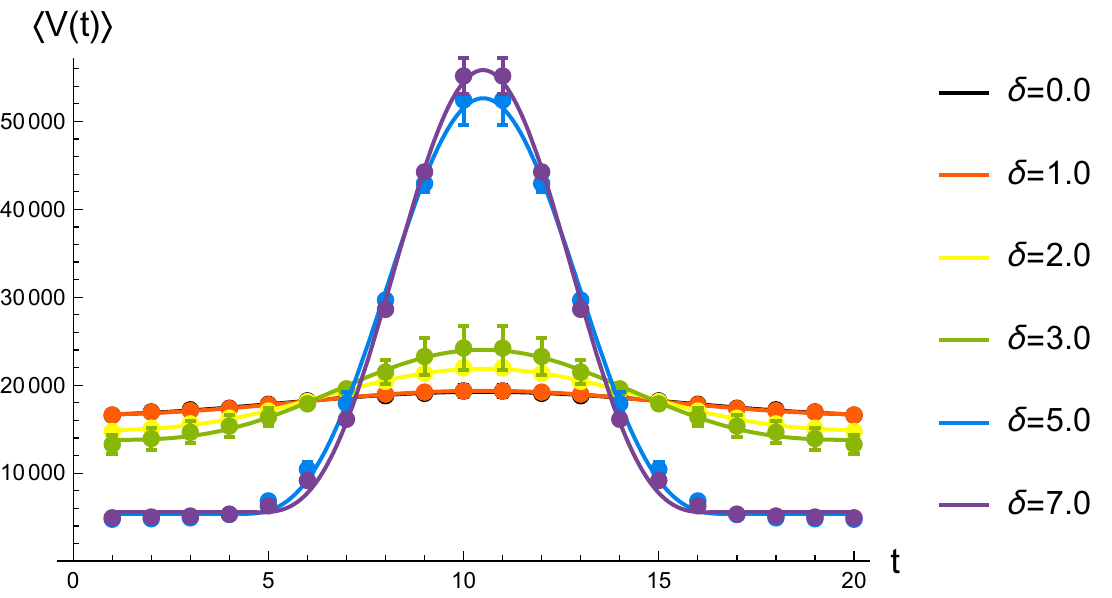}
\caption{Spatial volume t-profiles (averaged over many MC configurations) inside phase $C$ ($\kappa_0=4.0, \Delta=0.2$) for $T=20$ and  $N_{4,1} = 720\rmk$  with dynamical scalar fields with jumps of magnitude $\delta$ in all three spatial directions. Error bars for measured data points were estimated using single-elimination (binned) jackknife procedure. Solid lines are  fits of the function: $\max[c ,c+a\cos^3(b(t-t_0))]$ characteristic for the spherical CDT de Sitter solution.}\label{fig:VolJumpInSpaceBig}
\end{figure}

{To summarize the above results,  numerical MC simulations {performed for $N_{4,1}=160\rmk$ and $T=4$} suggest that coupling quantum geometry to scalar fields with non-trivial boundary conditions can lead to a new type of a phase transition. If spacetime is globally hyperbolic with a toroidal spatial topology, and if the scalar fields have matching topological boundary conditions, then for a sufficiently strong coupling (sufficiently large $\delta$ in our model) one observes a transition leading to an effective change of topology (from a toroidal to a simply connected one). This is the natural extrapolation of what is observed in numerical data presented above and what is schematically illustrated in figure~\ref{fig:ballon}, i.e., the dominating toroidal part with many non-trivially correlated (almost) spherical outgrowths changes into the dominating spherical part with many non-trivially correlated spherical outgrowths and a single toroidal outgrowth of cut-off size (which is needed due to the global topological restrictions imposed). The occurrence of such a phase transition seems to be independent of the number of fields with a jump as each such field pinches geometry in all spatial directions.} {These results are further supported by  analysis of  larger systems with $N_{4,1}=720\rmk$, $T=20$ and three scalar fields with jumps in all spatial directions.\footnote{{These data were measured for a different location of CDT bare couplings in the  $(\kappa_0,\Delta)$ parameter space, but the new location is also inside the semiclassical $C$ phase region.}} Contrary to configurations with small time extent,  spatial volume $t$-profiles are now visibly different for  small and  large values\footnote{{For the larger system, the critical value $\delta_c$ is now larger than for the smaller system discussed before. The terms small / large value mean here $\delta<\delta_c$ or  $\delta>\delta_c$, respectively.}} of the jump magnitude $\delta$, as presented in figure \ref{fig:VolJumpInSpaceBig}, where we plotted $\langle V(t) \rangle$, the $t$-profiles  averaged over many MC configurations. It is remarkable that for $\delta>\delta_c\approx 2.0$, where the pinching, i.e., the phase transition leading to the  effective change of the spatial topology from the toroidal to the spherical one, takes place, one can observe the volume profiles with a 'stalk' and the 'blob' part, exactly as it was observed  in  the pure gravity  spherical CDT,  where spherical spatial topology was put in by hand. What is more, for $\delta\gg \delta_c$ the averaged spatial volume $t$-profiles  $\langle V(t) \rangle$ seem to be quite universal, changing only a little with $\delta$, and, even more remarkably, well fitted by the $cos^3$ curves characteristic for the spherical CDT de Sitter solution observed in phase~$C$. In that case, the difference between the pure gravity spherical CDT (with imposed spherical spatial topology) and the toroidal CDT coupled to scalar fields with jumps (causing the effective spatial topology change)  lies in a different behaviour of the 'stalk' part. In the original spherical CDT, the 3-volume of the 'stalk' was of the cutoff size, and now, in the toroidal CDT with the effective topology change, it is significantly larger. This is partly explained by the size of the minimal three-dimensional toroidal triangulation, which is much bigger than the minimal spherical three-dimensional triangulation \cite{c-phase2}, resulting in much larger cutoff, but in the later case the 3-volume of the stalk is still two orders of magnitude larger than the minimal possible volume of the three-dimensional torus. Probably, the very nontrivial change of the effective spacetime topology: $T^4\to S^3\times T^1$  requires much larger triangulations than the minimal possible ones. At any rate, the existence of the 'stalk' is a discretization / finite size effect related to the fixed spacetime topology conditions imposed in the MC simulations, which cannot change regardless of the effective topology change, and it becomes negligible in the large volume limit. Therefore, the results presented above strongly support our conjecture that the newly observed phase transition leads to an effective spatial topology change.}

\section{Conclusions}\label{sec:conclusions}

The size of a typical CDT universe that can be studied on a computer is no larger than 10-20 Planck's lengths \cite{physrep}. While one could perhaps have expected that all that can be observed at such short scales is just wild quantum fluctuations, in fact this is not the case. The measurement of the spectral dimension indicates a fractal structure of the studied spacetimes \cite{spectral},
the scale-dependent spectral dimension seemingly being a result of the underlying quantum fluctuations, but the scale factor (i.e., the spatial volume profiles as a function of time) of the universe behaves surprisingly semiclassically \cite{c-phase1}. Those results were obtained by averaging over many independent field configurations. Understanding the nature of typical geometries, leading, after performing the average in the path integral, to both semiclassical and quantum phenomena, would be a step towards explaining the nature of quantum gravity (or at least what we can call four-dimensional quantum geometry).

In general, a single configuration in the path integral of a quantum theory is not physical. It can be measured on the computer but not in the real world because of the quantum nature of the theory. What is defined in a quantum theory is a value of an observable suitably averaged over the configurations of the path integral. This does not necessarily mean that a single ``typical'' configuration of the path integral is uninteresting. On the contrary, in some situations and for certain observables, the correct answer (up to finite-size corrections) can be obtained by calculating the value of the observable on a single ``typical'' configuration provided it be sufficiently large to be representative for the whole ensemble.
In principle, both the scale factor and the spectral dimension mentioned above could have been determined that way. Thus, it would be advantageous to understand the nature of an individual configuration in the path integral: it might be used to calculate certain observables even if it does not qualify as an observable itself.

As already mentioned above, CDT configurations are presented to us on the computer as geometries that are coordinate free in the spatial directions.
While this seems desirable from a GR point of view, it is well known that one should be careful what one wishes for. The reason that we were able to construct an effective action for the scale factor was precisely that we had at our disposal a coordinate in the time direction. Indeed, coordinates can be very useful, and in this article we tried to construct them also along the spatial directions in order better to understand the geometry of the configurations and to address the question of formulating an effective action that would include all the spacetime directions.

The geometries we extract from the path integral are not regular in the spatial directions, and it is not clear how to introduce ``good'' coordinates when the topology of the space is that of $S^3$. However, if it is $T^3$, then one can take advantage of the periodic structure of the piecewise linear manifold to introduce three scalar fields satisfying Laplace's equation and use them as spatial coordinates.\footnote{{Such coordinates are a close analogue of the harmonic coordinate condition used in the context of GR, but here we use them for non-classical and highly non-trivial geometries.}} The same can be done in the time direction if the CDT time $t$ is made periodic. The comparison of the time defined by the scalar field with the original $t$ can serve as a check of how well this prescription works.

Our starting point was a path integral triangulation ${\cal T}$ with four non-contractible hypersurfaces, the so-called boundaries, labelled by $x,y,z$ and $t$ and impossible to be continuously deformed into each other. The $t$ hypersurface was chosen as the spatial slice corresponding to some value $t_0$ of the CDT time $t$. Basing on these hypersurfaces, we found four harmonic maps $\tilde \phi^\mu_i$, $\mu =x,y,z,t$ from ${\cal T}$ onto $S^1$. These four maps now served as our new coordinates, and constant values of $\tilde \phi^\mu_i = \alpha_\mu$ defined hypersurfaces $H(\alpha_\mu)$.
Using the new $\alpha_t$ coordinate, we defined and measured the volume profiles $V(\alpha_t)$, i.e., the number of tetrahedra in each hypersurface $H(\alpha_t)$, and the volume-volume correlator $C(\Delta \alpha_t)$  between volumes of hypersurfaces whose $\alpha_t$ coordinate differs by $\Delta \alpha_t$, as defined in eq.~\rf{covariance}. The important point here is that the calculations proceed as well when using the $\alpha_t$ coordinate as when using the original $t$ coordinate. The measurement of $C(\Delta \alpha_t)$ is particularly promising since this correlator can be used to reconstruct an effective action (see \cite{physrep} for details). Analogously, we measured the volume profiles $V(\alpha_\mu)$, $\mu = x,y,z$ (see figure~\ref{fig:CXprofile}). The results are encouraging yet not as good as for the $V(\alpha_t)$ profiles. 
As discussed above, the precision is constrained to what can be obtained from a single configuration, since in principle we introduce a new coordinate system for each configuration, but the practicability of making superpositions coming from several configurations is not precluded. This idea, which we have yet to investigate and perhaps couple with generating even larger triangulations, would be especially useful to improve the results in the spatial directions. Anyhow, it would be really exciting to be able to measure the correlators $C(\Delta \alpha_\mu)$, $\mu=x,y,z$ with good precision.

Let us now turn to other observations made using the new harmonic coordinates. As explained in Section \ref{sub:maps}, the use of harmonic coordinates is well suited to record in a density plot the outgrowths of a triangulation. In the case of configurations from phase $C$, which is undoubtedly the most interesting one from the physical point of view, the projections of densities to $\mu - \nu$ planes (figure~\ref{fig:cosmic_C}) show what we denote, because of the visual similarity to pictures of the well-known structures in the real Universe, as cosmic voids and filaments. In our computer-generated spacetimes, the filaments are not matter content but regions where some of the harmonic fields $\phi_i^\mu$ vary slowly. In terms of geometry, those regions can most likely be associated with outgrowths sharing a small boundary with the rest of the triangulation. However, the fact that they have a filament structure instead of being randomly conglomerated indicates structures of a certain ``duration'' rather than what is shown in figure~\ref{fig:outgrowths} and realized in 2D Liouville quantum gravity \cite{ab}. This ``duration'' is particularly pronounced in the time direction in the upper left picture of figure~\ref{fig:cosmic_C}. That this situation is nontrivial (and not fully understood) is illustrated by plotting the same configurations in the $\beta$-coordinates rather than the $\alpha$-coordinates. As readily seen in figure~\ref{fig:beta_C}, a filament structure persists, despite the fact that the $\beta$-coordinates were specifically designed to be complementary to the $\alpha$-coordinates and thus sensitive to possible outgrowths.

The classical scalar fields $\phi_i^\mu$ used as coordinates do not influence the geometry of the manifold (the triangulation) on which they are defined, but their important aspect, which makes them independent of the hypersurface used to define them, is that they were mapped to $S^1$ and not to $\mathbb{R}$. Let us then turn to the examination of a genuine dynamical matter-gravity system, where the scalar field can influence the geometry. As mentioned in the introduction to Section \ref{sec:dynamical}, we did not observe a substantial effect on the geometry when we studied ordinary scalar fields, taking values in $\mathbb{R}$, coupled to gravity. This may be surprising since matter is supposed to have a dramatic effect on geometry in GR, but we have to remember that the configurations are Wick-rotated to Euclidean spacetimes, where gravity in some sense is repulsive, and also that, e.g., black hole solutions are completely regular solutions to Einstein's equations, and the mass $M$ appears in them just as a parameter. However, what we observe if we compel the scalar field to take values in $S^1$ and to wind around $S^1$ when moving around a non-contractible loop on the manifold (the triangulation) where it is defined is that the matter action is minimized if the geometry of the manifold deforms in such a way that it is almost pinched, and the scalar field makes all its winding just when passing the pinch, as explained in Section  \ref{sec:dynamical}. In the path integral, there is a competition between the matter action and the geometric Regge (Einstein-Hilbert) action, which in turn is minimized for non-pinched geometries. The result seems to be a phase transition occurring when the change of the scalar field winding around $S^1$ is forced to be sufficiently large. In the new phase, the geometry is ``squeezed'' in some regions. This kind of squeezing can lead to an effective topology change from a toroidal to a simply connected one. The precise nature of this phase transition is still unknown but clearly interesting to investigate since it is the first phase transition in higher-dimensional CDT caused by matter.

\section{Acknowledgements}

J.G.-S. acknowledges support of the grant UMO-2016/23/ST2/00289 from the National Science Centre Poland. J.J. acknowledges support from the grant 2019/33/B/ST2/00589 from National Science Centre Poland. Z.D. acknowledges support from the grant 2019/32/T/ST2/00390 from National Science Centre Poland (NCN). D.N. acknowledges support from NCN grant 2019/32/T/ST2/00389.

\begin{appendices}
\renewcommand{\theequation}{A-\arabic{equation}}
\setcounter{equation}{0}

\section*{Appendix 1: Harmonic functions and dipole sheets }
\label{sec:dipolesheets}

Let us consider $n$-dimensional flat space, $\mathbb{R}^n$. 
The dipole moment of two opposite point charges $\pm q$ is defined as $\delta^{\mu} = q d^{\mu}$ where $d^\mu$ is the vector between the two point charges. The dipole limit is obtained when $q$ goes to infinity and the length of $d^{\mu}$ goes to zero keeping $\delta^{\mu}$ fixed. 
A dipole sheet is a hypersurface $S$ with a surface dipole density $\delta(s)$,
i.e., to an infinitesimal area $\rmd S$ centered at any point $s^{\mu}$ on the surface corresponds the dipole moment given by $\rmd \delta^{\mu} (s) = \delta(s) n^{\mu} (s) \rmd S$. Let us write Poisson's equation in the form 
\begin{equation}\label{ja1}
\Delta_x \phi (x) = -\rho(x),\quad \phi(x) = \int \rmd^n y \; G(x,y) \rho(y),\quad \Delta_x G(x,y) = -\delta^n(x-y).
\end{equation}
Here $G(x,y)$ is defined for $n > 2$ as the Green function that goes to zero as $|x-y|$ goes to infinity. The dipole density is obtained as the limit where the charge density $\rho(y)$ is located in two infinitesimal sheets of charges on the opposite sides of the hypersurface $S$. Let $s^\mu$ be a point at the hypersurface and $n^\mu(s)$ the normal to 
the hypersurface. Then $\rho(s - \epsilon \,n(s)) = -\rho(s + \epsilon\, n(s))$, for $\epsilon$ infinitesimal, and in the dipole limit
 \begin{equation}\label{ja2}
 \rmd^n y \, \rho(y) \, G(x,y) \to \rmd S(s) \,\delta(s)\, n^\mu (s) \frac{\partial}{\partial y^\mu} G(x,y) \Big|_{y = s}, 
 \end{equation}
and from eq.~(\ref{ja1}) we obtain the corresponding dipole potential
\begin{equation}\label{ja3}
\phi (x) = \int_S \rmd S(s) \;\delta(s) \, n^\mu(s) \frac{\partial}{\partial y^\mu} G(x,y)\Big|_{y = s}, 
\end{equation}
where the integral is over the hypersurface $S(s)$. 
An important property of $\phi(x)$, following from the divergence theorem, is that it jumps by the amount $\delta (s)$ when one crosses the surface $S$ at the point $s$ in the direction of the dipole, i.e, in the direction of the normal to the surface $n(s)$.

Let us now consider the case where the space is a torus $T^n$ with volume $V$, and where the hypersurface $S$ is connected and closed. The constant mode is a zero mode of the Laplacian, and to invert the Laplacian it has to be projected out. Thus $\Delta_x G(x,y) = -\delta^n(x-y) +\frac{1}{V}.$ Given a dipole sheet, this $G(x,y)$ will produce a $\phi(x)$ orthogonal to the constant mode. However, $\phi(x)$ itself is only determined up to the constant mode from the defining Poisson equation, (\ref{ja1}), and it is more convenient in the following to fix $\phi(x)$ not by orthogonality to the constant mode but by being zero at a fixed point $x_0$. With this choice, $\phi(x)$ is given by 
\begin{equation}\label{ja4}
\phi (x) = \int_S \rmd S(s) \;\delta(s) \, n^\mu(s) \frac{\partial}{\partial y^\mu} (G(x,y)-G(x_0,y))\Big|_{y=s}.
\end{equation}
Let us now assume that the dipole density $\delta(s)$ is constant. If we deform the hypersurface $S$ in the direction of the normals $n_i(s),~s \in S$, to another hypersurface $S'$ not intersecting $S$ and let $V(S,S')$ denote the enclosed region, then the two potentials $\phi_S(x)$ and $\phi_{S'}(x)$, calculated by (\ref{ja4}) using dipole sheets $S$ and $S'$, respectively, will agree or differ by $\pm \delta$, depending on how $x_0$ and $x$ are located relatively to $V(S,S')$. More precisely, we have 
\begin{equation}\label{ja5}
x,x_0 \in V(S,S')\quad \textrm{or} \quad x,x_0 \notin V(S,S'): \quad \phi_S(x) = \phi_{S'}(x),
\end{equation}
\begin{equation}\label{ja6}
x_0 \in V(S,S'), \quad x \notin V(S,S'): \quad \phi_S(x) = \phi_{S'}(x)-
\delta,
\end{equation}
\begin{equation}\label{ja7}
x_0 \notin V(S,S'), \quad x \in V(S,S'): \quad \phi_S(x) = \phi_{S'}(x)+
\delta.
\end{equation}
This follows from the divergence theorem, which leads to 
\begin{eqnarray*}
&&\phi_S(x) -\phi_{S'}(x) = \\
&& \hspace{4mm}\delta \!\int_{S} \rmd S \; n^\mu {\partial_\mu} \big(G(x,y) - G(x_0,y)\big) 
 -\delta \!\int_{ S'} \rmd S \; n^\mu {\partial_\mu} \big(G(x,y) - G(x_0,y)\big) =\\
&& -\delta \!\int_{ V(S,S')} \hspace{-4mm} d^n z \; \Delta_z \big(G(x,z) \!-\!G(x_0,z)\big) =
 \delta\! \int_{V(S,S')} \hspace{-4mm}d^n z \; \big(\delta^n(x\!-\!z)-\delta^n(x_0\!-\! z)\big).
\end{eqnarray*}
The relation between $\phi_S(x)$ and $\phi_{S'}(x)$ is not only valid in flat space but also for a compact Riemannian manifold since it only depends on the divergence theorem, which for a Riemannian manifold reads (for our purpose):
$\int_S \rmd S(s)\,n^\mu(s) \frac{\partial}{\partial y_\mu} G(x,y)\Big|_{y=s} = \int_{V(S)} d^n z \sqrt{g(z)}\; \Delta_z (x,z)$, where $V(S)$ is the region enclosed by the hypersurface $S$, $\rmd S(s)$ is the volume element on $S$ induced from the metric $g_{ij}(y)$ on the Riemannian manifold, $n^\mu(s)$ is the normal vector to the hypersurface $S$ at $s$, and $\Delta = \frac{1}{\sqrt{g}} \partial_i \sqrt{g} g^{ij} \partial_j, \quad \Delta_x G(x,y) = -\frac{1}{\sqrt{g}} \delta^n (x,y) + \frac{1}{V}.$ Let us now view the field $\phi(x)$ as taking values in $S^1$ with circumference $\delta$ rather 
than in $\mathbb{R}$. We can implement this in a simple way, while still keeping the $\mathbb{R}$ values of $\phi(x)$ by defining 
\beql{ja8}
\phi(x) \equiv \phi(x) + n \, \delta, \quad n \in \mathbb{Z}.
\eeq
We see from eqs.~\rf{ja5}-\rf{ja7} that the redefined $\phi(x)$
is unchanged when we change the boundary, i.e., we have the option of viewing the dipole sheet as unphysical and in fact non-existent, and $\phi(x)$ as a harmonic map (i.e., a function which satisfies Laplace's equation) between our Riemannian manifold and the manifold $S^1$. Our setup for the triangulations considered in the article is a discretization of such a dipole situation. The field $\phi_i$ can be viewed as sitting in the center of each four-simplex $i$. We have a hypersurface $S$ build of tetrahedra $s_{ij}$ shared by four-simplices $i$ and $j$, and the field $\phi_i$ changes to $\phi_j = \delta+ \phi_i$ when we cross from $i$ to $j$ via the hypersurface at $s_{ij}$.
The link connecting the centers of the two four-simplices $i$ and $j$ can be viewed as proportional to the normal $n$ to $S$ at $s_{ij}$, and $B_{ij}$ plays the role of $n\, \rmd S$. Viewing the dipole associated with area element $\rmd S$ as two charges of opposite sign 
separated by a small distance $d$, as in eqs.~(\ref{ja1})-(\ref{ja3}) above, we see that $\delta \cdot b_i = \delta \cdot \sum_j B_{ij}$ can be viewed as the sum of charges associated with the dipoles that cross from the simplex $i$ to the simplices $j$. Then eqs.~(\ref{def:jump_matrix}), 
(\ref{def:jump_vector}) and (\ref{eq:poisson}) are the discretized versions of the continuum eqs.~(\ref{ja1})-(\ref{ja3}), and the solution $\bar{\phi}_i$ is the discretized version of $\phi(x)$ in (\ref{ja4}) on a Riemannian manifold. It is remarkable that the discretized versions of eqs.~(\ref{ja5})-(\ref{ja7}) are still valid on a triangulation without a need to take a continuum limit. 

\section*{Appendix 2: Solution of the discrete Laplace equation }
\label{sec:technicalities}

In this section, we describe the technical issues related to solving the discrete Laplace equation
(\ref{eq:poisson})
\[
	\bL \ph	= b.
\]
Although the computations have to be done for all four scalar fields, each field can be treated separately. Therefore, for simplicity, we will consider a single field $\ph$. Methods applicable for solving (\ref{eq:poisson})
must be suitable for sparse matrices because of the large size of the considered Laplacian matrix. They can be divided into two basic types:
\emph{direct} methods and \emph{iterative} methods.
Below we describe the methods of both types.
Wherever possible, all methods used gave similar results up to the machine precision.

Following equation (\ref{eq:phi_mean_b}), we tested the accuracy of the computed solution
by calculating the residual sum of squares,
\[ \mathrm{RSS}[\ph] \equiv \sum_i \left(\ph_i - \bar{\ph}_i\right)^2, \quad \mathrm{where} \ \bar{\ph}_i \equiv \frac{1}{5} \left(b_i + \sum_{j \to i} \ph_j \right). \]
For a perfect solution, $\mathrm{RSS}[\ph] = 0$, by definition.

\subsection*{A2.1. Direct methods}
\label{sub:direct_methods}

\paragraph{The Cholesky decomposition.}

After the modification (\ref{eq:laplacian_mod}),
the Laplacian matrix $\bL$ becomes a real positive-definite symmetric matrix and can be decomposed into the product
\begin{align}\label{eq:cholesky_decomposition}
\bL = \bP^{T} \cdot \bH \cdot \bH^{T} \cdot \bP,
\end{align}
where $\bH$ is a lower-triangular matrix and $\bP$ is a permutation matrix.
{This is} known as the Cholesky decomposition.
The permutation increases the sparsity of $\bH$.
The system of linear equations (\ref{eq:poisson}) can now be solved simply 
by \emph{forward} and \emph{back substitution}.
We used the \texttt{CHOLMOD} library to perform the sparse Cholesky decomposition
\cite{cholmod,suitesparse,Julia-2017}.

Surprisingly, the method was too computationally and time consuming for configurations in phases $B$ and $C_b$
but did particularly well in phases $A$ and $C$.
On the other hand, the iterative methods described below did not work so well in the $A$ phase.

\subsection*{A2.2. Iterative methods}
\label{sub:iterative_methods}

We tested various iterative methods and obtained the best results, both from the point of view of speed and accuracy, for a method that we called \emph{Parallel Preconditioned Conjugate Gradient method with Symmetric Successive Over-Relaxation and Approximate Inverse} (PPCG-SSOR-AI).

\paragraph{Conjugate gradient method.}

The \emph{conjugate gradient} method (CG) {was} designed for solving \emph{symmetric positive-definite} linear systems. Theoretically, it is a direct method, however, it is very sensitive to round-off errors
and is often used as an iterative method since it provides monotonically improving approximations to the exact solution. At each step, the approximate solution is improved by searching for a better solution in \emph{the conjugate gradient direction},
which is $\bL$-orthogonal to all previous search directions (thus avoiding repeated searches).
The conjugate gradient method {usually} converges much faster than standard iterative methods,
such as \emph{Jacobi's method}, \emph{Gauss–Seidel method}, or \emph{successive over-relaxation}.

\paragraph{Preconditioned conjugate gradient method.}

Unfortunately, the problem to be solved is \emph{ill-conditioned}, i.e., the condition number of matrix $\bL$ is large, $\kappa(\bL) = \frac{\vert\lambda_{\mathrm{max}}(\bL)\vert}{\vert\lambda_{\mathrm{min}}(\bL)\vert} \gg 1$. The idea of preconditioning is to substitute the original problem $\bL \ph = b$ with a preconditioned system \[ \bC^{-1} \bL \ph = \bC^{-1} b \] that has the same solution and much lower condition number. A particular choice of a preconditioner is the so-called symmetric successive overrelaxation (SSOR),
\[
\bC		= \left( \frac{\bD}{\omega} + \bH \right) \frac{\omega}{2 - \omega} \bD^{-1} \left( \frac{\bD}{\omega} + \bH^T \right),
\]
where $\bD$ and $\bH$ are the diagonal and lower-triangular parts of $\bL$, respectively, with  $\bL = \bH + \bD + \bH^T$. The preconditioner is chosen such that $\kappa(\bC^{-1} \bL) \ll \kappa(\bL)$
(i.e., $\bC \approx \bL$) and $\bC x = b$ can easily be solved. Calculating $x = \bC^{-1} b$ can be done using \emph{forward} and \emph{back substitution},
hence the name successive relaxation;
and since $\bC$ has a symmetric form and $\omega$ can be different from $1$,
the preconditioner is named symmetric successive overrelaxation.

\paragraph{Parallel preconditioned conjugate gradient method with symmetric successive over-relaxation and approximate inverse.}

The preconditioned version is much more stable than the original conjugate gradient method,
but cannot easily be parallelized. To solve this issue, the method can be further improved by approximating the inverse of the preconditioner $\bC^{-1}$.
For $\bD = \bI$ (we normalize the Laplacian matrix) and $\omega = 1$, we have
\begin{align}
	\bC 		&= \left( \bI + \bH \right) \left( \bI + \bH^\tau \right), \nonumber\\
	\bC^{-1} 	&= \left( \bI + \bH^\tau \right)^{-1} \left( \bI + \bH \right)^{-1}, \nonumber\\
	\bC^{-1} 	&\approx \bK = \left( \bI - \bH^\tau \right) \left( \bI - \bH \right).
\end{align}
Now we solve $\bK \bL \ph = \bK b$ using a slightly modified conjugate gradient method.

The PPCG-SSOR-AI method is fully parallelizable but also stable (due to preconditioning) and fast-convergent (conjugate gradient method).
It is also suitable for GPU \cite{ppcgssoraigpu:2012}.
We took advantage of multiple CPU cores and used the \emph{OpenMP framework} to gain a significant boost.

\section*{Appendix 3: Minisuperspace model with pinching }
\label{minisuperspacemodel}

Let us consider the situation where our universe is periodic in the time direction. With the use of the original CDT time coordinate $t$, the 
spatial volume $V(t)$ is now defined at discrete times $t_n$, and there exists a simple effective action describing the average
of $V(t)$ and its fluctuations \cite{semiclassical,c-phase1,c-phase2}. The continuum version of  this action  is very similar to the minisuperspace action
of Hartle and Hawking \cite{hh}, and the leading terms read:
\begin{equation}\label{jax1}
S[V] = \int \rmd t \, \left[ \frac{1}{G} \frac{\dot{V}^2}{V}  + \alpha V^{1/3} +\lambda \, V \right], 
\end{equation}
where $\dot{V}$ denotes the time derivative of $V(t)$. 
Here the discrete time has been replaced by a continuous one. In the Hartle-Hawking minisuperspace action, because of the assumption of
homogeneity and isotropy, the scale factor $a(t,x)$ is a function of time only. In CDT no such 
assumption is made, but nevertheless the functional form of the effective action in terms of $V(t)$ is the same as the 
Hartle-Hawking minisuperspace model if we write $V(t) \propto a^3(t)$. If the spatial topology is $S^3$, then the constant $\alpha$ is 
different from zero, and if the spatial topology is $T^3$, then $\alpha =0$. In both cases there exist corrections to 
the terms shown in \rf{jax1}, but they are small, and we will ignore them. The $\lambda$ in \rf{jax1} is not really
the cosmological constant but a Lagrange multiplier ensuring that the four-volume of the universe is fixed at $V_4$ in 
order to agree with the computer simulations where the total four-volume is kept constant. Furthermore, the time integration
is from $-T/2$ to $T/2$, as the CDT time of the universe is fixed to be $T$, and, finally, periodicity in the 
time direction is assumed, again to agree with the setup of the computer simulations. $G$ can be viewed as proportional to the gravitational
constant. 

We now consider the toroidal case, i.e., $\alpha =0$. Clearly, the minimum of the action is achieved for the constant spatial volume 
profile $V(t) = V_4/T$. Let us now couple a scalar field to the geometry and assume, in the spirit of a minisuperspace
action based on homogeneity and isotropy, that  $\phi$ only depends on $t$. Moreover, we assume that $\phi(t)$ has winding 
number one and changes by $\delta$ when going around the universe in the time direction. A minisuperspace action that
takes that into account can be written as
\begin{equation}\label{jax2}
S[V,\phi] = \int_{T/2}^{T/2}  \rmd t \, \left[ \frac{1}{G} \frac{\dot{V}^2}{V}  + V \, \dot{\phi}^2 +\lambda \, V + \kappa \, \dot{\phi} \right], 
\end{equation}
where $\kappa$ and $\lambda$ are Lagrange multipliers that introduce the constraints for $\phi(T/2)$ to equal $\phi(-T/2) + \delta$ and for the four-volume to be $V_4$, respectively. The corresponding Euler-Lagrange equations are 
\beq\label{jax3}
\frac{1}{G} \left(2\frac{\ddot{V}}{V} - \frac{\dot{V}^2}{V^2}\right)  - \,\dot{\phi}^2 - \lambda  =0, \qquad    \frac{d}{dt}( V\, \dot{\phi}) =0.
\eeq
They are easily solved by introducing $f(t) = \sqrt{V(t)}$, and the first integrals are 

\beq\label{jax4}
 V\,  \dot{\phi} = K_1\qquad \frac{\dot{V}^2}{G \,V} + \frac{K^2_1}{V} + \lambda V= K_2.
 \eeq
The only twice differentiable periodic solutions for $V(t)$ and $\phi(t)$ where $\phi(T/2) = \delta + \phi(-T/2)$ and where $V(t) > 0$ for all $t$ are of the form 
\beq\label{jax5}
V(t) = \frac{V_4}{T},\qquad \phi(t) = \const + \delta\cdot t/T,
\qquad S[V,\phi] = \delta^2 \frac{V_4}{T^2},
\eeq
except for $\delta = 2\pi n /\sqrt{G}$ where there are additional solutions. For simplicity we consider here 
only the case $n=1$:
\bea\label{jax6}
V(t) &=& a -b \cos \Big( 2\pi \, t/T\Big),\qquad 
a =\frac{V_4}{T} > |b|, \\
\phi(t) &=& \frac{\delta}{\pi}
\left( \arctan \left[ \sqrt\frac{{a+b}}{{a-b}}\; 
\tan \Big( \frac{\pi t}{T} \Big) \right] + \phi(-T/2)\right).
\label{jax6b}
\eea
For any $b$ such that $|b| \leq a$ the value of the action is
\beq\label{jax30}
S_{\rm critical} = \frac{4\pi^2V_4}{G \, T^2}, \qquad \delta = 
\frac{2\pi}{\sqrt{G}},
\eeq
which is the same value one obtains when using in the action the constant solution for $\delta = 2\pi/\sqrt{G}$.
When $\delta > 2\pi /\sqrt{G}$,
\rf{jax6}-\rf{jax6b} is no longer a solution to \rf{jax3}
for $|b| < a$, but 
for $|b| = a$ we have a special situation since $V(t)$ can be zero, for $b=a$ at $t=0$ and for $b=-a$ at $t= \pm T/2$. Let
us consider $b=a$. It is seen from \rf{jax6}-\rf{jax6b} that for $b \to a$ we obtain the solution
\beq\label{jax6a}
V(t) = \frac{V_4}{T}\Big(1 - \cos \Big( 2\pi t/T\Big)\Big),
\qquad \phi(t) = \delta \cdot \theta(t) + \phi(-T/2).
\eeq
The change of $\phi(t)$ is a jump of $\delta$ at $t=0$ where $V(t) =0$. The constant $K_1$ in \rf{jax4} is zero and the term $V(t) \dot{\phi}^2(t)$ in the action \rf{jax2} is identical to zero for all $t$. What is special 
about the situation $a=|b|$ is that \rf{jax6a} is a solution for all values of $\delta$, not only for 
$\delta = 2\pi/\sqrt{G}$, as for $|b| < a$.
The reason for this is that $\phi$ is  decoupled from
$V(t)$ since $V(t) \dot{\phi}^2(t)$ is identically zero, as mentioned. Thus the action is independent of $\delta$ for the solution \rf{jax6a}.

The value of the action for a given configuration (which is not necessarily a solution to eq.~\rf{jax3}) is 
\beq\label{jax7}
S[V,\phi] = \int_{T/2}^{T/2}  \rmd t \, \left[ \frac{1}{G} \frac{\dot{V}^2}{V}  + V \, \dot{\phi}^2\right].
\eeq
For the solutions \rf{jax5} and \rf{jax6a}, which we denote
the constant solution and the ``blob'' solution we have 
\beq\label{jax8}
S[V,\phi]\Big|_{\rm const}=  \delta^2 \,\frac{V_4}{T^2},
\qquad S[V,\phi]\Big|_{\rm blob}=\frac{4\pi^2 V_4}{T^2 G}
\eeq
Thus the constant solution \rf{jax5} has the lowest 
action when $\delta < 2\pi/\sqrt{G}$, while the blob-solution 
has the lowest action (independent of $\delta$) 
for $\delta > 2\pi/\sqrt{G}$.

In our computer simulations we do not allow $V(t) =0$. In fact there is a cut-off $V_\mathrm{min}$, which is the minimum number of tetrahedra needed to build a triangulation of 
a spatial slice $T^3$. Thus, to compare with computer results
we should solve the minisuperspace model with the additional requirement that $V(t) \geq V_\mathrm{min}$. For $\delta < 2\pi/\sqrt{G}$ \rf{jax5} is the solution. For 
$\delta > 2\pi/\sqrt{G}$ we have a generalized solution, which is a combination of the constant $V(t)$ like in \rf{jax5} and the ``blob''
$V(t)$ as in \rf{jax6}. Write 
\beq\label{jax11}
\delta = \delta_{\rm blob}+\delta_{\rm const}, \qquad 
\delta_{\rm blob}= \frac{2\pi}{\sqrt{G}}, \qquad
\delta_{\rm const} = \delta-\delta_{\rm blob} = 
\delta-\frac{2\pi}{\sqrt{G}}.
\eeq
We now use 
\bea\label{jax9}
V(t) &=& 
\tilde{a}- \tilde{b} \cos \Big( \frac{2\pi (|t|-\tau/2)}{\tilde{T}}\Big), \qquad 
\frac{\tau}{2} \leq |t| \leq T/2, 
\qquad \tilde{T} = T-\tau\\
V(t) &=& V_\mathrm{min}= \tilde{a}-\tilde{b}, \qquad
\tilde{a} \tilde{T} = V_4 - \tau V_\mathrm{min} \qquad |t| \leq \frac{\tau}{2}.\label{jax10}
\eea
In principle we could have used any $V \in [V_\mathrm{min}, V_4/T]$ in the Ansatz
\rf{jax9}-\rf{jax10}. However as will be clear from the solution, the 
corresponding action will be decreasing with decreasing $V$, 
and we have thus 
chosen the smallest possible $V$, i.e.\ $V_\mathrm{min}$, from the beginning.
The solution has a ``stalk'' of time extent $\tau$ and spatial volume $V_\mathrm{min}$, located  around $t=0$. This $V(t)$ satisfies 
\rf{jax3} except in the points $t=\pm \tau/2$ where 
$\ddot{V}(t)$ jumps. However, $\dot{V}(t)$ is continuous and
one still has the first integrals \rf{jax4}, with 
different $K_2$'s in the two regions, but the same $K_1$
which should then be used to calculate $\phi(t)$ and thus 
$\delta_{\rm blob}$ and $\delta_{\rm const}$. We find
\beq\label{jax12}
\delta_{\rm blob} = 
\frac{K_1 \tilde{T}}{\sqrt{{\tilde a}^2-{\tilde{b}}^2}}= 
\frac{2\pi}{\sqrt{G}},\qquad \quad
\delta_{\rm const}= \frac{K_1 \tau}{\tilde{a}-\tilde{b}}
= \frac{2\pi}{\sqrt{G}}\; 
\sqrt{\frac{\tilde{a}+\tilde{b}}{\tilde{a}-\tilde{b}} }
\; \frac{\tau}{\tilde{T}}.
\eeq
We thus obtain
\beq\label{jax13}
\left(\delta - \frac{2\pi}{\sqrt{G}}\right)^2 = 
\frac{4\pi^2}{G} \; \frac{\tau^2}{{\tilde{T}}^2} \; 
\frac{\tilde{a}+\tilde{b}}{\tilde{a}-\tilde{b}}
\qquad {\rm or} \qquad \bar{\delta}^2 \bar{V} = 
\frac{\bar{\tau}^2}{(1-\bar{\tau})^3} 
\Big( 2-\bar{V} (1+\bar{\tau}) \Big), 
\eeq
where we have introduced the dimensionless quantities 
\beq\label{jax13a}
\bar{\tau} = \frac{\tau}{T},\qquad 
\bar{\delta}= 
\frac{\delta-\frac{2\pi}{\sqrt{G}}}{\frac{2\pi}{\sqrt{G}}},
\qquad \bar{V} = \frac{T\,V_\mathrm{min} }{V_4}
\eeq
For given $\delta$, $V_4$, $T$ and $V_\mathrm{min}$ this is a third
order equation for $\bar{\tau}$, the extension of the stalk. 
Rather than giving the general solution, let us just give lowest order expression in $\bar{\delta}$ and $\bar{V}$:
\beq\label{jax14}
\bar{\tau}  = \bar{\delta} \;\sqrt{\bar{V}/2} \Big( 1+
 O\big(\bar{V},\bar{\delta} \sqrt{\bar{V}}\big)\Big)
\eeq
The qualitative results are thus as follows:
the smaller $\bar{V}$, the smaller $\bar{\tau}$ and 
$\bar{\tau} \to 0$ in the limit where $\bar{V} \to 0$ and 
we recover \rf{jax6a}. For fixed $\bar{V}$ and increasing 
$\bar{\delta}$, $\bar{\tau}$ will increase, starting at 
$\bar{\tau} =0$ for 
$\bar{\delta} =0$, i.e.\ $\delta= 2\pi/\sqrt{G}$, and for 
$\bar{\delta}\to \infty$ $\bar{\tau} \to 1$, i.e.\ the stalk basically covers the whole $t$-range and the ``blob'' becomes very narrow and very high. This is qualitatively in agreement with what we observe in the actual Monte Carlo simulations.

\begin{figure}[t]
\centering
\includegraphics[width = 0.7\textwidth]{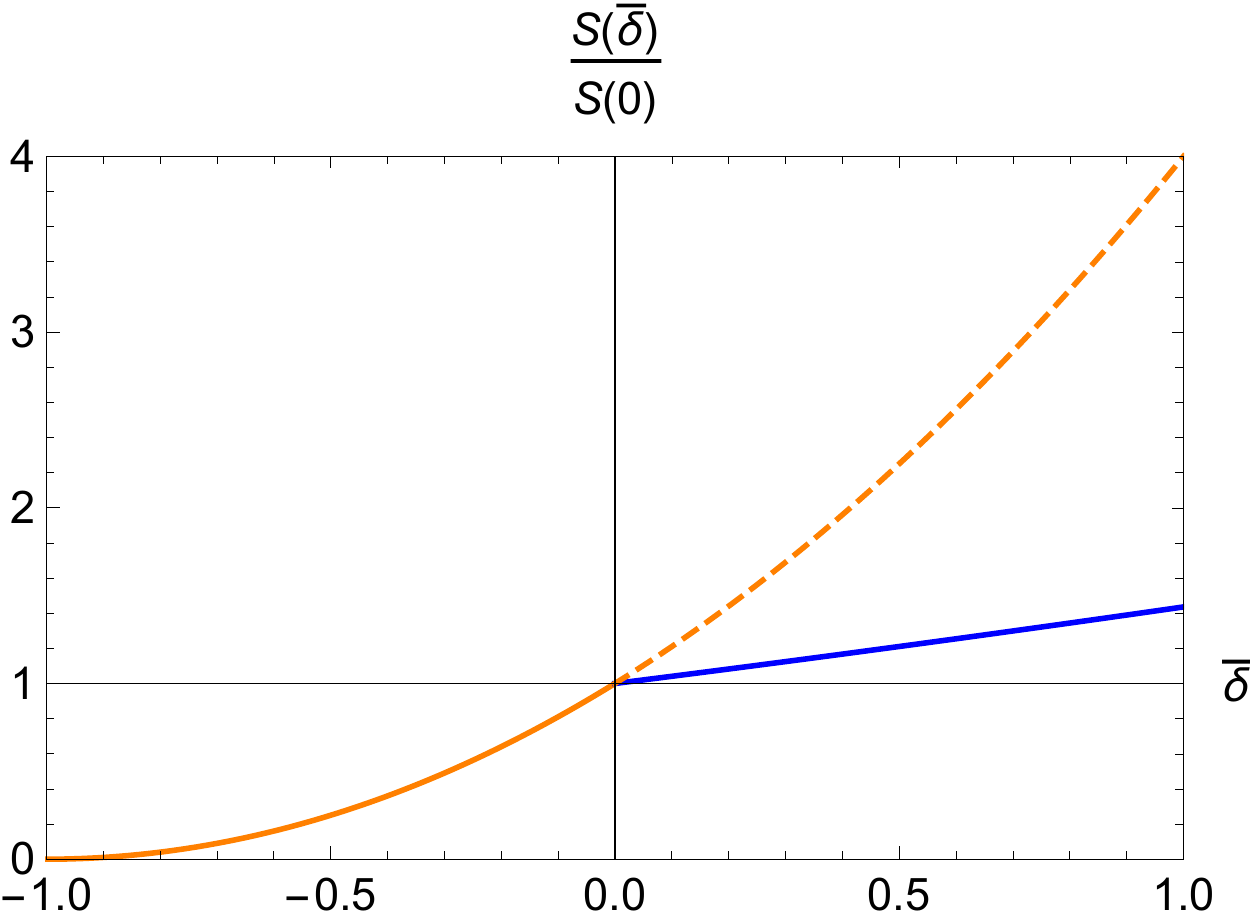}
\caption{$S[\bar{\delta}]/S[0]$ plotted as a function of $\bar{\delta}$. The orange curve is the constant solution, (the dashed part for $\bar{\delta} \geq 0$), while the blue curve shows the action \rf{jax15} for $\bar{\delta} \geq 0$ and $\bar{V} = 0.02$. The smaller is $\bar{V}$, the more horizontal the curve will be, and in the limit $\bar{V}\to 0$ the curve is the constant 1 and the solution $V(t)$ is precisely \rf{jax6a}. }\label{Plotjan}. 
\end{figure}

The action of the solution \rf{jax9}-\rf{jax10} follows 
from \rf{jax8}:
\beq\label{jax15}
S[V,\phi] = \delta_{\rm blob}^2 
\frac{V_4 -\tau V_\mathrm{min}}{\tilde{T}^2} +
\delta_{\rm const}^2 \frac{\tau V_\mathrm{min}}{\tau^2}=
\frac{4 \pi^2}{G} \; \frac{V_4}{T^2} 
\left[ 
\frac{1 +\bar{\tau}-
2\bar{V}\bar{\tau}}{(1-\bar{\tau})^3}\right],
\eeq
where $\bar{\tau}$ is a function of $\bar{\delta}$ and $\bar{V}$ given by \rf{jax13} or \rf{jax14}. If we consider 
$V_\mathrm{min}$ as fixed $S[V,\phi]$ becomes a function of 
$\bar{\delta}$, and we have (to lowest order in 
$\bar{\delta} > 0$ and also assuming $\bar{V} \ll 1$)
\bea\label{jax16}
S[\bar{\delta}] &=&S[0]\;\Big(1-\bar{\delta}\Big)^2 
\quad  {\rm for } \quad -1 \leq\bar{\delta} \leq 0,
\quad S[0]=S_{\rm critical}\\
S[\bar{\delta}] &=&S[0]\; \Big( 1+ \sqrt{8 \bar{V}} \; 
\bar{\delta}+ O(\bar{\delta}^2)\Big)\quad {\rm for}\quad \bar{\delta} \geq 0.
\label{jax16a}
\eea
The behavior of $S[\bar{\delta}]$ is shown in 
figure\ \ref{Plotjan}. First we note that for $\bar{\delta} > 0$ it is 
an increasing function of $\bar{V}$. As already mentioned this is the 
reason we, from the beginning, used the value $V_\mathrm{min}$ in the Ansatz 
\rf{jax9}-\rf{jax10}. While the curve for $S[\bar{\delta}]$ looks approximately linear for $\bar{\delta} > 0$ on the plot, this ceases 
to be true for large $\bar{\delta}$ where we have 
\beq\label{jax40}
S[\bar{\delta}] =   \frac{V_\mathrm{min}}{ T} \; \delta^2 + 
O\Big( \delta^{4/3}\Big),\qquad \delta \gg \frac{2\pi}{\sqrt{G}}.
\eeq
The leading contribution in \rf{jax40} comes from the stalk, which for large
$\delta$ fills almost all the $t$-range and is precisely of the 
form given in \rf{jax5}, except that $V_4/T$ has been replaced by $V_\mathrm{min}$. Also the squeezed ``blob'' has 
an action going to infinity with increasing $\delta$, but only as 
$\delta^{4/3}$.

The derivative of $S[\bar{\delta}]/S[0]$ with respect to $\bar{\delta}$ jumps at 0 from the value 2 to the much smaller value $\sqrt{8\bar{V}}$.
Consequently the simple minisuperspace model predicts a  first order 
phase transition as a function of $\bar{\delta}$.

\end{appendices}

\end{document}